\documentclass[12pt,a4paper]{article}
\pdfoutput=1

\usepackage{amsmath,amssymb}
\usepackage{hyperref}
\usepackage{subfigure}
\usepackage[numbers,compress]{natbib}

\usepackage{multirow}
\usepackage{bigdelim}

\makeatletter
\font\manfnt=manfnt
\def\Watchout{\@ifnextchar [{\W@tchout}{\W@tchout[1]}}
\def\W@tchout[#1]{{\manfnt\@tempcnta#1\relax%
  \@whilenum\@tempcnta>\z@\do{%
    \char"7F\hskip 0.3em\advance\@tempcnta\m@ne}}}
\def\@W@tchout#1{\W@tchout[#1]}
\def\dubious{\@ifnextchar[{\@dubious}{\@dubious[1]}}

\def\@dubious[#1]{%
  \setbox\@tempboxa\hbox{\@W@tchout#1}
  \@tempdima\wd\@tempboxa
  \list{}{\leftmargin\@tempdima}\item[\hbox to 0pt{\hss\@W@tchout#1}]}
\newif\if@preliminary
\@preliminaryfalse
\def\preliminary{\@preliminarytrue}
\def\preprintno#1{\def\@preprintno{#1}}
\def\address#1{\def\@address{#1}}
\def\email#1#2{\thanks{\tt #1@{}#2}}
\def\abstract#1{\def\@abstract{#1}}
\renewcommand\abstractname{ABSTRACT}
\newlength\preprintnoskip
\newlength\abstractwidth
\@titlepagetrue
\renewcommand\maketitle{\begin{titlepage}%
  \let\footnotesize\small
  \hfill\parbox{\preprintnoskip}{%
  \begin{flushright}\@preprintno\end{flushright}}\hspace*{1cm}
  \vskip 60\p@
  \begin{center}%
    {\Large\bf\boldmath \@title \par}\vskip 1cm%
    {\sc\@author \par}\vskip 3mm%
    {\@address \par}%
    \if@preliminary
      \vskip 1cm {\large\sf PRELIMINARY DRAFT}%
    \fi
  \end{center}\par
  \@thanks
  \vfill
  \begin{center}%
    \parbox{\abstractwidth}{\centerline{\abstractname}%
    \vskip 3mm%
    \@abstract}
  \end{center}
  \end{titlepage}%
  \setcounter{footnote}{0}%
  \let\thanks\relax\let\maketitle\relax
  \gdef\@thanks{}\gdef\@author{}\gdef\@address{}%
  \gdef\@title{}\gdef\@abstract{}\gdef\@preprintno{}
}%
\def\@citex[#1]#2{\if@filesw\immediate\write\@auxout{\string\citation{#2}}\fi
  \def\@citea{}\@cite{\@for\@citeb:=#2\do
    {\@citea\def\@citea{,\penalty\@m}\@ifundefined
       {b@\@citeb}{{\bf ?}\@warning
       {Citation `\@citeb' on page \thepage \space undefined}}%
\hbox{\csname b@\@citeb\endcsname}}}{#1}}
\def\citerange{\@ifnextchar [{\@tempswatrue\@citexr}{\@tempswafalse\@citexr[]}}
\def\@citexr[#1]#2{\if@filesw\immediate\write\@auxout{\string\citation{#2}}\fi
  \def\@citea{}\@cite{\@for\@citeb:=#2\do
    {\@citea\def\@citea{--\penalty\@m}\@ifundefined
       {b@\@citeb}{{\bf ?}\@warning
       {Citation `\@citeb' on page \thepage \space undefined}}%
\hbox{\csname b@\@citeb\endcsname}}}{#1}}
\long\def\@makecaption#1#2{%
  \vskip\abovecaptionskip
  \sbox\@tempboxa{#1: \emph{#2}}%
  \ifdim \wd\@tempboxa >\hsize
    #1: \emph{#2}\par
  \else
    \hbox to\hsize{\hfil\box\@tempboxa\hfil}%
  \fi
  \vskip\belowcaptionskip}
\def\fmslash{\@ifnextchar[{\fmsl@sh}{\fmsl@sh[0mu]}}
\def\fmsl@sh[#1]#2{%
  \mathchoice
    {\@fmsl@sh\displaystyle{#1}{#2}}%
    {\@fmsl@sh\textstyle{#1}{#2}}%
    {\@fmsl@sh\scriptstyle{#1}{#2}}%
    {\@fmsl@sh\scriptscriptstyle{#1}{#2}}}
\def\@fmsl@sh#1#2#3{\m@th\ooalign{$\hfil#1\mkern#2/\hfil$\crcr$#1#3$}}
\makeatother

\newcommand{\GeV}{\mathrm{GeV}}
\newcommand{\TeV}{\mathrm{TeV}}
\newcommand{\re}[1]{\mbox{\boldmath$#1$}}
\newcommand{\rb}[1]{\mbox{\boldmath$\overline{#1}$}}
\newcommand{\vev}[1]{\langle#1\rangle}
\usepackage{graphicx}
\usepackage{amsmath}
\usepackage{subfigure}
\allowdisplaybreaks
\begin{document}


\baselineskip20pt   
\preprintno{SI-HEP-2013-17\\[0.5\baselineskip] January 2014}
\title{%
 Multiple Scales in Pati-Salam Unification Models
}
\author{%
 F.~Hartmann\email{hartmann}{physik.uni-siegen.de}$^{,a}$,
 W.~Kilian\email{kilian}{physik.uni-siegen.de}$^{,a}$,
 K.~Schnitter\email{schnitter}{tp1.physik.uni-siegen.de}$^{,a,b}$,
}
\address{%
$^a$Universit\"at Siegen, Department Physik,
      57068~Siegen, Germany\\
$^b$Degenfeldstra\ss e 2, 76131 Karlsruhe, Germany
}
\abstract{We investigate models where gauge unification (GUT) proceeds in steps
  that include Pati-Salam symmetry.
  Beyond the Standard Model, we allow for a well-defined set of
  small representations of the GUT gauge group.  We show that all
  possible chains of Pati-Salam symmetry breaking can be realized in
  accordance with gauge-coupling unification.
  We identify, in particular, models with unification near the Planck scale,
  with intermediate left-right or $SU(4)$ quark-lepton symmetries that are
  relevant for flavor physics, with new colored particles at accessible
  energies, and with an enlarged electroweak Higgs sector.  We look both at
  supersymmetric and non-supersymmetric scenarios.}

\maketitle
\flushbottom

\section{Introduction}
\label{sec:intro}

The Standard Model (SM) of particle physics organizes all known
particles, including the Higgs boson, in a linear representation of
the space-time and gauge symmetries.  This representation is reducible
and decomposes into several irreducible multiplets.  Therefore, the SM
Lagrangian contains a considerable number of free parameters.  A
unified theory (GUT) where the SM gauge symmetry is embedded in a
larger gauge group can reduce the number of free parameters.
Furthermore, there is a general expectation that a unified model of
matter and interaction could explain some of the puzzles of particle
physics, such as quark and neutrino flavor mixing, the smallness of
neutrino masses, or dark matter.

Particularly promising GUT models are left-right symmetric extensions
\cite{Mohapatra:1974gc,Senjanovic:1975rk}, the Pati-Salam model
\cite{Pati:1974yy,Pati:1973uk}, the $SU(5)$ theory by Georgi and
Glashow \cite{Georgi:1974sy}, or the combination of both in
an $SO(10)$ \cite{Fritzsch:1974nn} or $E_6$ \cite{Gursey:1975ki} gauge theory.
In these GUTs, matter fields unify within each generation, separately.  The SM
Higgs sector is enlarged.
Further fields, symmetries, or geometrical structure are required to
explain the breaking of GUT symmetries and the emergence of the SM (or
its minimal supersymmetric extension, MSSM) at energies directly
accessible to us.

In this paper, we investigate various classes of Pati-Salam (PS)
models.  PS symmetry, with a discrete left-right symmetry included, is
the minimal symmetry group which collects all matter fields of a single
generation in one irreducible multiplet.  As an extension of the SM,
it is particularly interesting because it naturally accommodates
multiple energy thresholds or scales that fill the large gap between
the SM (TeV) scale and the scale of ultimate unification, presumably
the Planck scale. Such models have been widely discussed in case of both
supersymmetric
\cite{Malinsky:2005bi,Arbelaez:2013hr,DeRomeri:2011ie,Calibbi:2009wk,
Calibbi:2009cp,Biggio:2010me,Biggio:2012wx,Buckley:2006nv,Esteves:2009qr,
Esteves:2010ff,Esteves:2011gk,Majee:2007uv,Braam:2009fi,Kilian:2006hh,
Howl:2007hq,Dev:2009aw} and non-supersymmetric models
\cite{Lindner:1996tf,Arbelaez:2013nga,Siringo:2012bc}.
PS models do not just exhibit multiple scales associated with symmetry breaking,
but they can generate additional mass scales due to see-saw type patterns in the
spectrum \cite{Aulakh:1999pz,Aulakh:2000sn}.

We want to study the relation of symmetry breaking and intermediate
mass scales in PS models in a suitably generic way while keeping the
sector of extra (Higgs) fields below the GUT scale as compact as
possible.  To this end, we impose conditions on the spectrum of new
fields, so we can describe the chain of symmetry breaking via a staged
Higgs mechanism with only small representations of the GUT group.
In this setup, we scan over the complete set of models and investigate
the allowed ranges and relations of the various mass scales they can
provide.

There are several specific questions that one can address by such a
survey of models.  First of all, some GUT models are challenged by the
non-observation of proton decay, so we may ask whether complete
unification, and thus the appearance of baryon-number violating
interactions, can be delayed up to the Planck scale.  Secondly,
whether the origin of neutrino masses and flavor mixing could be
associated with a distinct scale, decoupled from complete unification.
Thirdly, whether there can be traces of unification, such as exotic
particles, a low-lying left-right symmetry, or an enlarged Higgs
sector, that may be observable in collider experiments or indirectly
affect precision flavor data.

By restricting our framework to economical spectra with only small
representations (see section~\ref{sec:model}), we focus the study on
phenomenologically viable models that look particularly attractive.
Nevertheless, they span a large range of intermediate-scale
configurations between the TeV and Planck scales.  We do not impose
extra technical conditions on the gauge-coupling running and matching
(\cite{Arbelaez:2013hr}, \cite{Calibbi:2009cp}, \cite{Arbelaez:2013nga}), so we
can cover also
corner cases where scale relations are tightly constrained.

After a short review of PS models in Section~\ref{sec:PS}, we lay out the
detailed framework of our study in Section~\ref{sec:model}. In
Section~\ref{sec:Higgs} we take a brief look at the staged Higgs
mechanism, before we discuss the overall spectrum patterns in
Section~\ref{sec:spectra}.  In Section~\ref{sec:unification} we implement
the conditions of gauge coupling unification.  Finally,
Section~\ref{sec:SUSY} and Section~\ref{sec:non-SUSY} present generic
properties and interesting details for the complete survey of PS
models, divided in the two classes of supersymmetric and
non-supersymmetric models respectively, before we draw conclusions.

\section{Pati-Salam Symmetry}
\label{sec:PS}

By the label of Pati-Salam gauge group, we refer to the symmetry group
\begin{equation}
  SU(4)_C\times SU(2)_L\times SU(2)_R\times Z_\text{LR}\,.
\end{equation}
The $SU(4)_C$ factor is an extension of QCD with lepton number as
fourth color, and there is a right-handed $SU(2)_R$ symmetry analogous
to the weak interactions.  We impose an exact discrete symmetry
$Z_\text{LR}$ which exchanges $SU(2)_L$ with $SU(2)_R$, and
left-handed with right-handed matter fields, respectively.

If we impose this symmetry on any generation of quarks and
leptons, they combine to a single irreducible representation
$\Psi_{L/R}$. Regarding the continuous gauge group only, this is a
direct sum $\Psi_L\oplus \Psi_R$, but it is rendered irreducible by the
left-right symmetry\footnote{Throughout this paper, we indicate such
  $Z_2$-irreducible direct sums by the subscript L/R, or simply write
  $\Psi$ for brevity.}.  In particular, this
representation necessarily includes a right-handed neutrino field.
The explicit embedding of the matter fields in the PS multiplet is as
follows:
\begin{align}
   \Psi_L = (\re{4},\re{2},\re{1}) &= \begin{pmatrix} u_r  & u_g  & u_b &\nu
            \\ d_r & d_g & d_b & e \end{pmatrix}_L \,,
   &\qquad
   \Psi_R = (\rb{4},\re{1},\re{2}) &= \begin{pmatrix} u^c_r  & u^c_g  & u^c_b
            &\nu^c \\ d^c_r & d^c_g & d^c_b & e^c \end{pmatrix}_R \,.
\end{align}
Here the subscript (r,g,b) are explicit color indices.  This notation
is compatible with the usual conventions for supersymmetric GUT
models, where we will denote by $\Psi_{L/R}$ the corresponding
superfield.

The set of gauge bosons of a PS gauge theory contains the eight QCD
gluons and the weak $W_L$ triplet.  On top of that, there are a $W_R$
triplet and seven extra gauge bosons of $SU(4)_C$.  The latter consist
of six (vector) leptoquarks and one SM singlet that gauges the
difference of baryon and lepton number, $B-L$.  The SM hypercharge
gauge boson emerges as a linear combination of the two neutral PS
gauge bosons.  The discrete $Z_\text{LR}$ symmetry guarantees that the
interactions of left-handed and right-handed fields coincide for all
representations.

The PS group is a subgroup of $SO(10)$, but does not contain $SU(5)$.
It is well known that the gauge couplings unify for
a pure $SU(5)$ GUT model, if
supersymmetry is imposed.  However, the recent discoveries about
neutrino masses and mixing (see review article in
\cite{Beringer:1900zz}) suggest additional new physics at high
energies.  Concrete estimates tend to place the neutrino
mass-generation scale \emph{below} the GUT scale (either as direct
mass scale \cite{Weinberg:1979sa} or via a see-saw mechanism
\cite{Minkowski:1977sc}). We should expect multiple thresholds
where the pattern of multiplets and symmetries changes.  This idea
fits well in the context of a PS symmetry and its breaking down to the
SM.

Therefore, we investigate scenarios where a PS gauge symmetry is valid
above some high energy scale.  Below the PS scale, we allow for further
thresholds where only a subgroup of PS is realized.
One such subgroup is the minimal left-right symmetry $SU(3)_C\times
SU(2)_L\times SU(2)_R\times U(1)_{B-L}\times Z_\text{LR}$.  By construction,
left-right
symmetry breaking is associated with flavor physics, both in the quark
and lepton sectors.\footnote{In a left-right symmetric model, up- and
  down-type masses are degenerate, and all mixing angles can be rotated
  away.}  Above the PS scale, we maintain the possibility of further
unification to a GUT model.  Examples are $SO(10)$ \cite{Fritzsch:1974nn} or a
larger group such as $E_6$~\cite{Gursey:1975ki}.

PS models provide a partial explanation for the observed stability of
the proton.  $SU(4)_C$ contains the $U(1)_{B-L}$ subgroup, so $B-L$ is
conserved.  In analogy with QCD, lepton number as the fourth color --
and thus baryon number -- is conserved by all interactions that do not
involve the totally antisymmetric tensor $\epsilon_{abcd}$ with four
color indices.  Counting field dimensions, this excludes baryon-number
violating interactions of fermions in the fundamental or adjoint
representations of $SU(4)_C$, at the renormalizable level.  In a
minimal PS extension of the SM, proton-decay operators are thus
excluded or, at least, naturally suppressed.  In particular, the terms
of the MSSM which mediate proton decay via squark fields are
inconsistent with PS symmetry.\footnote{We note that supersymmetric PS
  models need not include $R$ parity for eliminating
  proton-decay operators. However, for more general extensions of the
  MSSM we have to discuss this issue (see
  section~\ref{sec:Fmultiplet}). For the sake of simplicity we imply $R$
  parity for all our models.}

A field condensate that spontaneously breaks $SU(4)_C$ to $SU(3)_C$
carries lepton number, but no baryon number.  Hence, PS breaking does
not induce proton decay either, at the level of renormalizable SM or
MSSM interactions.  On the other hand, breaking PS symmetry down to
the SM gauge groups can generate and relate the operators that provide
neutrino Majorana masses and mixing in the lepton sector, as well as
Yukawa terms and mixing in the quark sector.

\section{General Setup}
\label{sec:model}

In this paper, we consider both supersymmetric and non-supersymmetric
versions of a PS gauge theory.  We first investigate supersymmetric
models, where the model space is more constrained.  For this purpose,
we assume $N=1$ supersymmetry to hold over the full energy range above
the $\TeV$ scale.  Later we will also discuss the running of the gauge
couplings in non-supersymmetric versions of the models.

Following a top-down approach, we put the highest scale under
consideration at the Planck scale (we use $M_{\text{Planck}}=
10^{18.2}\,\GeV$).  In that energy range, gravitation is strong and a
perturbative quantum field theory in four space-time dimensions is
unlikely to be an appropriate description.  Below this scale,
gravitation becomes weak, and an effective weakly coupled
four-dimensional gauge theory can emerge.  This gauge symmetry may be
the PS group.  We suspect that at the highest energies there is a
larger local symmetry (GUT) that contains PS as a subgroup.  By
$M_{\text{GUT}}$ we denote the energy scale where this larger symmetry
is broken down to PS\@.  We note that $M_{\text{GUT}}$ may coincide
with $M_{\text{Planck}}$, and that the GUT theory need not be a
perturbative field theory.

It is important to study possible mechanisms that can break a
Planck-scale GUT symmetry down to the PS group.  Unfortunately, few
reliable and generic results are available.  For instance, the
survival principle that restricts the pattern of multiplets in the
low-energy effective theory~\cite{Georgi:1979md,del
  Aguila:1980at,Mohapatra:1982aq,Dimopoulos:1984ha} does not hold in
supersymmetric
theories~\cite{Aulakh:1999pz,Aulakh:2000sn,Aulakh:1982sw}.  In the
vicinity of the Planck scale, all conceivable effects from string
states, gravitation, extra dimensions, strong interactions, and
possibly completely unknown principles need to be taken into account
in the GUT theory.  This is beyond the scope of this paper.
Therefore, we will leave unspecified the field content above
$M_{\text{GUT}}$ and the mechanism responsible for GUT breaking down
to the PS group.  

On the other hand, at energies significantly below the GUT scale,
there are good reasons to expect the effective theory as a (possibly
supersymmetric) weakly interacting four-dimensional quantum field
theory, where further symmetry breakings are realized by a
conventional Higgs mechanism.  We will take this as a working
hypothesis.  Without solid knowledge about the available PS multiplets
below the GUT scale, we could allow for any set of group
representations as the field content of the effective theory.
Instead, we follow a simple phenomenological approach and restrict our
analysis to a set of small representations.  This set is chosen such
that it is just sufficient to realize all possible chains of further
symmetry breaking.

Below $M_{\text{GUT}}$, we are thus dealing with a PS model which we specify
in more detail.  In the supersymmetric case, we
list all the PS-symmetric supermultiplets that are contained in the
spectrum.  All such supermultiplets can interact and contribute to the
running of the gauge couplings.  The effective Lagrangian may contain
all symmetric renormalizable interactions of these superfields.  It
will also contain, in general, non-renormalizable interactions of
arbitrary dimension which are suppressed by powers of the relevant higher mass
scales, i.e., GUT or Planck scales.  While non-renormalizable terms, as it turns
out, are not required for the symmetry-breaking chains that we discuss below,
they can
contribute significantly to flavor physics in the low-energy effective
theory.  We will not discuss non-renormalizable effects in
this paper in any detail, so we do not attempt a complete description
of flavor.  The necessary inclusion of all kinds of subleading terms is
beyond the scope of this paper.

For the supersymmetric PS model and the effective theories down to the
SM, we state the following simplifying assumptions:
\begin{enumerate}
\item
  The model is effectively a perturbative quantum field theory in four
  dimension at all scales below $M_{\text{GUT}}$.
\item
  The breaking of the PS symmetry and its subgroups is due to 
  Higgs mechanisms that involve the superfields contained in the given
  PS spectrum.
\item 
  SUSY is broken softly by terms in the TeV energy range.
\item
  The effective scale-dependent gauge couplings
  coincide at $M_{\text{GUT}}$, consistent with a GUT symmetry that
  contains $SO(10)$.
\item
  The chiral supermultiplets all fit in the
  lowest-lying (\re{1}, \re{27}, and \re{78}) representations of $E_6$
  or, equivalently, 
  in \re{1}, \re{10}, \re{16}, \rb{16}, and \re{45} representations of
  $SO(10)$.
\end{enumerate}
The last point is the most important restriction that we impose on our
models.  (For a recent survey with different assumptions,
see~\cite{Arbelaez:2013hr}, \cite{Arbelaez:2013nga}).  As we will discuss below,
this minimal
field content can realize all possible chains of PS sub-symmetry
breaking and addresses the phenomenological questions that we raised in
the Introduction above.  Simultaneously, a model with only small
representations can easily be matched to a more ambitious theory of
Planck-scale symmetry breaking.  For instance, all multiplets that we
consider, occur in the fundamental \re{248} representation of
$E_8$~\cite{Slansky:1981yr}.

\begin{table}[tbp]
\centering
 \begin{tabular}{|c|clc|}
 \hline &&& \\[-2ex]
  Field			& $\bigl(SU(4),SU(2)_L,SU(2)_R\bigr)$	& $SO(10)$
& $E_6$	
\\[1ex]
  \hline &&& \\[-2.5ex]\hline &&& \\[-1.5ex]
  $\Sigma$		& $(\re{15},\re{1},\re{1})$
     &\rdelim\}{3}{1.8cm}[$\;\re{45}$] & \rdelim\}{7}{1.8cm}[$\;\re{78}$]
\\
  $T_{L}\oplus T_{R}$	& $(\re{1},\re{3},\re{1}) \oplus (\re{1},\re{1},\re{3})$
     & &
\\
  $E$			& $(\re{6},\re{2},\re{2})$	& 		&	
\\[2mm]
  $\Phi_{L}\oplus \Phi_R$	& $(\re{4},\re{2},\re{1})\oplus
    (\rb{4},\re{1},\re{2})$	& \hspace{2.7ex}$\re{16}$ &
\\[2mm]
  $\overline{\Phi}_L\oplus \overline{\Phi}_R$ & $(\rb{4},\re{2},\re{1})\oplus
    (\re{4},\re{1},\re{2})$	&\hspace{2.7ex}$\rb{16}$	&	
\\[2mm]
  $S_{78}$		& $(\re{1},\re{1},\re{1}) $ & \hspace{3.5ex}$\re{1}$&
\\[3mm]
  $\Psi_{L}\oplus \Psi_{R}$	& $(\re{4},\re{2},\re{1})\oplus
(\rb{4},\re{1},\re{2})$
    & \hspace{2.7ex}$\re{16}$	& \rdelim\}{5}{1.8cm}[$\;\re{27}$]
\\[3mm]
  $h$			& $(\re{1},\re{2},\re{2})$	& 
    \rdelim\}{2}{1.8cm}[$\;\re{10}$]	&	\\		
  $F$			& $(\re{6},\re{1},\re{1})$	&  		&	
\\[2mm]
  $S_{27}$		& $(\re{1},\re{1},\re{1}) $ & \hspace{3.2ex}$\re{1}$ &
\\[1ex]
\hline
 \end{tabular}
 \caption{Chiral superfield multiplets of the PS models considered in
   this paper.  The fields are classified by their gauge-group quantum
 numbers; the discrete $Z_\text{LR}$ symmetry renders the multiplets
 left-right symmetric and thus irreducible.}
 \label{tab:fieldcontent}
\end{table}

We now state explicitly the field spectrum that is consistent with our
assumptions and display all possible irreducible PS multiplets in
table~\ref{tab:fieldcontent}. The matter fields of the MSSM are
contained in $\Psi_{L/R}$ multiplets. There are three copies of this
representation, each one including a
right-handed neutrino superfield.  The other supermultiplets are new.  They
contain the MSSM Higgs bidoublet, various new Higgs superfields, and extra
``exotic'' matter. Let us take a closer look at these superfields:
\begin{enumerate}
\item
  $h$ could directly qualify as the MSSM Higgs bi-doublet.
\item
  $\Sigma$ and $T_{L/R}$ are chiral multiplets in the adjoint
  representation.  They have the same quantum numbers as the gauge
  fields of $SU(4)_C$ and $SU(2)_{L,R}$, respectively.
\item
  $\Phi_{L/R}$ and $\overline\Phi_{L/R}$ are
  extra multiplets with matter quantum numbers and their
  charge-conjugated images, respectively.  Both must coexist in equal
  number, otherwise additional chiral matter generations would be
  present at the $\TeV$ scale\footnote{A fourth chiral generation is not
  finally excluded, but unlikely in view of present data.}.
\item
  $E$ and $F$ contain only colored fields.  Under $SU(3)_C$ they
  decompose into triplets and anti-triplets, so they can be viewed as
  vector-like quarks and their scalar superpartners.  Depending on
  their couplings, they may also behave as leptoquarks or
  diquarks.  The $F$ multiplet may have both leptoquark and diquark
  couplings and thus can violate baryon number (cf.\
  section~\ref{sec:Fmultiplet}).  For the $E$ multiplet, this is not possible.
  Note that $\Sigma$ also provides vector-like quark and antiquark
  superfields, together with a color-octet and a color-singlet
  superfield.  The $\Sigma$ couplings conserve baryon number.
\item
  The singlet fields $S_{78}$ and $S_{27}$ (and any further singlet
  fields that originate from $SO(10)$ or $E_6$ singlets) can couple to any
  gauge-invariant quadratic polynomial.
\end{enumerate}
Chiral superfields which are not copies of the above are excluded by
our basic assumptions.

An explicit (i.e., PS invariant) mass term in the superpotential can
be either present in the superpotential \emph{ab initio}, or generated
by a singlet condensate.  Without further dynamical assumptions, mass
terms and mass thresholds are thus constrained only by matching
conditions on the coupling parameters, and may exhibit a hierarchical
pattern. 

Some of the above mentioned multiplets contain an electroweak singlet and thus
qualify as Higgs superfields for various steps of PS symmetry breaking, if the
corresponding scalar component acquires a vacuum expectation value (vev). The
associated vevs may be denoted as
$\vev{T(1,1,3)}$, $\vev{\Sigma(15,1,1)}$, $\vev{\Phi(4,1,2)}$, and
$\vev{h(1,2,2)}$, corresponding to the Higgs fields introduced below.
They induce the following breaking patterns:
\begin{subequations}
\begin{align}
 SU(4)_C &\xrightarrow{\vev{\Sigma}} SU(3)_C \times U(1)_{B-L} \,,\\[1.5ex]
 SU(2)_R &\xrightarrow{\vev{T}} U(1)_R \,,\\[1.5ex]
 SU(4)_C \times U(1)_R &\xrightarrow{\vev{\Phi}} SU(3)_C \times U(1)_Y \,,\\
 SU(2)_R \times U(1)_{B-L} &\xrightarrow{\vev{\Phi}} U(1)_Y \,,\\
 U(1)_R  \times U(1)_{B-L} &\xrightarrow{\vev{\Phi}} U(1)_Y \,,\\[1.5ex]
 SU(2)_L\times U(1)_Y &\xrightarrow{\vev{h}} U(1)_E \,.
\end{align}
\end{subequations}

\begin{figure}[tbp]
 \centering
 \includegraphics[width=.9\textwidth]{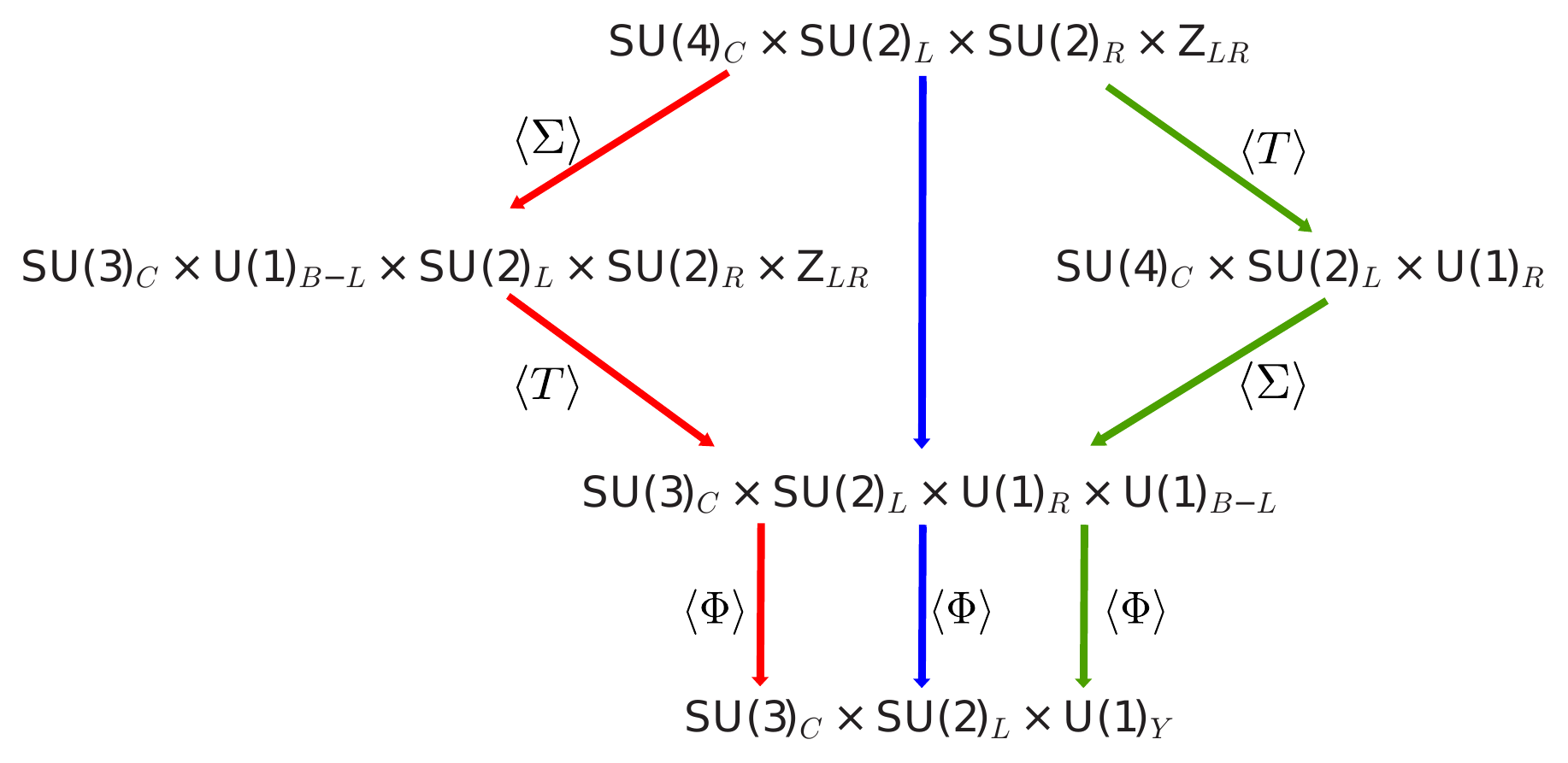}
 \caption{Graphical illustration of the different breaking paths
   depending on the different classes. Class B is shown in red, C in
   green, and D in blue.}
 \label{fig:breakingchaines}
\end{figure}

If there is a hierarchy between the vevs, we observe a cascade
of intermediate symmetries. Thus, we have the possibility for multi-step
symmetry breaking. In the following, we distinguish six classes of PS models by
their hierarchy patterns.  We denote them as follows:
\begin{quote}
\begin{description}
 \item[A:] $\vev\Phi \sim \vev{T} \sim \vev\Sigma$  (one scale),
 \item[B:] $\vev\Phi \ll \vev{T} \ll \vev\Sigma$ (three scales),
 \item[C:] $\vev\Phi \ll \vev\Sigma \ll \vev{T}$ (three scales),
 \item[D:] $\vev\Phi \ll \vev{T} \sim \vev\Sigma$  (two scales),
 \item[E:] $\vev\Phi \ll \vev\Sigma$ and $\vev{T}=0$ (two scales),
 \item[F:] $\vev\Phi \ll \vev{T}$ and $\vev\Sigma=0$ (two scales).
\end{description}
\end{quote}
The symmetry breaking chains associated to these classes are shown in
Fig~\ref{fig:breakingchaines} (see also \cite{Aulakh:2000sn}).

The scale $\vev\Phi$ is always the lowest symmetry-breaking scale (above the
electroweak scale), since this vev breaks all symmetries.  The other vevs break 
the PS symmetry only partially and are only relevant if they lie above
$\vev\Phi$. Nevertheless they are needed because we require the symmetry
breaking to result from a renormalizable potential, which does not exist for
$\Phi$ alone. Class A and D are limiting cases of the
other and will not be discussed in greater detail.
The number of matter multiplets ($\Psi_{L/R}$) is fixed to
three.  All other multiplets may appear in an arbitrary number of
copies.  We will specifically study scenarios where the multiplicity
is either zero, one, or three.  This already provides a wide range of
possibilities for spectra and hierarchies and appears most natural in view
of the generation pattern that is established for SM matter.

Let us further categorize models, constructed along these lines, that
we will study below.  Each model type may fit into any of the above
classes.  It contains the three generations of MSSM matter superfields
$\Psi_{L/R}$ together with the following extra superfields:
\begin{description}
\item[Type m (Minimal Model):] Within a given class, the minimal model is the
  one with minimal field content that realizes the corresponding
  symmetry-breaking chain.  In classes A to D, this model contains a
  single copy of each of the multiplets $\Phi$, $\Sigma$ and
  $T_{L/R}$.  In classes E (F), $T_{L/R}$ ($\Sigma$) is
  omitted, respectively.
\item[Type s (SO(10)-like Model):]
  In this model type, the multiplets can be composed to complete $SO(10)$
  representations.  All
  multiplets listed in table~\ref{tab:fieldcontent} are present as a
  single copy.
\item[Type e ($E_6$-like Model):]
  In this type of model, all fields of table~\ref{tab:fieldcontent} are
  present.  The multiplets $h$, $F$, and $S_{27}$ appear in three
  copies since they combine with the matter multiplets.  For the other
  multiplets, we set the multiplicity to one.
\item[Type g (Generic Model):]
  Here we classify all models that do not qualify as either of the
  above three types.  We will denote these models by their class and a unique
  number. The numbering scheme is described in Appendix~\ref{sec:fieldsetlist}.
\end{description}

\section{Higgs Mechanism and Supersymmetry}
\label{sec:Higgs}

Some of the superfields that we introduce beyond the MSSM matter
fields, should act as Higgs fields and break the gauge symmetry, step
by step, down to the SM gauge group.  For each breaking step, we
have to verify that the scalar potential has a local minimum for a
non vanishing field configuration that breaks the gauge symmetry in the
appropriate way.  The value of the superpotential at this field
configuration should be zero, so supersymmetry is maintained in the
ground state.

To achieve this, we construct a generic renormalizable superpotential
for the multiplets that we include in a specific model.
Non-renormalizable terms could be added, but we will see that
the various routes of gauge symmetry breaking result from
renormalizable terms alone, so they can be ignored in a first
approach.  The specific models derive from a generic superpotential
that includes all allowed superfields and their renormalizable
interactions.

\subsection{Generic Superpotential}

We start at the GUT scale with the most general renormalizable
PS-invariant superpotential for the superfields shown in
table~\ref{tab:fieldcontent} and impose parity ($Z_{L/R}$) as a symmetry at this
scale. The superpotential can be broken down in several parts which we
organize in view of the model types m, s, and e as introduced above.
\begin{equation}
W = W_{\Phi/\Sigma/T} + W_\text{h/F} + W_\text{E} +  W_\text{S} +
W_\text{Yukawa}
\label{eq::Superpotential}
\end{equation}

The terms within $W_{\Phi/\Sigma/T}$ generate the Higgs potential for
all steps in the staged Higgs mechanism.  This superpotential consists
only of those fields that are allowed to get a vev.  Depending on the
desired vev structure (case A to F), we may set some terms to zero, to
obtain a minimal superpotential.

The superpotential $W_{h/F}$ is present in the models that contain the
fields $h$ and $F$.  Similarly, the field $E$ comes with the terms
$W_\text{E}$.  The potential $W_\text{S}$ contains all interactions
of the singlets with the other Higgs fields, where in each term, $S$
indicates an arbitrary linear combination of all PS singlets present
in the model.

If $h,F$ are present, there is the $W_\text{Yukawa}$ superpotential.
This part, which is the only renormalizable superpotential involving
(two) matter superfields\footnote{Only $\Psi_L$ and $\Psi_R$ are
  considered as matter superfields, containing all fermions of the
  SM.}, implicitly contains generation indices.  Analogously,
generation indices are implied for all superfields that occur in more
than one copy.  We note that $Y$ contains only symmetric Yukawa couplings
that are universal across leptons, neutrinos, up- and down-type quarks.
\begin{subequations}
\begin{align}
W_{\Phi/\Sigma/T} =& 
	    - m_{\Phi}\; \left(\rb{\Phi}_L\,\re{\Phi}_L +
\rb{\Phi}_R\,\re{\Phi}_R\right)
	    \nonumber \\
	&   
	    - \tfrac{1}{2} m_{\Sigma}\; \re{\Sigma}^2
	    + \tfrac{1}{3} l_\Sigma\; \re{\Sigma}^3
	    + l_{\Sigma\Phi}\; \left(\rb{\Phi}_L\,\re{\Sigma}\,\re{\Phi}_L +
\rb{\Phi}_R\,\re{\Sigma}\,\re{\Phi}_R\right)
	    \nonumber \\
	&
	    - \tfrac{1}{2} m_T\; \left(\re{T}_L^2 + \re{T}_R^2\right)
	    + l_{T\Phi}\; \left(\rb{\Phi}_L\,\re{T}_L\,\re{\Phi}_L +
	    \rb{\Phi}_R\,\re{T}_R\,\re{\Phi}_R\right)
    	    \\[2ex]
W_\text{h/F} =& 
	    - \tfrac{1}{2} m_{h}\; \re{h}^2  
	    - \tfrac{1}{2} m_{F}\; \re{F}^2
	    + l_{h\Phi}\; \re{h}\,\left(\re{\Phi}_L\,\re{\Phi}_R +
              \rb{\Phi}_R\,\rb{\Phi}_L\right)
	    \nonumber \\
	&
	    + l_{F\Phi}\; \re{F}\,\left(\re{\Phi}_L\,\re{\Phi}_L +
              \re{\Phi}_R\,\re{\Phi}_R\right)
	    + l_{F\bar{\Phi}}\; \re{F}\,\left(\rb{\Phi}_L\,\rb{\Phi}_L +
              \rb{\Phi}_R\,\rb{\Phi}_R\right)
	    \nonumber \\
	&  
	    + l_{\Sigma F}\; \re{F}\,\re{\Sigma}\,\re{F}
	    + l_{Th}\; \re{h}\,\left(\re{T}_L + \re{T}_R\right)\,\re{h}
	    \\[2ex]
W_\text{E} =& 
	    - \tfrac{1}{2} m_E\; \re{E}^2
	    + l_{TE}\; \re{E}\,\left(\re{T}_L + \,\re{T}_R\right)\,\re{E}
	    \nonumber \\
	&
	    + l_{\Sigma E}\; \re{E}\,\re{\Sigma}\,\re{E}   
	    + l_{FEh}\; \re{F}\,\re{E}\,\re{h}
	    \\[2ex]
W_\text{S} =&
	     - \tfrac{1}{2} m_{S}\,\re{S}^2 + \tfrac13 l_{S}\,\re{S}^3 
\nonumber \\
	&
	    + s_{\Phi}\, \re{S} \; \left(\rb{\Phi}_L\,\re{\Phi}_L +
	      \rb{\Phi}_R\,\re{\Phi}_R \right)
	    + s_T\, \re{S} \; \left(\re{T}_L^2 + \re{T}_R^2\right)
	    + s_{\Sigma}\, \re{S} \; \re{\Sigma}^2 
	    \nonumber \\
	& 
	    + s_h\, \re{S} \; \re{h}^2  
	    + s_F\, \re{S} \; \re{F}^2
	    + s_E\, \re{S} \; \re{E}^2
	    \\[2ex]
W_\text{Yukawa} =&
	    \,Y\;\re{\Psi}_L\,\re{h}\,\re{\Psi}_R 
	    + Y_F\;\re{F}\,\left(\re{\Psi}_L\,\re{\Psi}_L +
	    \re{\Psi}_R\re{\Psi}_R \right)
\end{align}
\end{subequations}

\subsection{Higgs Mechanism and Supersymmetry}

For our supersymmetric models, we want to keep SUSY unbroken down to
the $\TeV$ (MSSM) scale.  To satisfy this constraint, we have to
verify that all $F$ and $D$ terms vanish \cite{Nilles:1983ge,Haber:1984rc} after
inserting the vacuum
expectation values of the fields\footnote{We note that these
  conditions are trivially satisfied if all expectation values vanish.
  Without specifying further (possibly non-perturbative) dynamics, for
  a supersymmetric Higgs mechanism there is the alternative of no
  symmetry breaking, degenerate in energy.  A complete theory should
  include a mechanism for supersymmetry breaking that lifts this
  degeneracy.}:
\begin{equation}
 \left. F_i\right|_\text{vev} = 0 \quad \text{and} \quad \left.
D_a\right|_\text{vev} = 0 \,.
\end{equation}

We explicitly calculate the scalar potential for the generic superpotential
shown in (\ref{eq::Superpotential}).  This covers all models under
consideration; for any specific model, we set the couplings to zero
that are not involved.

The fields that transform under PS but contain a singlet under the SM
gauge group, are $\Phi_R$, $\overline\Phi_R$, $\Sigma$, $T_R$, and the matter
superfields $\Psi_R$.  In particular, the $SU(4)_C$ singlets in $\Phi_R$
and $\overline\Phi_R$ simultaneously break $SU(2)_R$ and thus induce direct
breaking $\text{PS}\to\text{MSSM}$.  However, there is no nontrivial
renormalizable superpotential of $\Phi_R,\overline\Phi_R$ alone, so we
include the $\Sigma$ and $T_R$ multiplets in the discussion.  The vevs
$\vev{\Sigma}$ and $\vev{T_R}$ are particularly relevant to the
symmetry-breaking chain if they are significantly larger than
$v_\Phi$, since they leave subgroups of PS intact (cf.
figure~\ref{fig:breakingchaines}).  We introduce the following notation:
\begin{subequations}
\begin{align}
   \langle \re{\Phi}_R \rangle &= \Phi^R_{4,1,1} = v_{\Phi} \,,\\
   \langle \rb{\Phi}_R \rangle &= \overline{\Phi}^R_{4,1,1} =
v_{\overline{\Phi}} \,,\\
   \langle \re{\Sigma} \rangle &= \Sigma_{15,1,1} = v_{\Sigma} \,,\\
   \langle \re{T}_R \rangle &= T_{1,1,3} = v_{T} \,.
\end{align}
\end{subequations}
Here the subscript denotes the field component (or generator) that
acquires the vev.  These are all vevs we allow for in this paper.

For flavor physics, the vevs $v_\Phi$ and $v_T$ are of particular interest,
because they break the discrete $Z_\text{LR}$ symmetry.  As long as the
left-right symmetry is unbroken, there are no terms that distinguish
right-handed up-type
from down-type quarks.  Therefore, the energy scale below which flavor
mixing appears in the renormalizable part of the effective theory, is
given by $v_\Phi$ or $v_T$.  On the other hand, both $v_\Sigma$ and $v_\Phi$
separate quarks from leptons, so only in class-F models where
$v_\Sigma=0$, we expect a direct relation between lepton-flavor and
quark-flavor mixing.

The ground-state values of D terms are zero if and only if the vevs of
mutually conjugate fields exist simultaneously and coincide in value.
For $T$ and $\Sigma$ the D-terms automatically vanish since their generators are
traceless. Therefore, we must have
$\langle\Phi_R\rangle=\langle\overline\Phi_R\rangle$:
\begin{equation}
  v_{\Phi} \equiv v_{\overline{\Phi}}.
\end{equation}

The vacuum expectation values of the F terms also have to vanish.
Inserting the vevs, we obtain the necessary and sufficient conditions
\begin{subequations}
  \label{eq:Ftermconditions}
  \begin{align}
    0 = F_{\Phi_4} &= v_{\Phi} \left(m_{\Phi} - l_{\Sigma\Phi}\, v_{\Sigma} -
l_{T\Phi}\,v_T\right) \,,\\
    0 = F_{\Sigma_{15}} &= v_\Sigma \left( m_{\Sigma} - l_{\Sigma}\, v_{\Sigma}
\right) - l_{\Sigma\Phi}\, v_{\Phi}^2 \,,\\
    0 = F_{T^R_{3}} &=  m_T\,v_{T} - l_{T\Phi} v_{\Phi}^2 \,,\\
    0 = F_{S} &= s_{\Phi}\,v_\Phi^2 + s_{\Sigma}\,v_{\Sigma}^2 +s_{T}\,v_T^2\,. 
  \end{align}
\end{subequations}

Solving for the mass parameters of $\Phi$, $T$ and $\Sigma$ as well as for one
of the
singlet couplings, we obtain
\begin{subequations}
\begin{align}
  m_{\Phi}   &= l_{\Sigma\Phi}\, v_{\Sigma} + l_{T\Phi}\, v_T \,,\\
  m_T	     &= \frac{l_{T\Phi}\, v_{\Phi}^2}{v_T} \,,\\
  m_{\Sigma} &= \frac{l_{\Sigma\Phi}\,v_{\Phi}^2}{v_{\Sigma}} +
      l_{\Sigma}\, v_{\Sigma} \,,\\
  s_{\Phi}     &= -\frac{s_{\Sigma}\,v_\Sigma^2 + s_T\,v_T^2}{v_\Phi^2} \,.
  \end{align}
\end{subequations}

We have verified that these vev configurations also minimize
the scalar potential while maintaining supersymmetry.  Thus,
depending on their mutual hierarchy, they realize the
symmetry-breaking chains of the model classes A to D.

For the models classes E and F, we assume a vanishing vev of $T$ or
$\Sigma$, respectively.  Nevertheless, PS symmetry is broken
completely down to the MSSM gauge symmetry.  A vanishing vev results
if the corresponding multiplet does not couple to
$\Phi_R/\overline{\Phi}_R$.  This can be realized by either omitting
the multiplet entirely, or by setting $l_{\Sigma\Phi}$ ($l_{T\Phi}$) to
zero, respectively. Otherwise, SUSY would be broken.

As long as we stay in the regime of renormalizable superpotentials, it
is not allowed to set $v_{\Phi}= 0$, because in this case the solution
of the minimization condition would be $v_T \equiv 0$, so $SU(2)_{R}$
would remain intact.  The reason is the absence of trilinear terms for
$T_R$ in the renormalizable superpotential: $SU(2)$ has no cubic
invariant. In addition there would be no breaking of $U(1)_R\times
U(1)_{B-L}$ down to hypercharge without $v_{\Phi}$.

\section{Spectra and Phenomenology}
\label{sec:spectra}

The various classes of models that we introduced above, lead to a
great variety in the observable spectra. In the current work, we
aim at a qualitative understanding, so we refrain from detailed
numerical estimates or setting up benchmark models.  Nevertheless, we
observe a few characteristic patterns which can have interesting
consequences for collider and flavor physics.  We discuss these
patterns and their phenomenological consequences below.

\subsection{Mass Matrix}

Having verified that SUSY remains unbroken down to the TeV scale where
soft-breaking terms appear, we consider the mass matrix of the
scalars.  We do not intend to derive quantitative results here, but
just assign mass scales to the individual multiplets which depend on
the hierarchy of the vevs. In table~\ref{tab:masstable}, we list the
results for all fields and the interesting model classes B, C, E and F. The
classes A and D are limiting cases. Therefore they are not listed separately.

We should keep in mind that this table refers only to the mass
contributions that result from symmetry breaking.  All superfields
except for the matter multiplets, may carry an individual PS-symmetric
bilinear superpotential term.  These mass terms are completely
arbitrary and, with supersymmetry, do not cause naturalness problems
\cite{Wess:1973kz,Grisaru:1979wc}.
Furthermore, a vev in any PS singlet field may contribute a similar
term, again unrestricted by symmetries.  Hence, by including a
multiplet in the spectrum of the effective theory in a particular
energy range, we implicitly assume that the sum of all mass terms for
this field is either negligible compared to the energy, or fixed by
the conditions that determine the vacuum expectation values.

On the other hand, in many models, not all of the fields $E$, $\Sigma$
and $T$ receive masses from PS and subsequent symmetry breaking, given
only the renormalizable Lagrangian terms above.  For these fields,
either the bilinear mass term or effective masses induced by
higher-dimensional operators play an important role and must be
included as independent parameters.  From a phenomenological viewpoint,
we are mainly interested in the lowest possible mass for each of those
fields.  We discuss this issue in section~\ref{sec:runningmass}.

\begin{table}
\centering
\setlength{\tabcolsep}{3ex}
  \begin{tabular}{|p{2ex} c| c| c| c| c|}
  \hline&&&&&\\[-2ex]
  field & $(SU(3)_c,SU(2)_L)_{Y}$ & B    & C    & E    & F  
  \\[.5ex]
  \hline &&&&&\\[-2.5ex] \hline &&&&&\\[-1.5ex]
  $\Sigma$ & $(\re8,\re1)_0$   &    $v_\Sigma$  & $v_\Sigma$   & $v_\Sigma$ 
    & ---
   \\[1.75ex]
  $E$    & $(\re3/\rb3,\re2)_{\pm\frac56}$ & --- & ---    & --- & ---
   \\[1.75ex]
  $E$    & $(\re3/\rb3,\re2)_{\pm\frac16}$ & --- & ---    & --- & ---
   \\[1.75ex]
  $\Phi_L/\overline{\Phi}_L$ & $(\re3/\rb3,\re2)_{\pm\frac16}$ & $v_\Sigma$  &
    $v_T$    & $v_\Sigma$  & $v_T$
   \\[1.75ex]
  $\Phi_R/\overline{\Phi}_R$  & $(\re3/\rb3,\re1)_{\pm\frac23}$ &  $v_\Sigma$  &
    $v_\Sigma$   & $v_\Sigma$ & $v_\Phi$  
   \\[.75ex]
  $\Sigma$   & $(\re3/\rb3,\re1)_{\pm\frac23}$ & $v_\Sigma$  & $v_\Sigma$ &
    $v_\Sigma$ & ---
   \\[1.75ex]
  $\Phi_R/\overline{\Phi}_R$ & $(\re3/\rb3,\re1)_{\pm\frac13}$ & $v_\Sigma$  &
    $v_T$    & $v_\Sigma$  & $v_T$
   \\[.75ex]
  $F$ & $(\re3/\rb3,\re1)_{\pm\frac13}$  & $\frac{v_\Phi^2}{v_\Sigma}$&
    $\frac{v_\Phi^2}{v_T}$  & $\frac{v_\Phi^2}{v_\Sigma}$&
    $\frac{v_\Phi^2}{v_T}$
   \\[1.75ex]    
  $T_L$ & $(\re1,\re3)_0$   & $\frac{v_\Phi^2}{v_T}$& $\frac{v_\Phi^2}{v_T}$ &
    ---  & $\frac{v_\Phi^2}{v_T}$
   \\[1.75ex]
  $\Phi_L/\overline{\Phi}_L$ & $(\re1,\re2)_{\pm\frac12}$ & $v_T$ & $v_T$ &
    $v_\Phi$ & $v_T$ 
   \\[.75ex]
  $h$ & $(\re1,\re2)_{\pm\frac12}$ & $\frac{v_\Phi^2}{v_T}$ &
    $\frac{v_\Phi^2}{v_T}$ & $v_\Phi$ & $\frac{v_\Phi^2}{v_T}$
   \\[1.75ex]
  $\Phi_R/\overline{\Phi}_R$ & $(\re1,\re1)_{\pm1}$   & $v_T$ & $v_T$    &
    $v_\Phi$ & $v_T$ 
   \\[.75ex]
  $T_R$ & $(\re1,\re1)_{\pm1}$   & $v_T$ & $v_T$    & --- & $v_T$ 
   \\[1.5ex]
  $T_R$ & $(\re1,\re1)_0$ & $v_\Phi$  & $v_T$ & --- & $v_\Phi$  
   \\[.75ex]
  $\Sigma$ & $(\re1,\re1)_0$ & $v_\Sigma$  & $v_T$ & $v_\Sigma$ & --- 
   \\[.75ex]
  $\Phi_R/\overline{\Phi}_R$ & $(\re1,\re1)_0$ & $v_\Phi$  & $v_\Sigma$   &
    $v_\Phi$ & $\frac{v_\Phi^2}{v_T}$
   \\[.75ex]
  $S_{27} / S_{78}$  & $(\re1,\re1)_0$ & $\frac{v_\Sigma^2}{v_T}$&
    $\frac{v_\Sigma^3}{v_T^2}$  & --- & ---
   \\[.75ex] 
  $S_{27} / S_{78}$  & $(\re1,\re1)_0$ & $\frac{v_\Sigma^2}{v_T}$&
    $\frac{v_\Sigma^3}{v_T^2}$ & $v_\Phi$
    & $\frac{v_\Phi^2}{v_T}$ \\[1ex]
    \hline
  \end{tabular}
 \caption{Mass hierarchy of the scalar fields in the different classes of the
   complete model. If non is shown, there is no
   contribution from the vev and their hierarchy is undefined. Class B:
   $v_\Phi \ll v_T \ll v_\Sigma$; Class C: $v_\Phi \ll
   v_\Sigma \ll v_T$; Class E: $v_\Phi \ll v_\Sigma,\; v_T=0$; Class F: $v_\Phi
   \ll v_T,\; v_\Sigma=0$; The classes A and D are a limiting case of B and C.
   Class A can 
   be reached if one sets all vevs equal to a single vev $v$ and class D if one
   just sets $v_\Sigma=v_T=v$. The order is thus, that fields which mix are
   grouped together. Thus the mass eigenvalues are a linear combination
   of the listed fields. Massless components which are the Goldstone Bosons are
   explicitly not considered here.}
 \label{tab:masstable}
\end{table}

While some multiplets acquire masses proportional to either one of the
symmetry-breaking scales $v_T,v_\Sigma,v_\Phi$, there are various
cases where the mass becomes of order $v_\Phi^2/v_T$ or
$v_\Phi^2/v_\Sigma$, which can be significantly smaller than $v_\Phi$
if there is a hierarchy between the vevs.  In other words, there is an
extra see-saw effect, unrelated to the well-known neutrino
see-saw~\cite{Aulakh:1999pz,Aulakh:1997ba,Aulakh:1998nn}.  We denote
this induced mass scale by $M_\text{IND}$.  It is located below the
scale where PS is completely broken down to the MSSM symmetry group.
A generic expression is
\begin{equation}
\label{eq:Mind}
 M_{\text{IND}} \,\sim\, \frac{v_\Phi^2}{v_\Sigma + v_T}\,.
\end{equation}

Depending on the model class, some of the field multiplets $F$, $h$, or
$T_L$ become associated with $M_\text{IND}$\footnote{In class F, this
  applies also to the singlet part of $T_R$.}
(cf. table~\ref{tab:masstable}).  We thus get ``light''
supermultiplets consisting of scalars and fermions, which may be
colored, charged, or neutral, and acquire a mass that does not
coincide with either of the symmetry-breaking scales.  If the
hierarchy between the vevs is strong, $M_\text{IND}$ may be
sufficiently low to become relevant for collider phenomenology.  In
model classes B, C, and F, it provides a $\mu$ term for
$h$ and may thus be related to electroweak symmetry breaking.  In any
case, the threshold $M_\text{IND}$ must be taken into account in the
renormalization-group running of the gauge couplings.

\subsection{Goldstone Bosons}
\label{sec:goldstone}

Not all of the scalar fields are physical: since the broken symmetries
are gauged, nine of the scalar fields are Goldstone bosons that
provide the longitudinal modes of the PS gauge bosons that are
integrated out in the breaking down to the MSSM\@.  Six of them come from
$SU(4)_C\to SU(3)_C$ (\ref{eq:GB1b3},\ref{eq:GB4b6}), and two additional ones
implement $SU(2)_R\to U(1)_R$ (\ref{eq:GB7},\ref{eq:GB8}). The last one comes
from the breaking of the $U(1)$ subgroups $U(1)_{B-L}\otimes U(1)_{R}\to U(1)_Y$
(\ref{eq:GB9}). We identify these Goldstone bosons as
\begin{subequations}
  \begin{align}
    GB_{1,2,3} &= -\mathrm{i}\, \frac{\sqrt{3}}{2} \frac{v_{\Phi}}{v_{\Sigma}}
		  \,\re{\Phi}^R_{3} - \mathrm{i}\, \frac{\sqrt{3}}{2}
		  \frac{v_{\Phi}}{v_{\Sigma}} \,\rb{\Phi^{*}}^R_{3} +
		  \re{\Sigma}_{3} + \re{\Sigma^{*}}_{3}  
  \label{eq:GB1b3} \,,\\
    GB_{4,5,6} &= \mathrm{i}\, \frac{\sqrt{3}}{2} \frac{v_{\Phi}}{v_{\Sigma}}
		  \,\re{\Phi^{*}}^R_{\overline{3}} + \mathrm{i}\,
		  \frac{\sqrt{3}}{2} \frac{v_{\Phi}}{v_{\Sigma}}\,
		  \rb{\Phi}^R_{\overline{3}} + \re{\Sigma}_{\overline{3}} +
		  \re{\Sigma^{*}}_{\overline{3}}
  \label{eq:GB4b6} \,,\\
    GB_{7}     &= \re{T}^R_{1_1} + \re{T^*}^R_{1_1} -
		  \frac{\mathrm{i}}{\sqrt{2}}\,
		  \frac{v_\Phi}{v_T} \rb{\Phi}^R_{1_1} -
		  \frac{\mathrm{i}}{\sqrt{2}}\, \frac{\,v_\Phi}{v_T}
		  \re{\Phi^*}^R_{1_1}
  \label{eq:GB7} \,,\\
    GB_{8}     &= \re{T}^R_{1_{-1}} + \re{T^*}^R_{1_{-1}} +
		  \frac{\mathrm{i}}{\sqrt{2}}\, \frac{v_\Phi}{v_T}
		  \re{\Phi}^R_{1_{-1}} + \frac{\mathrm{i}}{\sqrt{2}}\,
 		  \frac{v_\Phi}{v_T} \rb{\Phi^*}^R_{1_{-1}}
  \label{eq:GB8} \,,\\ 
    GB_{9}     &= \text{Im}\left(\re{\Phi}^R_{1_0}\right) -
		  \text{Im}\left(\rb{\Phi}^R_{1_0}\right) 
  \label{eq:GB9} \,.
  \end{align}
\end{subequations}
Here $3$ and $\overline{3}$ are the $(\re{3},\re{1})_{\frac23}$ and
$(\rb{3},\re{1})_{\frac23}$ components of the Higgs fields
$\re{\Phi}_R$, $\rb{\Phi}_R$ and $\re{\Sigma}$, respectively. For vanishing
vevs, the corresponding fields do not mix into the GBs. Thus, if $v_\Sigma=0$
($v_T=0$), the Goldstone bosons $GB_{1-6}$ ($GB_{7/8}$) are only mixtures
of $\Phi$.

\subsection{MSSM Higgs}
\label{sec:MSSMHiggs}

Apart from the matter fields and Goldstone bosons, the chiral
superfield spectrum must provide the Higgs bi-doublet of the MSSM that
is responsible for electroweak symmetry breaking.  Above the TeV
scale, this multiplet should appear as effectively massless; mass
terms (which together set the EWSB scale) are provided by soft-SUSY
breaking parameters and the $\mu$ term which mixes both doublets.  We
note that the $\mu$ term may result from the vev of one of the
electroweak singlets in the model, i.e., the model may implement a
NMSSM-type solution for the $\mu$~problem~\cite{Ellis:1988er,Nilles:1982dy}.

The obvious candidate for the MSSM Higgs bi-doublet is the superfield
$h$.  From table~\ref{tab:masstable} we read off that the PS-breaking
contribution to the $h$ mass is see-saw suppressed in models with
$v_T\neq0$ (B, C, D and F). In these models, the electroweak hierarchy is
generated, at least partly, by a high-energy hierarchy in the PS
symmetry-breaking chain.

However, the $\Phi_L$ multiplets provide further Higgs candidates.  In
particular, while the right-handed doublets in $\Phi_R$ serve as
Goldstone bosons for the $SU(2)_R$ breaking and are thus unphysical,
the mirror images in $\Phi_L$ are to be considered as physical scalars
(above the EWSB scale).  Above the L-R breaking scale, these fields
are protected against mass terms, since all contributions to the mass are
absorbed in the minimization conditions\footnote{The mass of $\Phi_R$ is fixed
since it is proportional to the vev which we keep fixed}. Therefore they can
only acquire nonzero mass due to mixing effects proportional to
$l_{h\Phi}$ and
$l_{T\Phi}$ breaking LR-symmetry.
With respect to the SM gauge symmetry, they have the same quantum
numbers as $h$, and will be called $h_\Phi$ in the following.

In the limit $v_\Phi \ll v_T$, the mixing effect which provides a mass
for $h_\Phi$ becomes negligible, while there may still be an large
contribution to the $h$ mass.  For the $h/h_\Phi$ system, we obtain a
mass matrix with an approximate eigenvalue structure
\begin{subequations}
\begin{align}
 \mu &\approx m_{h} + \frac{l_{h\Phi}^2}{l_{T\Phi}}
\frac{v_{\Phi}^2}{v_T} \,,\\
 m_{h_{\Phi}}' &\approx  l_{T\Phi}\, v_T \,.
\end{align}
\end{subequations}
where we allow for $m_h$ as an independent $\mu$ term, not directly
related to PS breaking.  Both mass terms may be as low as the $\TeV$
scale where soft-breaking terms come into play.  In other words, the
MSSM Higgs bi-doublet (in particular, the Higgs boson that has been
observed) may belong to either $h$ or $h_\Phi$, or be a mixture of
both.

If $v_T$ vanishes (class E), the situation becomes more complicated.
Now, both masses get a contribution of the order
$l_{h\Phi}\,v_{\Phi}$. For
$m_h=0$, both mass eigenvalues are degenerate.  Conversely, if the mixing
between $h$ and $h_\Phi$ vanishes, they are maximally split ($m_{h}$,
$0$). For small mixing ($l_{h\Phi}\,m_h \ll v_\Phi $), we get an
additional see-saw effect and a factor $l_{h\Phi}^2v_\Phi^2/m_h$ for
the smaller eigenvalue, which generates the effective $\mu$ term.

In short, in various classes of PS models, the MSSM Higgs bi-doublet
may be naturally light, and it could actually originate
from the $\Phi_L$ superfields.

\subsection[The F multiplet]{The $F$ multiplet}
\label{sec:Fmultiplet}

The mass of the multiplet $F$ is generically see-saw suppressed and thus
comparatively light.  This appears as a common feature of all model
classes with more than one scale.  It can couple to matter via
$W_\text{Yukawa}$.  Since $F$ is a $SU(4)_C$ antisymmetric tensor, the
possible Yukawa couplings provide both diquark and leptoquark
couplings, explicitly breaking baryon number in the low-energy theory.
In fact, $F$ is the analog of the colored Higgs field which in $SU(5)$ GUT
models induces rapid proton decay unless it is very heavy.  

However, in PS models the Yukawa matrices $Y_F$ and $Y$ are not
related, so there is no doublet-triplet splitting problem
\cite{Kilian:2006hh,Howl:2007hq}. Hence, by omitting the coupling of $F$ to
$\Psi$, proton decay
is excluded in the presence of all gauge symmetries.  This can be
achieved, for instance, by a flavor symmetry or by an appropriate
discrete quantum number.

If the $F$ multiplet is sufficiently light, it may provide detectable
new particles at colliders.  Without the $Y_F$ Yukawa coupling, there
is no immediate decay to MSSM matter fields, but other terms in the
Lagrangian provide indirect decay channels.  In this situation, the
particles (color-triplet scalars and fermion superpartners) may become
rather narrow as resonances.

\subsection{SM Singlets}

The most complicated mass matrix belongs to the electroweak singlets that are
contained in the various PS multiplets.  Even if we do not consider PS
singlets, there are still five electroweak singlets which can mix
non-trivially.  In the general case, their masses cannot be calculated
in closed analytical form. To get a handle on these particles, we
computed the dependence on the different scales numerically.
Given one of the scale hierarchy patterns introduced above, we find
additional scales and new hierarchy patterns which may have
interesting consequences for flavor and Higgs physics.  However,
singlets do not contribute to the running of the gauge couplings at
leading-logarithmic level, so we do not attempt a detailed discussion
of the singlet sector in the present work.

\subsection{Matter Couplings}
\label{sec_Yukawa}

The renormalizable superpotential contains terms that couple $h$ to
matter $\Psi$, so matter fields can get masses via the $h$ vev, after
electroweak symmetry breaking.  However, there is no reason for flavor
physics to originate solely from the \emph{renormalizable}
superpotential.  In particular, if $h_\Phi$ turns out to be the MSSM
Higgs, there are no contributions from this superpotential at all.

Instead, we expect significant and non-symmetric contributions to
masses and mixing from non-renormalizable terms that induce effective
Yukawa couplings at one of the symmetry-breaking scales.  The
generated terms are proportional to powers of the various vevs in the
model, and suppressed by masses of heavier particles, by
$M_\text{GUT}$, or $M_\text{Planck}$.  There is ample space for flavor
hierarchy in the resulting coefficients.  For instance, if we consider
dimension-four terms in the superpotential, we identify the following
interactions that can affect matter-Higgs Yukawa couplings in the low-energy
effective theory,
\begin{align}
 W_\text{Yukawa}^\text{NLO}
  =& 
      \; \frac{Y_{h\Sigma}}{\Lambda}\;
         \re{\Psi}_L\,\re{h}\,\re{\Sigma}\re{\Psi}_R 
      +  \frac{Y_{\Phi}}{\Lambda}\;\re{\Psi}_L\left(
         \re{\Phi}_L\re{\Phi}_R + \rb{\Phi}_L\rb{\Phi}_R \right)\re{\Psi}_R
  \nonumber\\
  &
	+ \frac{Y_{EF}}{\Lambda}\;\re{\Psi}_L\,\re{E}\,\re{F}\,\re{\Psi}_R \,.
\end{align}
There are additional couplings to gauge singlets, which may also be
flavored, so overall there is great freedom in assigning masses and
generating hierarchies in the mass and mixing parameters.

\subsection{Neutrino Mass}

As a left-right-symmetric extension of the SM, PS models contain
right-handed neutrinos and allow for a Dirac neutrino mass term.
Furthermore, the extra fields that are present in our setup can induce
any of the three neutrino see-saw mechanisms for mass generation
\cite{Mohapatra:1979ia,GellMann:1980vs}.  The fields $\Phi_R$ and
$T_R$ provide singlets with a vev that can couple to right-handed
neutrinos, generating a Majorana mass term proportional to $v_\Phi$ and
$v_T$, respectively.  Combined with the Dirac mass, this results in a
type I see-saw.  The field $T_L$ contains $SU(2)_L$-triplet scalars
and fermions and may thus induce a type-II or type-III see-saw
mechanism.

\section{Unification conditions}
\label{sec:unification}

Within the framework of PS model classes that we have defined in the
previous sections, we now impose unification conditions on the gauge
couplings.  As stated above, we require complete unification for all
gauge groups to a GUT symmetry that contains $SO(10)$.  The
unification scale $M_\text{GUT}$ where this should happen is not fixed
but depends on the spectrum.  At each threshold below this where the
spectrum changes, i.e., particles are integrated out, we state the
appropriate matching conditions.

We work only to leading-logarithmic level, where non-abelian running
gauge couplings are continuous in energy, and matching conditions
depend only on the spectra\footnote{At next-to-leading order, the
  superpotential parameters enter the running.  However, given the
  great freedom in choosing a model in the first place, there is
  little to be gained from including such effects in our framework.}.
Abelian gauge couplings do exhibit discontinuous behavior as an
artifact of differing normalization conventions in different effective
theories.

The leading-logarithmic running of a gauge couplings between fixed
scales $\mu_1$ to $\mu_2$ is given by
\begin{equation}
  \frac{1}{\alpha_i\left(\mu_2\right)} = \frac{1}{\alpha_i\left(\mu_1\right)} -
\frac{b_i}{2\,\pi}\,\ln\left(\frac{\mu_2}{\mu_1} \right) \,.
\end{equation}
Here, the $b_i$ are group theoretical factors that can be calculated from the
representations of the particles  \cite{Jones:1981we}.

Inserting the intermediate mass scales (case B for definiteness) as discussed
above, the
complete running is a sum of multiple terms,
\begin{align}
\label{eq:runninggaugecoupling}
  \frac{1}{\alpha_i\left(M_\text{GUT})\right)} =
    \frac{1}{\alpha_i\left(M_Z\right)} 
  & - \frac{b_i^{(1)}}{2\,\pi}\,\ln\left(\frac{M_\text{SUSY}}{M_Z}\right)
  - \frac{b_i^{(2)}}{2\,\pi}\,\ln\left(\frac{M_\text{IND}}{M_\text{SUSY}}\right)
  - \frac{b_i^{(3)}}{2\,\pi}\,\ln\left(\frac{v_{\Phi}}{M_\text{IND}}\right)
  \nonumber\\
  &  - \frac{b_i^{(4)}}{2\,\pi}\,\ln\left(\frac{v_{T}}{v_{\Phi}}\right)
  - \frac{b_i^{(5)}}{2\,\pi}\,\ln\left(\frac{v_{\Sigma}}{v_{T}}\right)
  - \frac{b_i^{(6)}}{2\,\pi}\,\ln\left(\frac{M_\text{GUT}}{v_{\Sigma}}\right)
\,.
\end{align}
where $M_\text{IND}$ denotes the additional see-saw induces scale introduced
in (\ref{eq:Mind}).  Since this scale depends on the numerical values of
superpotential parameters, which we do not determine, we treat
this as a free parameter.  Distinguishing different model classes with
their corresponding hierarchy patterns, we have to appropriately adapt
the ordering of scales and the definition of the $b_i^{(n)}$.

The calculation of the coefficients in this formula is straightforward
and can be found in Appendix~\ref{sec:runningcoefficients}.  For
simplicity, we always assume that the listed scales exhaust the
available hierarchies, and no further hierarchies from couplings
become relevant here.  Thus, we assume all additional scalar fields
to be integrated out at their ``natural'' mass scale, which is
determined by the considerations in the previous section.

In passing, we note that the calculation is
actually independent from a supersymmetry assumption.  The
supersymmetric and non-supersymmetric frameworks differ only in the
form of the $\beta$-function (cf.~(\ref{eq:runninggcoefficientsSM}),
(\ref{eq:runninggcoefficientsSUSY})).

Regarding $U(1)$ couplings with their normalization ambiguity, we have
to explicitly consider the unification condition for
\begin{equation}
   U(1)_R\ \otimes\ U(1)_{B-L} \; \longrightarrow \; U(1)_\text{Y} \,.
\end{equation}

To define the hypercharge coupling strength, we explicitly calculate
the unbroken direction and identify the charges of the larger
groups. This results in a relation between the group generators and
therefore between charges and couplings.  We obtain\footnote{We do not
  rescale $U(1)$ in order to match the $SU(5)$ normalization, as is often
  done in the literature.}
\begin{subequations}
\begin{align}
Y &=\, \frac{B-L}{2} \,+\, T_3^R \\
\label{eq:LRcondition}
\alpha^{-1}_Y\left(v_{\Phi}\right) &=\, \tfrac{2}{3}\
\alpha^{-1}_{B-L}\left(v_{\Phi}\right) \,+\,
\alpha^{-1}_{R}\left(v_{\Phi}\right)
\,.
\end{align}
\end{subequations}

For the non-abelian symmetry breaking steps, the unification
conditions just depend on the breaking pattern, i.e.,
\begin{subequations}
\begin{align}
 \text{GUT} &\quad \longrightarrow \quad  SU(4)_C \;\otimes\; SU(2)_{L}
\;\otimes\;SU(2)_R
 \,,\\
 SU(4)_C &\quad \longrightarrow \quad  SU(3)_C \;\otimes\; U(1)_{B-L} \,,\\
 SU(2)_R &\quad \longrightarrow \quad U(1)_R \,.
\end{align}
\end{subequations}
where the GUT group contains $SO(10)$.  These three breaking patterns
result in the matching conditions\footnote{In case one of the breaking steps is
absent, the corresponding conditions apply at the next step below.}
\begin{subequations}
\begin{eqnarray}
\label{eq:GUTcondition}
& \alpha^{-1}_4\left(M_\text{GUT}\right) \;=\; \alpha^{-1}_{L}
  \left(M_\text{GUT}\right) \;=\; \alpha_R^{-1}\left(M_\text{GUT}\right)
  \;\equiv\; \alpha^{-1}_\text{GUT} \left(M_\text{GUT}\right)
\,,\\ \label{eq:PScondition}
& \alpha^{-1}_3\left(v_{\Sigma}\right) \;=\; \alpha^{-1}_{B-L}
  \left(v_{\Sigma}\right) \;\equiv\; \alpha^{-1}_4 \left(v_{\Sigma}\right)
\,,\\
& \alpha^{-1}_{U(1)_R}\left(v_{T}\right) \;=\; \alpha^{-1}_{R}
\left(v_{T}\right) \;=\; \alpha^{-1}_{L} \left(v_{T}\right) 
\end{eqnarray}
\label{eq:highercondition}
\end{subequations}
respectively.

In addition to the unification and matching conditions, we have the additional
constraint that the mass scales are properly ordered.  For instance,
for class B we have:
\begin{equation}
M_\text{SUSY}\,\leq\, M_\text{IND} \,\leq\, v_\Phi \,\leq v_T
\,\leq v_\Sigma \,\leq\, M_\text{GUT} \,\lesssim\, 10^{19}\,\GeV \,.
\label{eq:masscondition}
\end{equation}
Furthermore, the coupling strengths $\alpha_i$ have to be sufficiently
small and positive at all mass scales, so we do not leave the
perturbative regime.

Counting the number of conditions and free parameters (scales), we
observe that the models are still under-constrained.  Hence, we can
derive constraints for the mass scales and exclude particular models,
but not fix all scales completely.  Nevertheless, imposing unification
does restrict the model parameter space significantly, as we can show
in the following sections.

\section{Supersymmetric Pati-Salam Models}
\label{sec:SUSY}

In this section, we study the supersymmetric models that are
consistent with our set of assumptions.  We scan over all
models by varying the number of superfield
generations (0,1,3) that are present in each effective theory (i.e., between
the various symmetry breaking scales), independently for each gauge
multiplet. 

We force the masses of all superfields that are not part
of the low-energy (MSSM) spectrum, to coincide with one of the
thresholds $M_\text{IND}$, $v_T$, $v_\Phi$, etc., as explained in
section~\ref{sec:spectra}. In any case, order-one prefactors in the
mass terms would only enter logarithmically in the gauge-coupling
unification (\ref{eq:runninggaugecoupling}) and matching conditions
(GCU)
(\ref{eq:LRcondition},\ref{eq:highercondition},\ref{eq:masscondition}).
This is a minor uncertainty.  We should note, however, that additional
coupling hierarchies, as they exist in the flavor sector of the known
matter particles, are also possible and lead to a wider range of
possibilities which we do not investigate further.

With these conventions understood, the model scan will be exhaustive,
since we vary just discrete labels.  In total, there are 1078 distinct
configurations. Since we now are dealing with mass scales rather then
vev structures or specific mass eigenvalues, we change our notation
from vevs to mass scales (see Appendix~\ref{sec:notationswitch}).

For each model, we analytically calculate the numerical values of
mass scales for which the GCU conditions stated in the previous
section can be fulfilled.  Models where no solution is possible are
not considered further.  For the remaining models, we obtain
model-specific relations between the mass scales.  As a result, we can
express those scales as functions of one or two independent mass
parameters that we may choose as input.  Varying those within the
possible range, we obtain allowed ranges for all mass scales, within
each model separately.

For all numerical results, we fix the common soft SUSY-breaking scale
at $M_\text{SUSY}=2.5\,\TeV$.\footnote{We also considered a lower
SUSY-breaking scale of $M_\text{SUSY}=250\,\GeV$ which is
disfavored by LHC data; it turns out that the unification conditions
are generically easier to satisfy for the larger value of the soft
SUSY-breaking scale.}

\subsection{Extra mass parameters}
\label{sec:runningmass}

As discussed in the previous section, in some model classes the fields
$E$, $\Sigma$ and $T$ do not necessarily obtain a mass term from
symmetry breaking. Thus, their masses must be treated as extra free
parameters.  To get a handle on these scales, we considered all
possibilities for assigning the mass scales of these superfields to
the other mass scales in our framework, while keeping the GCU
conditions.  As a result, we can exclude the possibility that these
extra scales are at the lower end of the spectrum.

More specifically, in all model classes we find a lower bound for the
colored $E$ multiplet of about $m_E\gtrsim 10^8\,\GeV$.  Similar
results apply to $\Sigma$ and $T$, if we do not consider lowering the
GUT scale below about $10^{16}\,\GeV$.  In other words, the proton
stability constraint which limits the GUT scale, suggests that these
fields have rather large masses.  For the further scan over models, we
fix their mass scales, whenever they are not determined by the vevs,
at $M_\text{PS}$.

Any model has to provide a candidate for the electroweak Higgs boson.
This excludes a large invariant mass for the corresponding PS
multiplet.  We therefore do not include an explicit mass term for the
$h$ superfield.  We have checked to what extend $h_\Phi$ may serve as
the low energy MSSM-like Higgs.  This is possible, but only for class
E.  Thus, we must include at least one generation of $h$ in the
classes B, C and F. This reduces our scan to 828 configurations.

\subsection{General overview}

Before we discuss the various classes and types of models in more
detail, we summarize generic features and specific observations that
we can extract from the study of all 828 supersymmetric models.

Roughly one half of all models are capable of GCU\@. Except for class E,
all such configurations exhibit a unification scale $M_\text{GUT}>10^{16}\,
\GeV$ and are thus favored by the non-observation of proton decay. In class E
this is true for half of them. 

In contrast to the classes C, E and F, the allowed ranges in class B
are rather constrained, so in this class, models can be fixed in a
semi-quantitative way.

We now take a look at the low energy spectrum of the different classes. These
can be extra color triplets ($F$) or $SU(2)_L$ triplets ($T_L$).
In particular, we are interested in models where the lowest new
threshold $M_\text{IND}$ is already in the $\TeV$-range, so new
particles are potentially accessible at colliders, while full unification
occurs near the Planck scale.
114 models satisfy these conditions.  72 of them are categorized as class
E and as such contain only light color triplets. 34 of them are categorized as
class C. In class B, only a few models fulfill this condition, none in
class F.

One key feature of our configurations is that we allow for up to six
light $SU(2)_L$ scalar bidoublets $h$ and $h_\Phi$ (three each).  Most
configurations with more than one fall in class E. But also in class C
there is a handful of configurations. In class E most successful
configurations have more than one light bidoublet. This is because $h$
and $h_\Phi$ are taken as degenerate in mass. A more detailed
discussion will follow when we look explicitly at class E.

In class E the LR-breaking scale is allowed to be rather
light.  We find $120$ distinct models with $M_\text{LR}<10^5\,\GeV$. Also in
class C we find some configurations.

As a generic observation, $SU(2)$ triplets $T$ rarely get low mass,
and, if present in the intermediate range, tend to be associated with
lower GUT scales. One exception is class F with three generations of light
triplets $F$. Here, light $T$ are realized for a GUT scale near
$M_\text{Planck}$.
Light color triplets $F$, i.e., extra quarks and
their superpartners, are more common.  Actually, in class E they are
allowed over a large mass range for the GUT scale.  In classes A to D,
we may have color triplets around some $100\,\TeV$, as long as there is only one
generation of light $SU(2)$ triplets.  In class F, the fields $T$ and $F$ are
generically more heavy (cf.\ section~\ref{sec:classF}).

\begin{figure}[tbp]
 \centering
 \includegraphics[width=.95\textwidth]{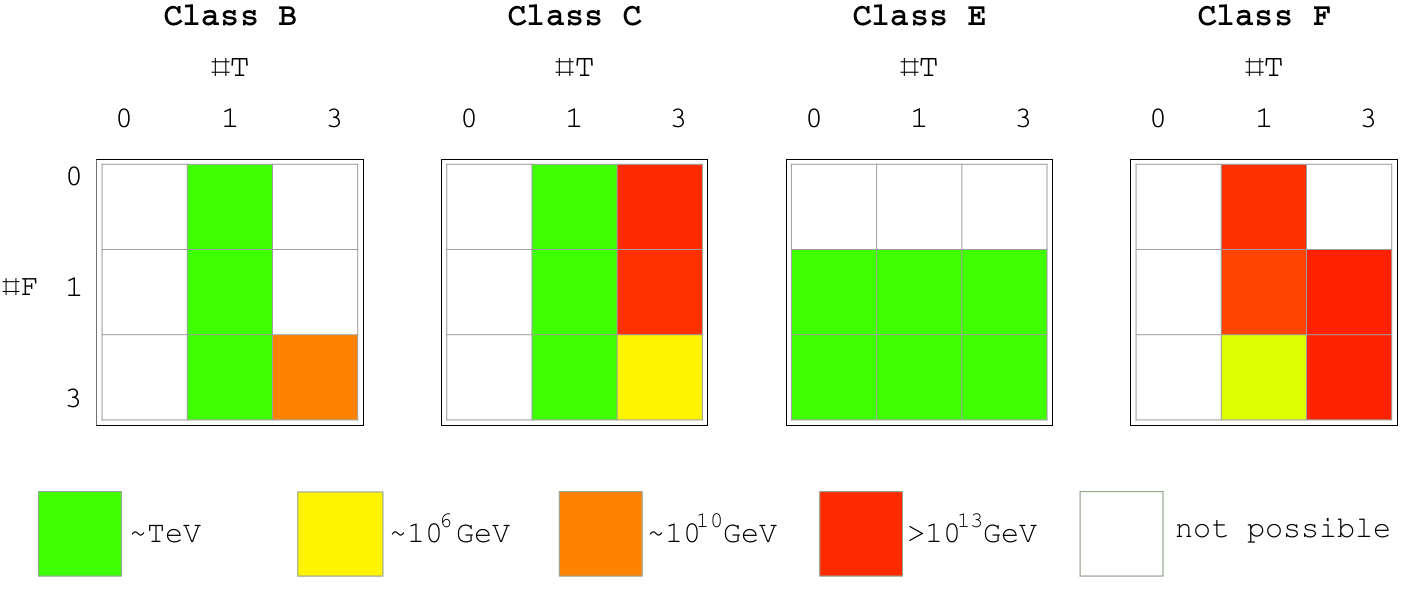}
 \caption{Graphical illustration of the lowest new mass scale,
   dependent on the multiplicity in the low-energy spectrum.  We vary
   the number of low-lying $SU(2)$ triplets $T$ (x axis) and low-lying
   color triplets $F$ (y axis), independently.  The colors indicate
   the lowest mass scale, ranging from green (SUSY scale) to red
   (Planck-scale).  White6 squares correspond to configurations not leading to
   GCU or which are inconsistent with the class definitions.}
 \label{fig:lowscalesoverview}
\end{figure}

We illustrate these results in figure~\ref{fig:lowscalesoverview}.  The
figure displays a considerable fraction of models where new matter is
possible at the lowest scales (green squares), so we should be
prepared to observe exotic particles, or at least their trace in
precision observables, at collider experiments.

Another observable of interest is the preferred mass range for
right-handed neutrinos.  In our setup, the Majorana mass parameter
should be of the order $\vev{\Phi}$ where all symmetries that protect
this term are broken.  Scanning all configurations with respect to this
scale, we find 184 with $10^{12}\,\GeV \lesssim
M_{N_R}\lesssim 10^{14}\,\GeV$. Actually, in classes B, C, three-quarter  and
in class E still half of all successful models fall in this category. Only in
class F, this scale is typically higher; only $10\%$ of the models result in a
value in this range.

We also find that a neutrino mass scale in this range is associated
with at most one generation of $h$ fields.  As an interesting
non-standard scenario, hinting at $E_6$ grand unification, three light
$h$ generations are also possible, but accompanied by heavier
right-handed neutrinos.  In classes B and C, we find a few successful
models with three $h$ generations, and none in class F.  However,
there is quite some space for this scenario in class E.

\begin{table}
\centering
 \begin{tabular}{|l|c|c|c|c|c|}
  \hline &&&&&\\[-2ex]
  & class B & class C & class E & class F & $\sum$ \\
  \hline &&&&&\\[-2.5ex] \hline &&&&&\\[-1.5ex]
  scanned& 144 & 144 & 324 & 216 & 828 \\[1ex]
  GCU & 18 & 57 & 254 & 29 & 358 \\[1ex]
  $M_\text{GUT}>10^{16}\,\GeV$& 18 & 57 & 131 & 29 & 235 \\[1ex]
  $M_\text{IND}<10\,\TeV \text{ and } M_\text{GUT}>10^{16}\,\GeV$ & 8 & 34 & 72
& 0 &
    114 \\[1ex]
  $M_\text{LR} < 100\,\TeV $ & 1 & 11 & 108 & 0 & 120
    \\[1ex]
  $10^{12}\,\GeV < M_{N_R} < 10^{14}\,\GeV $ & 16 & 42 & 123 & 3 &
    184 \\
  $ M_\text{IND} \in\,[0.1,10]\; \frac{v_{\Phi}^2}{v_{\Sigma}+v_{T}}$ & 14 & 20
    & 203 & 26 & 263 \\[1ex]  
    \hline
 \end{tabular}
 \caption{Number of configurations full filling certain conditions.}
 \label{tab:SUSYOverview}
\end{table}

So far, we treated $M_\text{IND}$ as a free parameter.  To obtain more
specific predictions, we imposed the restriction $M_\text{IND}\in
\,[0.1,10] \;\frac{v_{\Phi}^2}{v_{\Sigma}+v_{T}}$.  While this does
not significantly reduce the number of allowed models, it drastically
reduces the configurations with $\TeV$-scale new particles.  Most of
these models still allowing $\TeV$-scale new particles belong to class $E$.  In
this class, one quarter of all models allows for colored triplets below about
$10\,\TeV$. More generic statistics can be read off
table~\ref{tab:SUSYOverview}.

In some models, the GCU constraints pin down all scales to a narrow
range.  The most obvious case is standard $SO(10)$ coupling
unification at the GUT scale, i.e., all vevs are of the same order of
magnitude and located at $M_\text{GUT}$.  Clearly, this well-studied
model is contained in our scan as a limiting case.  We reproduce
the observation that for this case, the only light multiplet is the
MSSM Higgs $h$.  However, we also find a few models where scales are
essentially fixed, but the spectrum and unification pattern is
different. Those are all classified as class~E.


\subsection[Class E]{Class E: $v_T=0$ and $v_\Sigma \neq 0$}

From the overview above we can conclude that class E contains the
largest set of models with phenomenologically interesting features.
In this class, the ordering of new thresholds is, in
ascending order: the scale of soft SUSY breaking $M_\text{SUSY}$, the see-saw
induced scale $M_\text{IND}$, the left-right unification scale $M_\text{LR}$,
the scale where Pati-Salam symmetry emerges $M_\text{PS}$, and the scale of
complete gauge-coupling unification $M_\text{GUT}$.
\begin{align}
  &E:\ M_\text{SUSY} \leq M_\text{IND}  \leq M_\text{LR} \leq M_\text{PS} \leq
M_\text{GUT}
\end{align}

Two of those scales can be regarded as
free parameters; we may take the lowest ($M_\text{IND}$) and highest
($M_\text{GUT}$) scale for that purpose.  The other scales are then
fixed by the matching and unification conditions, if they can be
satisfied at all.

There are 324 class-E models.  In 254 configurations it is possible to
implement
gauge-coupling unification (GCU), i.e., satisfy all matching and unification
conditions.  131 of these configurations allow for a scale
$M_\text{GUT}>10^{16}\,\GeV$. 76 configurations are able to produce GCU at the
Planck scale.

As discussed in the previous section, in class E the superfield $h$ is not
necessarily contained in the spectrum. We found 77 configurations leading to
GCU in which the MSSM-like Higgs is $h_\Phi$.
As mentioned above, most of the configurations have more than one bidoublet at
the EWSB scale. Only 29 feature exactly one. There are roughly 50 sets with 1, 2
or 6 bidoublets each and 81 with 4. Zero or five are generally excluded by our
setup. We also find that a larger number of light bidoublets is correlated
with lower scales. Especially in the case of six bidoublets, we find that the
maximal value for the GUT scale is $M_\text{GUT}<10^{16}\,\GeV$.

Note that in some models of class E, the see-saw scale $M_\text{IND}$ is
not phenomenologically relevant (cf.\ table~\ref{tab:masstable}), since
they do not contain the colored superfield $F$.  Thus, we should break
down the model space according to the multiplicity of the $F$
multiplet: zero on the one hand (no see-saw scale), one or three on
the other hand.

If there is no field $F$, the lowest-lying threshold above the soft
SUSY-breaking scale is the scale of left-right-handed unification,
$M_\text{LR}$.  It turns out that, in some models, this scale can be
as low as the SUSY scale. At the other end of the spectrum, the
complete unification scale $M_\text{GUT}$ can vary in the range between
$10^{9}\,\GeV$ and $10^{19}\,\GeV$.

In the cases of one or three generations of $F$, the see-saw scale can
be as low as the soft SUSY-breaking scale, independent of the GUT
scale.  The upper bound for the see-saw scale is only fixed by the
requirement that it is the lowest-lying scale, and is approximately
$M_\text{IND}\lesssim10^{16}\,\GeV$.

As mentioned before, there are configurations fixing all scales. These posses
three generations of $h$, $\Phi$, $T_{L/R}$ and one generation of $\Sigma$. The
multiplicity of $E$ is not fixed. For three generations of $E$ also one or zero
generations of $T_{L/R}$ are possible. The LR-scale is in these configurations
fixed to $M_\text{LR}=7\times 10^{3}\,\GeV$ and the PS scale to be 
$M_\text{PS}=10^{9}\,\GeV$.

Let us now consider in somewhat more detail, the three particular
model types described in section~\ref{sec:model}.

\subsubsection*{Type Em: Minimal Model}

\begin{figure}[tbp]
\begin{minipage}[b]{.48\textwidth}
 \centering
 \subfigure[Possible scale variation leading to GCU\@. The GUT-scale is shown
in   black, the PS-scale in blue and the LR-scale in red. A IND-scale is not
present  in this type. The dots indicate the scales for the exemplary plot shown
in  (b)]{
 \includegraphics[width=.95\textwidth]{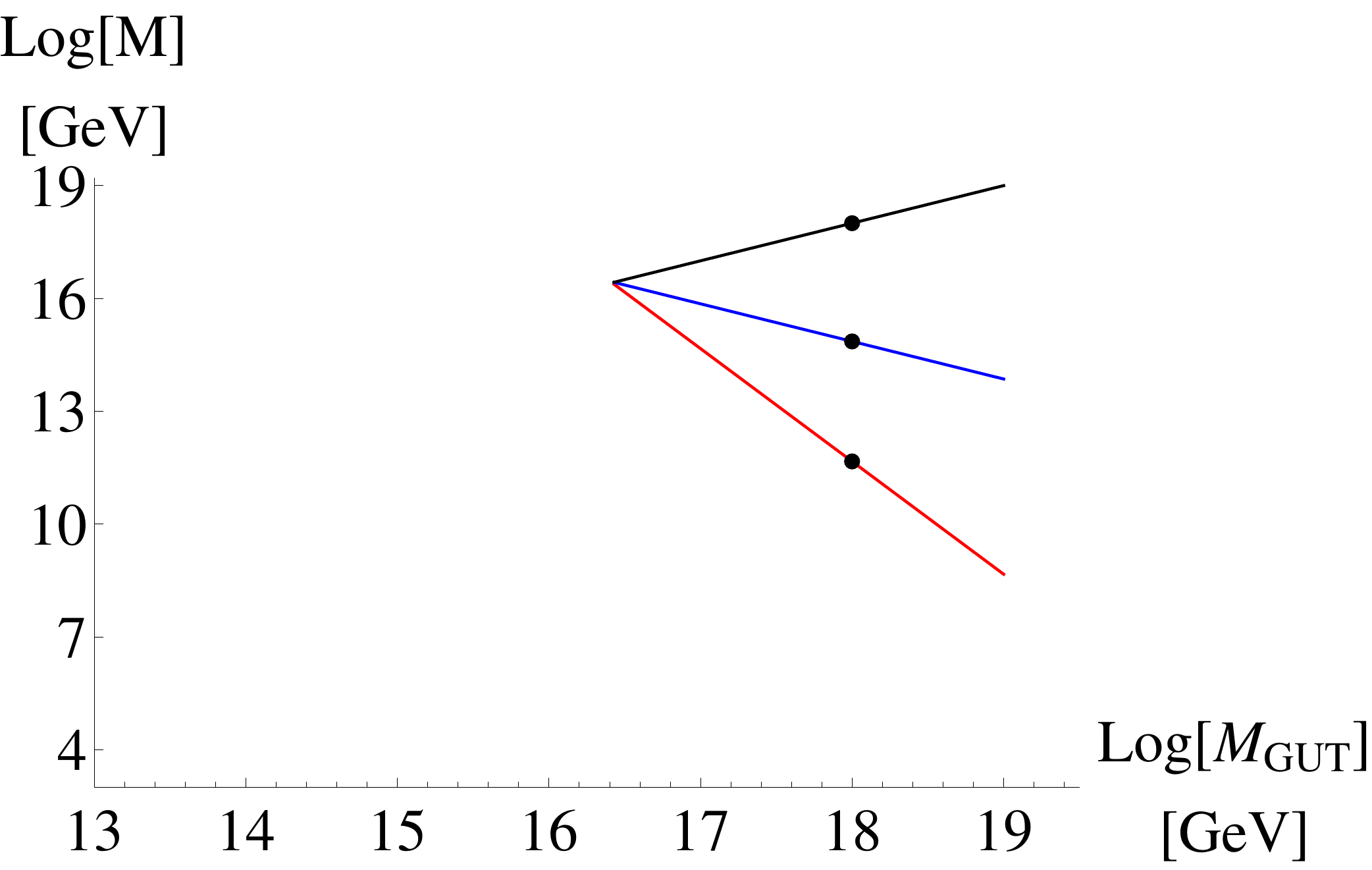}
 \label{fig:varscaleEm}}
\end{minipage}\hfill
\begin{minipage}[b]{.48\textwidth}
 \centering
 \subfigure[Exemplary running of the gauge couplings for complete unification
  at $M_\text{GUT}=10^{18}\,\GeV$. The hypercharge coupling is shown in red, the
  B-L in green, the weak in blue and the strong coupling in black.]{
 \includegraphics[width=.95\textwidth]{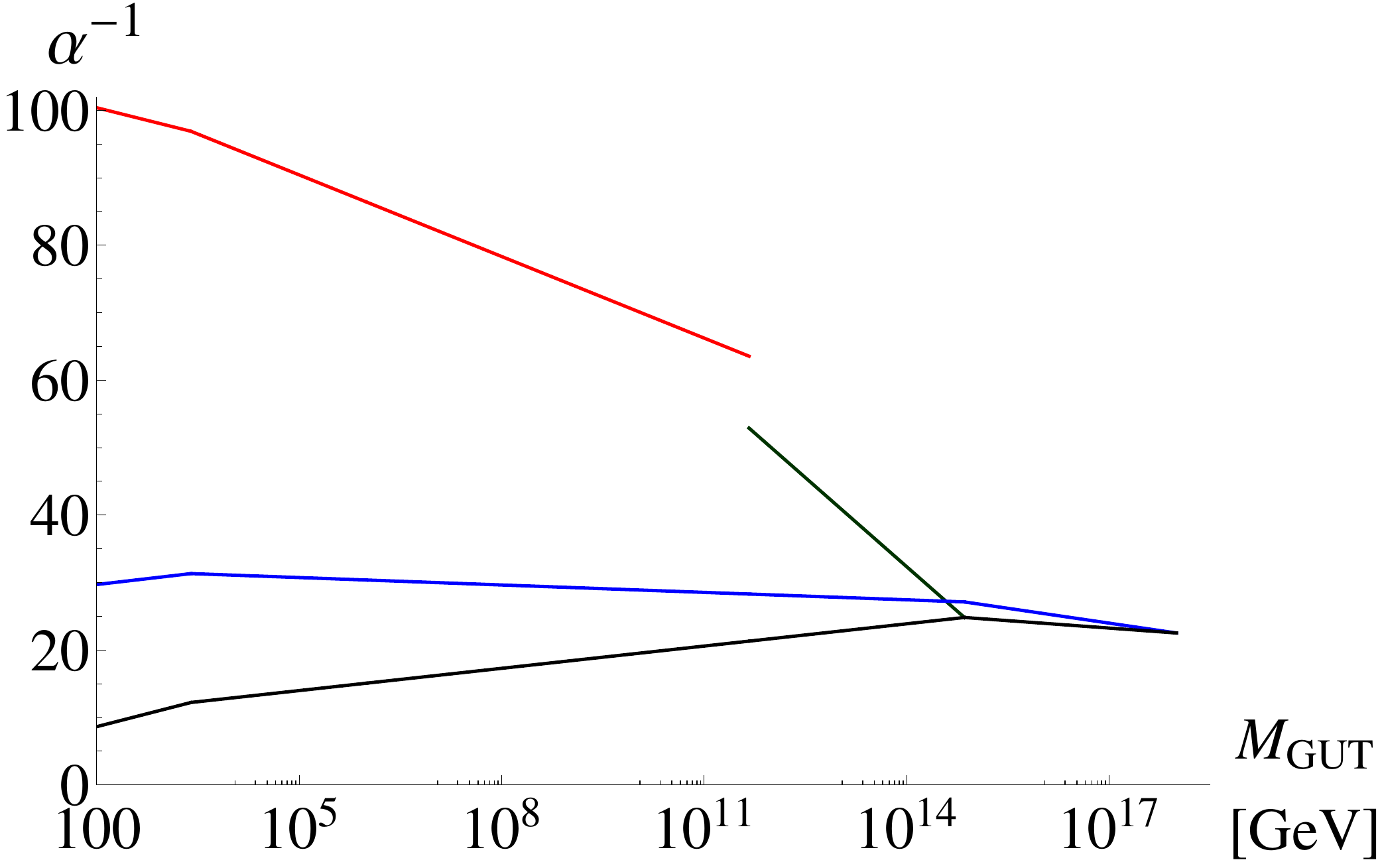}
\label{fig:unificationEm}}
\end{minipage}
\caption{Variation of the unification scales and exemplary running of the
gauge couplings for the type Em.}
\end{figure}

In class E, the minimal model is the standard MSSM without Higgs\footnote{the
electroweak Higgs is contained in $\Phi_L$}, supplemented only by
the additional fields $\Phi$ and $\Sigma$ above their respective
thresholds.  Looking at table~\ref{tab:masstable}, we see that this
setup does not provide a see-saw scale.  Hence, the sub-unification
scales depend only on one free parameter, which we take to be
$M_\text{GUT}$. Figure~\ref{fig:varscaleEm} shows the variation of the
three other scales as a function of $M_\text{GUT}$.

For the value of $M_\text{GUT}\approx3\times10^{16}\,\GeV$, all scales
approximately
coincide.  This is the minimal $M_\text{GUT}$ value for which GCU is
possible in this setup.  For this particular parameter point, the GUT
symmetry (e.g., $SO(10)$) directly breaks down to the MSSM by virtue
of $v_\Sigma=v_\Phi$, so this is actually the standard $SO(10)$
scenario.  If we demand a larger GUT scale, the Pati-Salam scale decreases but
never drops below $M_\text{PS}\gtrsim10^{14}\,\GeV$. The left-right
unification scale can vary in the range $10^{11}\,\GeV\lesssim
M_\text{LR}\lesssim 10^{16}\,\GeV$.  This is within the favored mass
range for right-handed neutrinos.  A sample unification plot is shown
in figure~\ref{fig:unificationEm}.

\subsubsection*{Type Es: SO(10)-like Models}

We now turn to a model with complete $SO(10)$ representations below
the GUT scale.  With this spectrum, we can vary independently
$M_\text{IND}$ and $M_\text{GUT}$, within a certain range. 

It turns out that $M_\text{GUT}$, in this type of model, cannot reach
the Planck scale.  The maximal allowed value for $M_\text{GUT}$
depends on $M_\text{IND}$ and decreases with increasing $M_\text{IND}$.
The value of $M_\text{IND}$, and thus the mass of the color-triplet
fields $F$, can be as low as the soft SUSY-breaking
scale.

\begin{figure}[tbp]
\begin{minipage}[b]{.48\textwidth}
 \centering
 \subfigure[Possible scale variation leading to GCU\@. The GUT-scale is shown
  in black, the PS-scale in blue and the LR-scale in red. The variation in the
  IND-scale is shown discrete with $M_\text{IND}=10^{3.4}\,\GeV$ as solid
  lines, $M_\text{IND}=10^{5.4}\,\GeV$  dashed $M_\text{IND}=10^{7.4}\,\GeV$
dotted. The
  dots indicate the scales for the exemplary plot shown in (b)]{
 \includegraphics[width=.95\textwidth]{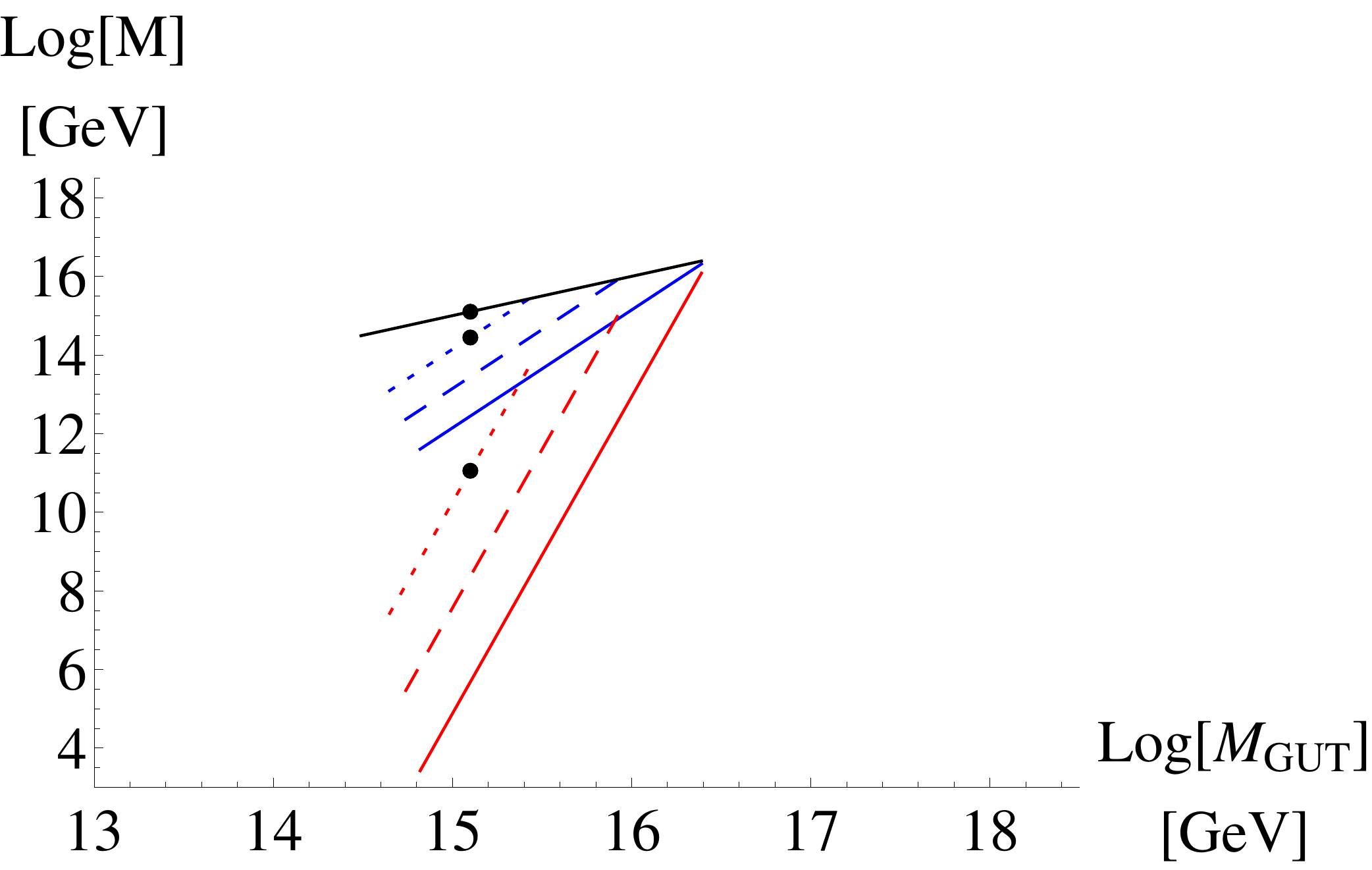}
 }
\end{minipage}\hfill
\begin{minipage}[b]{.48\textwidth}
 \centering
 \subfigure[Exemplary running of the gauge couplings for complete unification
  at $M_\text{GUT}=10^{15.1}\,\GeV$. The hypercharge coupling is shown in red,
  the B-L in green, the weak in blue and the strong coupling in black.]{
 \includegraphics[width=.95\textwidth]{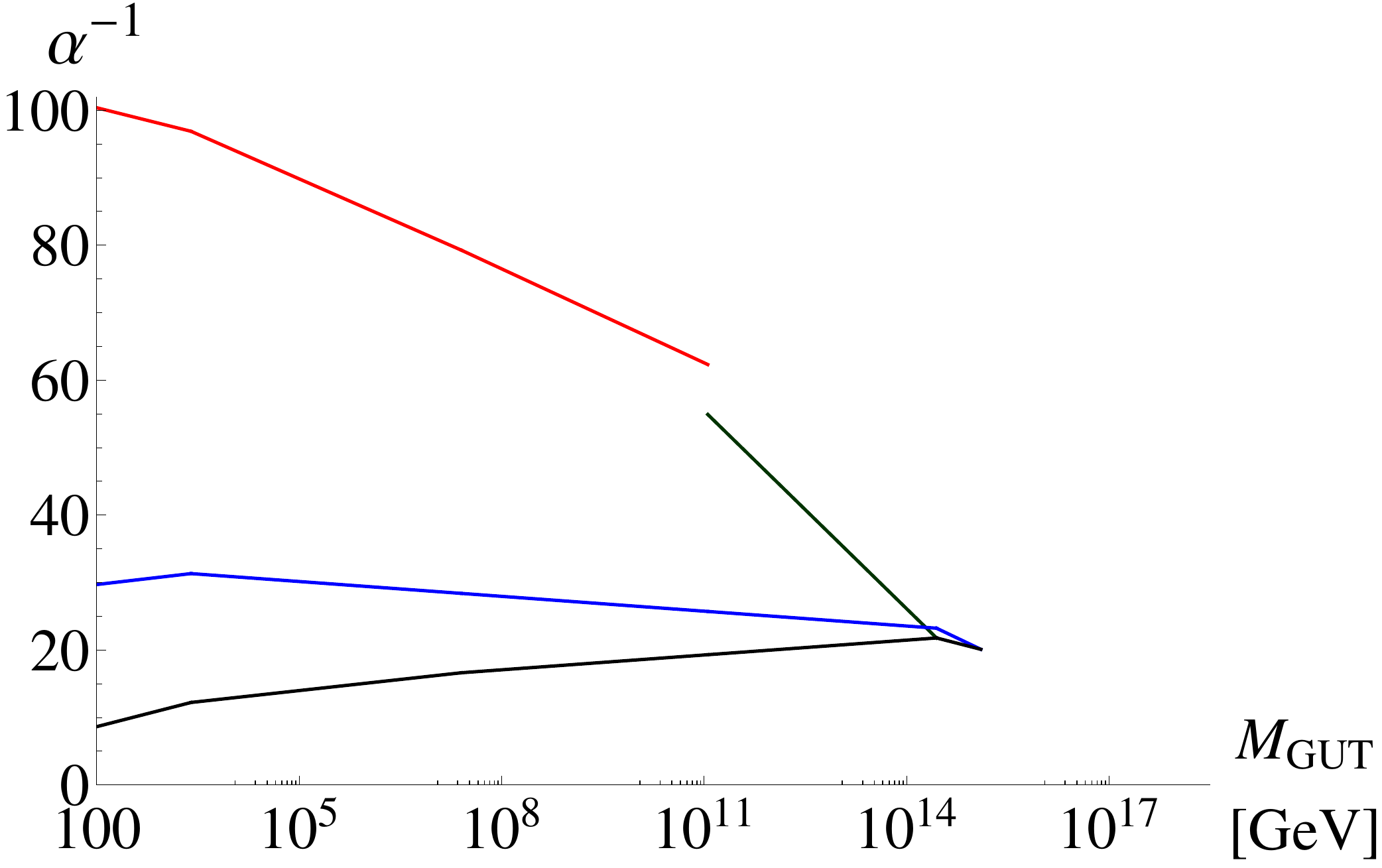}
 }
\end{minipage}
\caption{Variation of the unification scales and exemplary running of the
gauge couplings for the type Es.}
\label{fig:plotsEs}
\end{figure}

Another important difference is that the scales approach each other when
$M_\text{GUT}$ gets larger.  In figure\ref{fig:plotsEs}, we
plot the variation of the sub-unification scales as function of
$M_\text{GUT}$ for three fixed values of $M_\text{IND}$ (solid
$M_\text{IND}=10^{3.4}\,\GeV$, dashed $M_\text{IND}=10^{5.4}\,\GeV$ and dotted
$M_\text{IND}=10^{7.4}\,\GeV$).  For the lowest value of $M_\text{IND}$, it is
possible to have GCU without any sub-unification.  For larger $M_\text{IND}$
we see a gap opening between $M_\text{LR}$ and $M_\text{PS}$, but it is still
possible to achieve $M_\text{PS}=M_\text{GUT}$.

\subsubsection*{Type Ee: $E_6$-inspired Models}

In this model, we combine complete $SO(10)$ multiplets with
three generations of the ``MSSM-like'' Higgs fields $h$ and the
color-triplets $F$, so within each generation, matter fields unify
with the MSSM Higgs fields and an additional singlet each, to complete
$\re{27}$ representations of $E_6$.  In this scenario, GCU is possible
over a wide range of mass scales.

Like in the model Es discussed above, the separation between the
sub-unification scales decreases with increasing scale
$M_\text{GUT}$. Over the whole range of $M_\text{IND}$ it is possible to
have the PS and GUT unification coincide, $M_\text{PS}=M_\text{GUT}$.
Complete unification at a single scale is possible for
$M_\text{GUT}\approx10^{16.4}\,\GeV$ if the scale of light triplets is
equal to the soft SUSY-breaking scale, $M_\text{F}=M_\text{SUSY}
=2.5\times10^{3}\,\GeV$. This is the well known $SU(5)$ limiting case, since all
fields of the low energy spectrum can be grouped to complete $SU(5)$
representations.

\begin{figure}[tbp]
\begin{minipage}[b]{.48\textwidth}
 \centering
 \subfigure[Possible scale variation leading to GCU\@. The GUT-scale is shown
  in black, the PS-scale in blue and the LR-scale in red. The variation in the
  IND-scale is shown discrete with $M_\text{IND}=10^{3.4}\,\GeV$ as solid
  lines, $M_\text{IND}=10^{5.4}\,\GeV$  dashed $M_\text{IND}=10^{7.4}\,\GeV$
dotted. The
  dots indicate the scales for the exemplary plot shown in (b).]{
 \includegraphics[width=.95\textwidth]{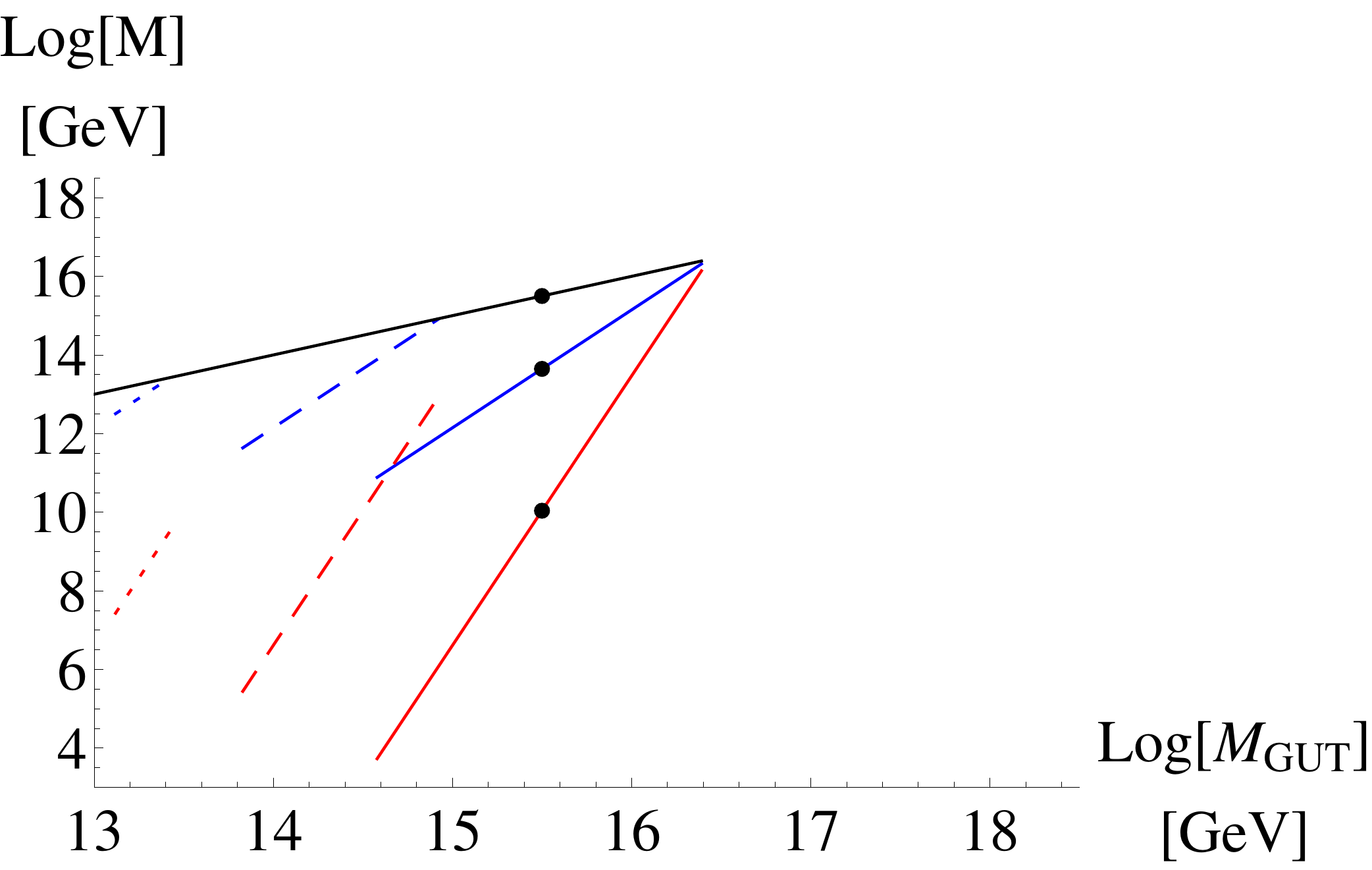}
 }
\end{minipage}\hfill
\begin{minipage}[b]{.48\textwidth}
 \centering
 \subfigure[Exemplary running of the gauge couplings for complete unification
  at $M_\text{GUT}=10^{15.1}\,\GeV$. The hypercharge coupling is shown in red,
  the B-L in green, the weak in blue and the strong coupling in black.]{
 \includegraphics[width=.95\textwidth]{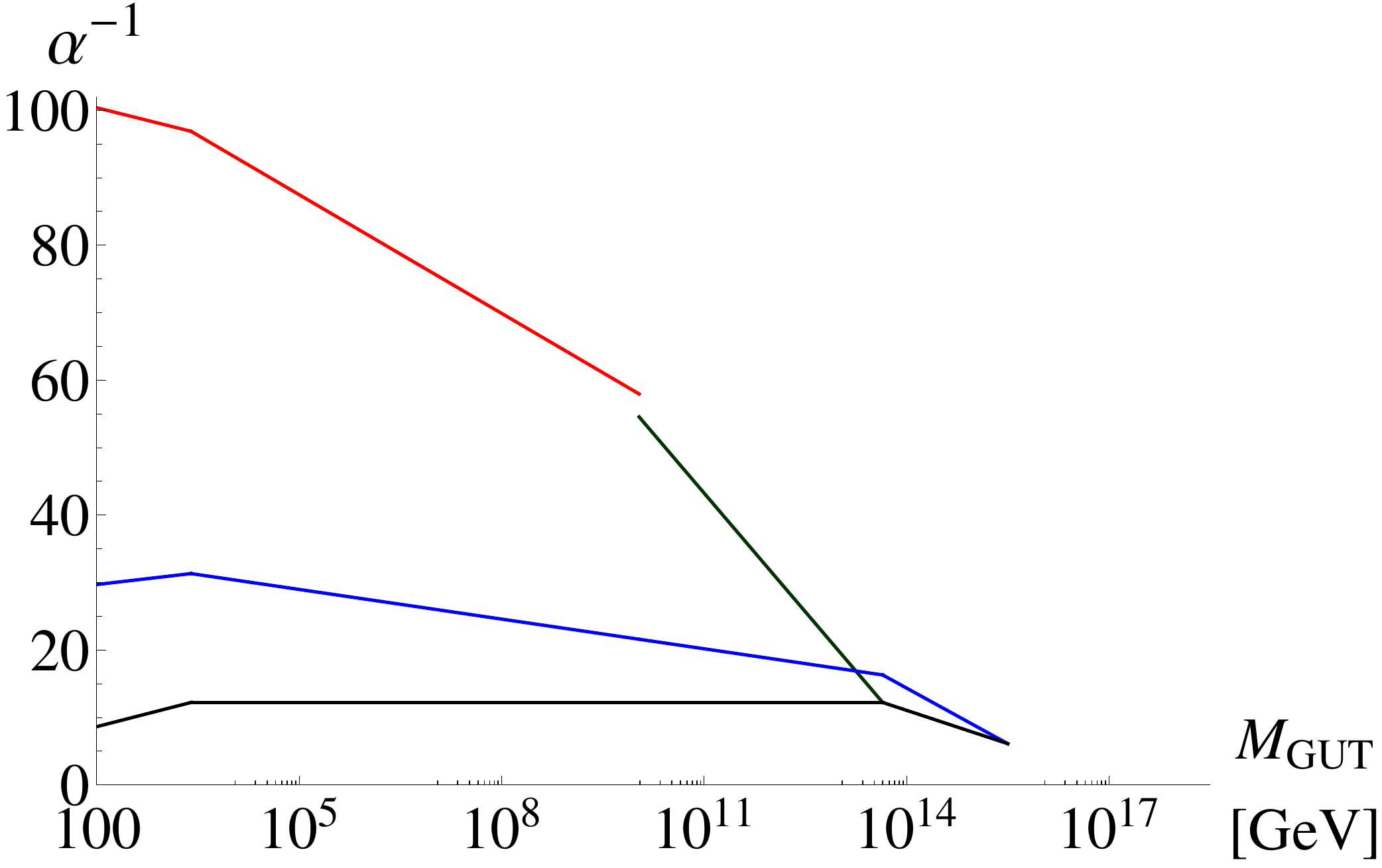}
 }
\end{minipage}
\caption{Variation of the unification scales and exemplary running of the
gauge couplings for the type Ee.}
\label{fig:plotsEe}
\end{figure}

Compared to the previous two model types, the gauge coupling at the
unification point $\alpha_\text{GUT}^{-1}$ is significantly lower and,
in some cases, touches the non-perturbative regime.  In
figure~\ref{fig:plotsEe}, we show the variation of scales and an
exemplary unification plot.

\subsection[Class F]{Class F: $v_T \neq 0$ and $v_\Sigma =0$}
\label{sec:classF}

This model class has a more restricted phenomenology.
Nevertheless, this class contain some models that exhibit GCU\@.

In class-F models, $SU(2)_R$ (and thus LR symmetry) is broken at
$v_T$, above the scale $v_\Phi$ where $SU(4)_C$ reduces to color.  We
therefore might expect closer relations between lepton-flavor and
quark-flavor mixing.  The relevant scales of this class are, in
ascending order: the see-saw induced scale $M_\text{IND}$, the quark-lepton
unification scale $M_\text{QL}$, the Pati-Salam scale $M_\text{PS}$,
and the unification scale $M_\text{GUT}$. 
\begin{align}
  &F:\ M_\text{SUSY} \leq M_\text{IND} \leq M_\text{QL} \leq M_\text{PS} \leq
M_\text{GUT}
\end{align}

Table~\ref{tab:masstable} indicates that all models in this class do have the
additional see-saw scale $M_\text{IND}$.  The intermediate scales tend to be
higher than in class E above.

Out of the 216 class-F models only 29 configurations are consistent with
GCU\@.  In all cases, $M_\text{GUT}$ can be as large as the Planck
scale.

As discussed in the overview, light degrees of freedom are not possible in this
class. As above, we break down the set of configurations according to their
content of light fields.  There are now 6 categories. In addition to
the ones mentioned above (zero, one, or three generations of $F$), we
have to distinguish cases of one or three generations of $SU(2)$
triplets~$T_L$.

The minimal $M_\text{IND}$ value is strongly dependent on the number of
$SU(2)$ triplets.  In the case of three triplets, it is strictly
larger than $10^{16}\,\GeV$, essentially independent of the number of
$F$ fields.  Thus, let us look at the configurations with only one
$T_{L/R}$ generation.

In these configurations, $M_\text{IND}$ is bound to be larger than
$M_\text{IND}\gtrsim10^{6}\,\GeV$. It is realized for three generations
of
$F$ and rises the less are included.

We conclude that in class F, the extra fields may play a
role for flavor physics in an intermediate energy range, but are
unlikely to be observable in collider experiments.

\subsubsection*{Type Fm: Minimal Model}

The minimal model of class F contains the superfields
$\Phi$ and $T_{L/R}$ in addition to the MSSM spectrum.

In models of this type, the lowest possible see-saw mass value is
$M_\text{IND}\approx 10^{12}$ and LR-scale is $M_\text{LR}\gtrsim 10^{15}$. Thus
these are ruled out, since the mass of the EWSB Higgs is associated to the
LR-scale.
A next to minimal setup explicitly including one generation of $h$ is
not able to produce GCU\@.

\subsubsection*{Type Fs/Fe}

In type Fs, there is no model consistent with GCU\@. This is because
$\alpha_3^{-1}$ grows to fast and overshoots $\alpha_2^{-1}$ before
the condition (\ref{eq:LRcondition}) for a possible QL-scale can be
fulfilled. A model of type Fe consistent with GCU is also not possible.

\subsubsection*{Type Ff: Flavor-symmetry inspired Model}

In the absence of the previous types, we take a look at a configuration which
might be viewed as $E_6$-inspired, but with
the additional condition that two of three Higgs bidoublets get heavy
(Planck scale) by some unspecified mechanism.  In this model,
unification is possible over a wide range of mass scales. There is a strong
correlation of $M_\text{IND}$ and $M_\text{GUT}$, so
the latter is the only relevant parameter.  $M_\text{IND}$ can vary
between $10^{7}\,\GeV\lesssim M_\text{IND}\lesssim10^{16}\,\GeV$.  For its
largest
allowed value all scales are approximately equal, which is the $SO(10)$
limiting case. Conversely, the
lowest possible $M_\text{IND}$ value corresponds to GUT unification
near the Planck scale. The QL, the PS and the GUT scales are nearly degenerate
in any case.

As mentioned above, there is no possible configuration leading
to GCU with three generations of the field $h$.


\subsection[Classes A to D]{Classes A to D: $v_T\neq 0$ and $v_\Sigma \neq 0$}

In these classes, we effectively combine model classes E and F.  There
are five different scales, two of which are
fixed by requiring GCU\@.  For concreteness, we also fix
$M_\text{GUT}=10^{18.2}\,\GeV$, i.e, we assume complete unification
at the Planck scale.  Still, we can choose two parameters
independently, so we will obtain allowed and forbidden regions, but no
one-to-one correspondences.

More specifically, we distinguish the cases $v_T \leq v_\Sigma$ (class
B) and $v_T > v_\Sigma$ (class C), where A and D appear as limits. The ordering
of scales in the two scenarios is
\begin{align}
  &B:\ M_\text{SUSY} \leq M_\text{IND} \leq M_\text{U1} \leq M_\text{LR} 
  \leq M_\text{PS} \leq M_\text{GUT}
   \,,\\
  &C:\ M_\text{SUSY} \leq M_\text{IND} \leq M_\text{U1} \leq M_\text{QL} 
  \leq M_\text{PS} \leq M_\text{GUT}\,.
\end{align}

Here, $M_\text{U1}$ indicates the mass scale where the extra $U(1)$ groups
break down to hypercharge. This is also the natural scale for a mass term of the
right-handed neutrinos. The spectrum below $M_\text{U1}$ still
contains the extra particles that are integrated out at the lower
see-saw scale $M_\text{IND}$, which in turn is located above the soft
SUSY-breaking scales. The labels LR and QL refer to left-right and quark-lepton
symmetry breaking, respectively. 

Of the 144 models in classes B and C, 18 (B) and 57 (C) are consistent with GCU,
respectively. We observe again that the number of $T_L$ generations has a strong
impact on phenomenology.  First, we look at the case of three $T_L$ generations.
 In class C, $M_\text{IND}$ depends on the number of generations of the field
$F$. It ranges from $M_\text{IND}\gtrsim10^{15}\,\GeV$ (no $F$) down to to
$M_\text{IND}\gtrsim10^{6}\,\GeV$ (three $F$ generations).  The situation in
class B is even worse. Here GCU is not possible with less then three generations
of the field $F$. 

If there is only one generation of $T$, the $M_\text{IND}$ value can approach
the SUSY scale, independent on the number of $F$.

\subsubsection*{Type Bm/Cm: Minimal Model}

In the minimal model, there is a single generation of each of
$\Phi$, $T_{L/R}$ and $\Sigma$.  We can achieve GCU for both Bm
and Cm.  The see-saw scale is stuck at rather high values,
$M_\text{IND}\gtrsim 10^{13}\,\GeV$ and $M_\text{IND}\gtrsim
10^{10}\,\GeV$ for type Bm and Cm, respectively. Since we again do not
explicitly add a EWSB Higgs, such models are ruled out because of the
nonexistence of light $SU(2)$ doublets. If we include one generation of $h$ in
addition, GCU is no longer possible.

\subsubsection*{Types Bs/Cs and Be/Ce}

These setups do not allow GCU\@.

\subsubsection*{Class-B/C Models with $M_\text{IND}<10\,\TeV$}

\begin{figure}[tbp]
\begin{minipage}[b]{.48\textwidth}
 \centering
 \subfigure[Possible scale variation leading to GCU\@. The PS-scale is shown in 
  black, the LR-scale in blue and the MSSM-scale in red. The variation in the
  IND-scale is shown discrete with $M_\text{IND}=10^{4}\,\GeV$ as solid
  lines, $M_\text{IND}=10^{7}\,\GeV$  dashed $M_\text{IND}=10^{10}\,\GeV$
dotted. The
  dots indicate the scales for the exemplary plot shown in (b).]{
 \includegraphics[width=.95\textwidth]{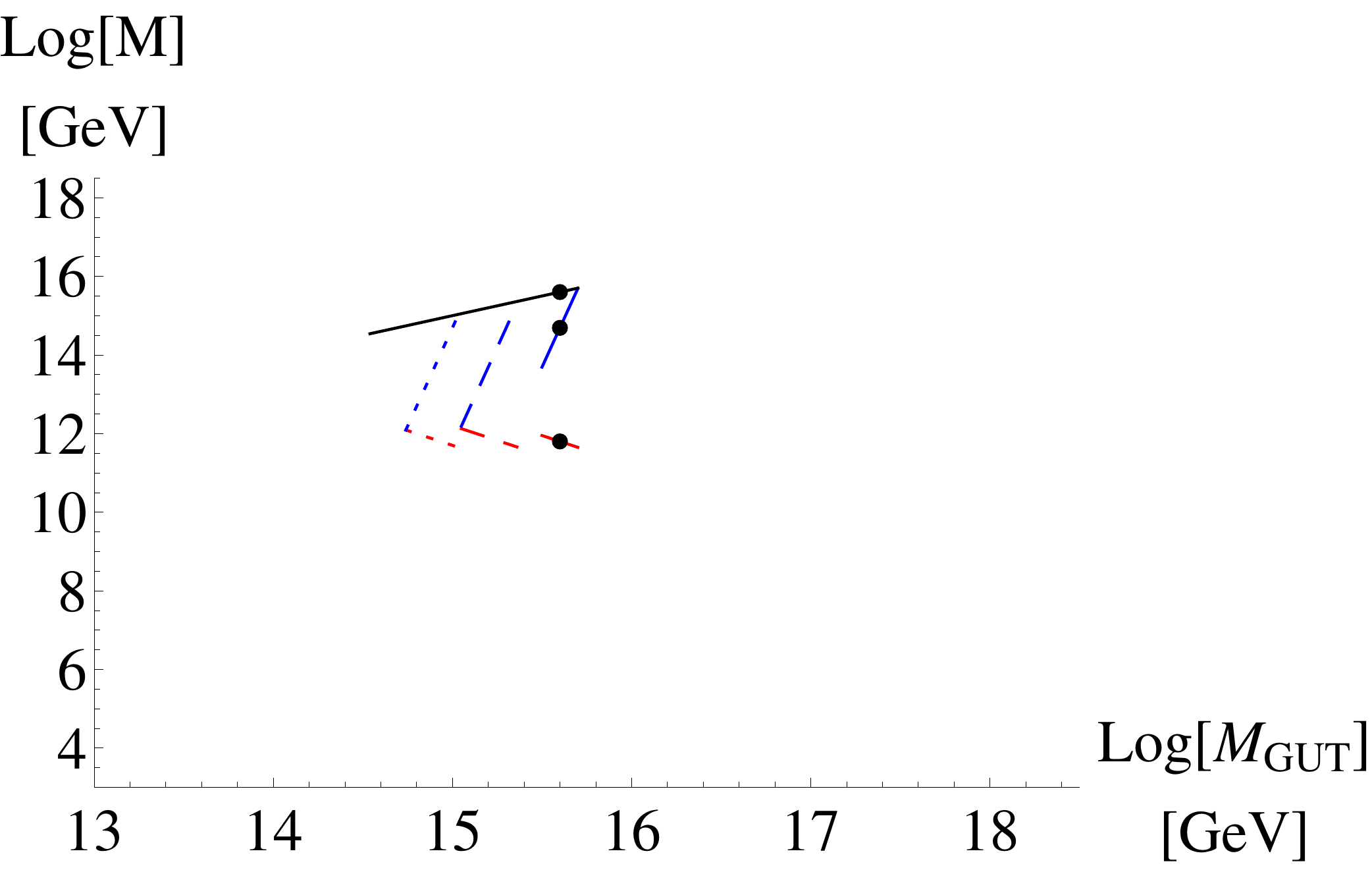}
 }
\end{minipage}\hfill
\begin{minipage}[b]{.48\textwidth}
 \centering
 \subfigure[Exemplary running of the gauge couplings for complete unification
  at $M_\text{GUT}=10^{18.2}\,\GeV$. The hypercharge coupling is shown in red,
  the $U(1)_R$ in brown, the B-L in green, the weak in blue and the strong
  coupling in black.]{
 \includegraphics[width=.95\textwidth]{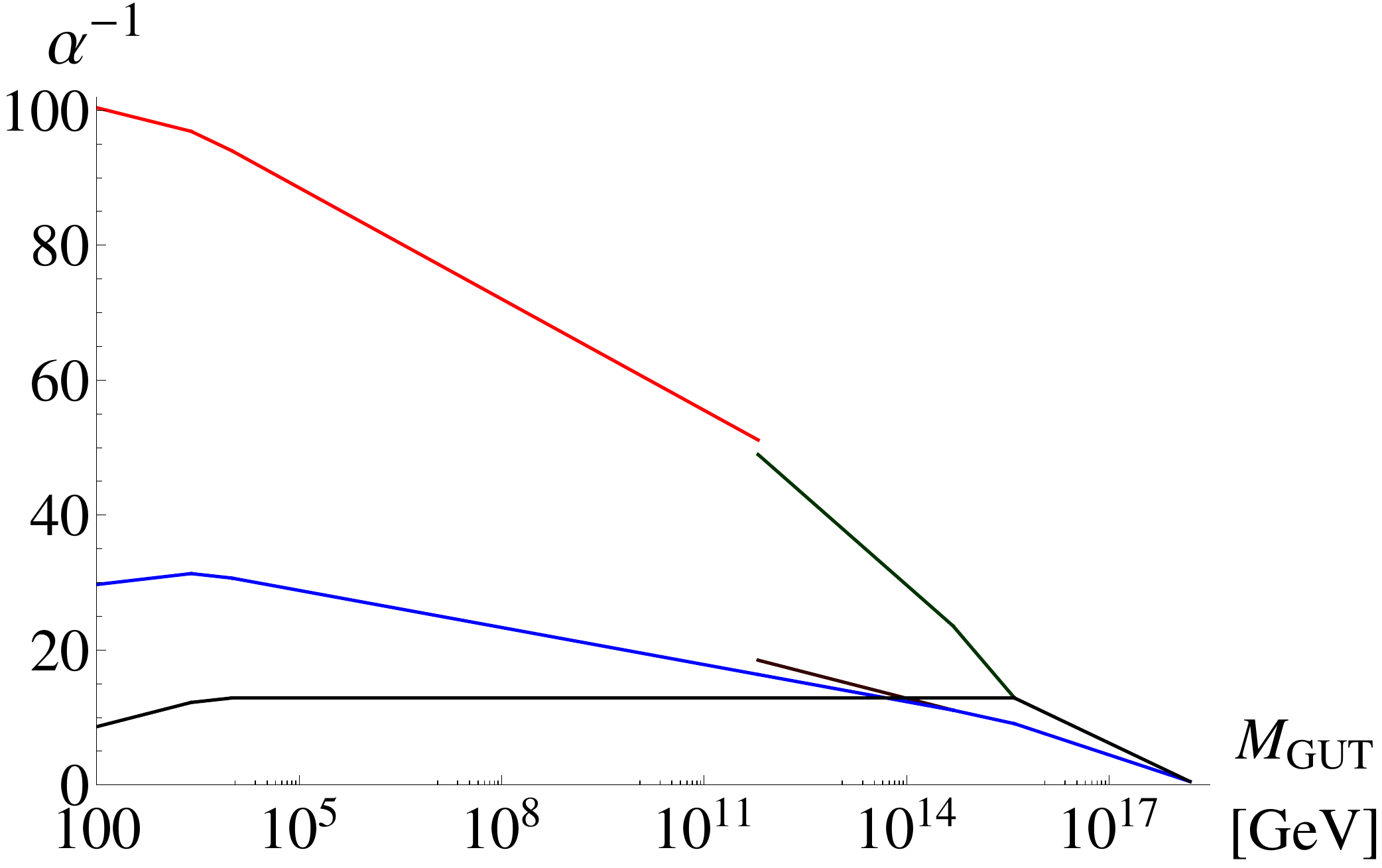}
 }
\end{minipage}
\caption{Variation of the unification scales and exemplary running of the
gauge couplings for the type B199.}
\label{fig:plotsB199}
\end{figure}

\begin{figure}[tbp]
\begin{minipage}[b]{.48\textwidth}
 \centering
  \subfigure[Possible scale variation leading to GCU\@. The QL-scale is shown
in 
  black, the PS-scale in blue and the MSSM-scale in red. The variation in the
  IND-scale is shown discrete with $M_\text{IND}=10^{4}\,\GeV$ as solid
  lines, $M_\text{IND}=10^{7}\,\GeV$  dashed $M_\text{IND}=10^{10}\,\GeV$
dotted. The
  dots indicate the scales for the exemplary plot shown in (b).]{
 \includegraphics[width=.95\textwidth]{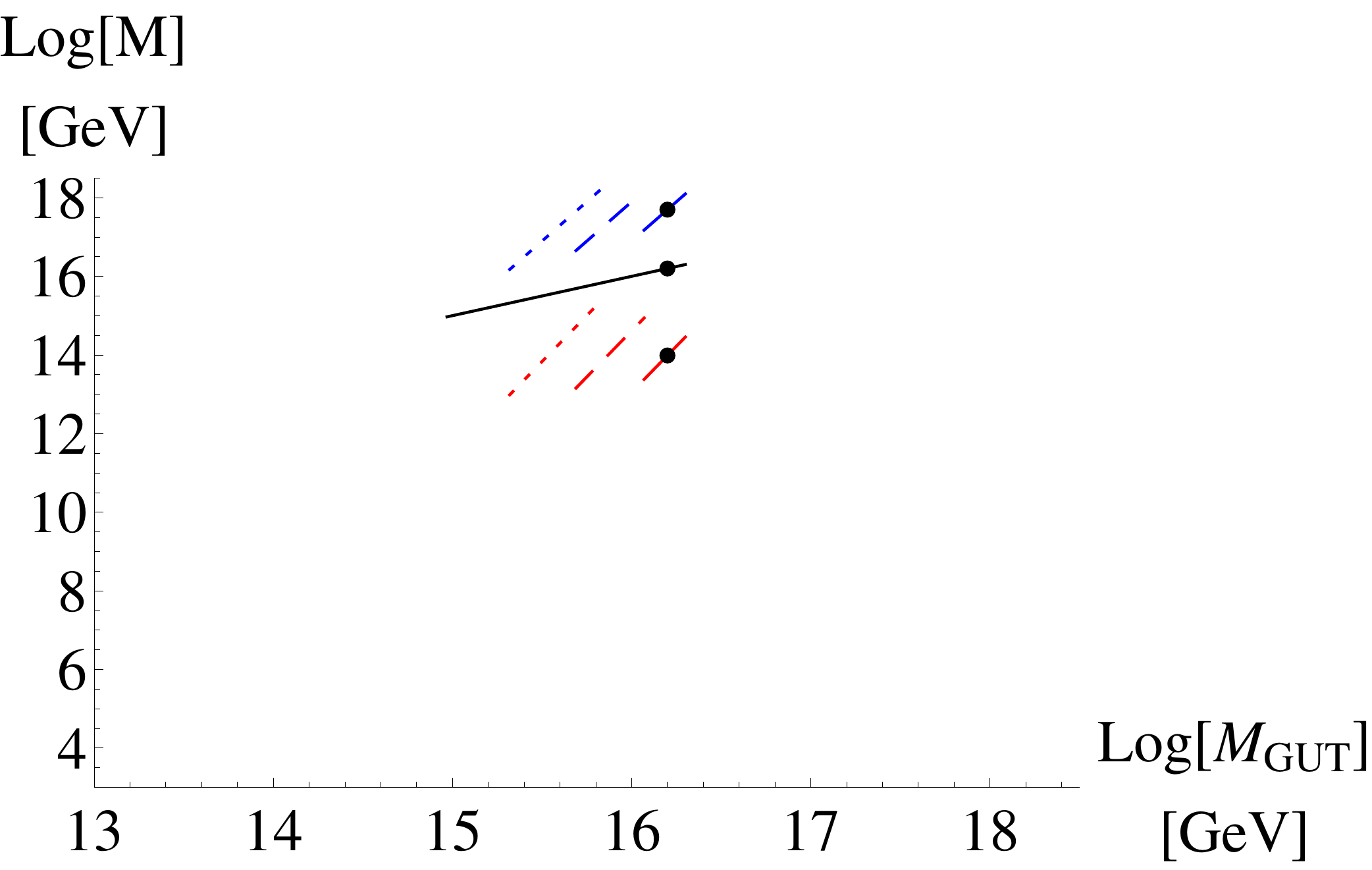}
 }
\end{minipage}\hfill
\begin{minipage}[b]{.48\textwidth}
 \centering
 \subfigure[Exemplary running of the gauge couplings for complete unification
  at $M_\text{GUT}=10^{18.2}\,\GeV$. The hypercharge coupling is shown in red,
  the $U(1)_R$ in brown, the B-L in green, the weak in blue and the strong
  coupling in black.]{
\includegraphics[width=.95\textwidth]{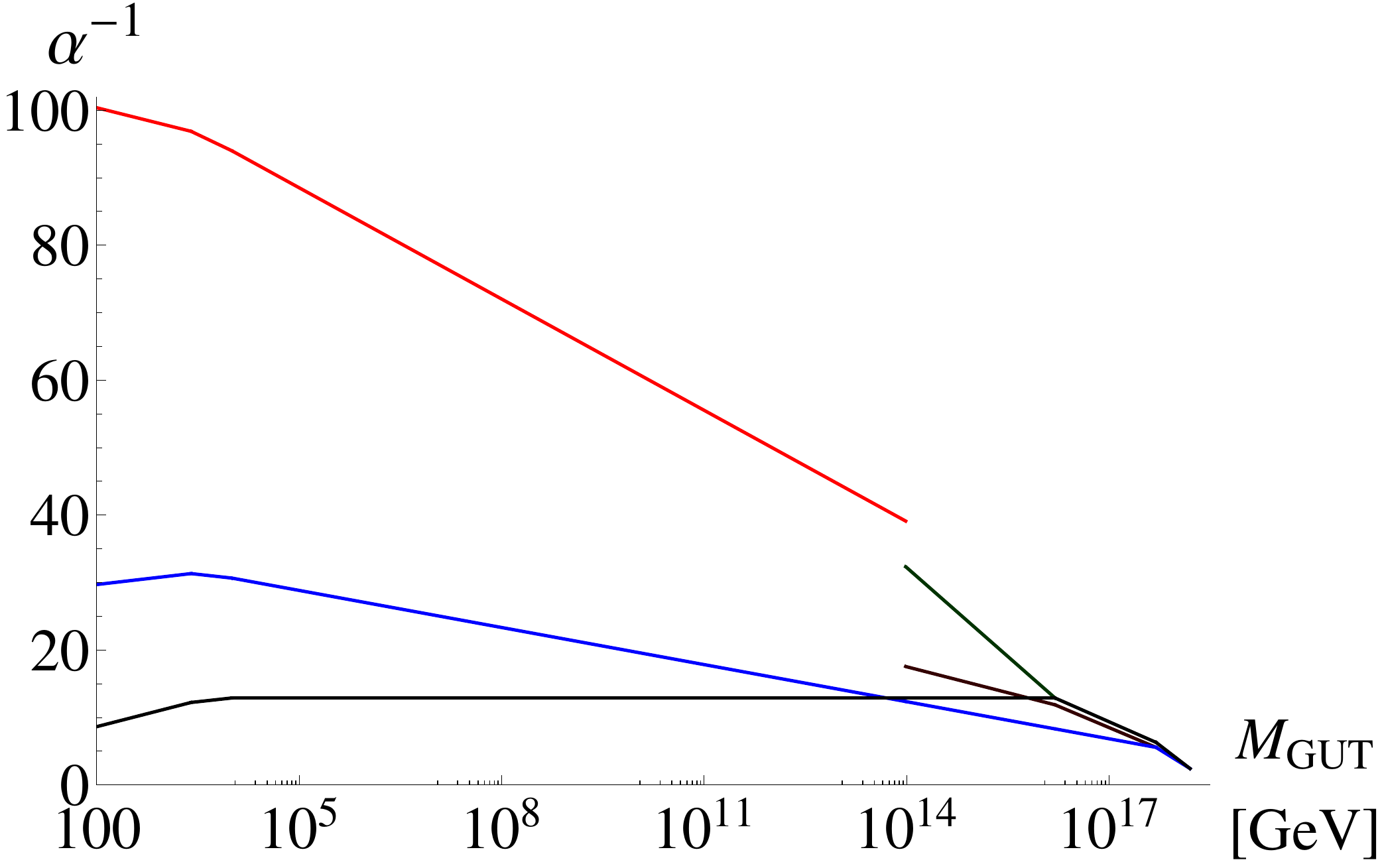}
 }
\end{minipage}
\caption{Variation of the unification scales and exemplary running of the
gauge couplings for the type C211.}
\label{fig:plotsC208}
\end{figure}

We may ask for models where the see-saw scale is sufficiently low (say,
$M_\text{IND}<10\,\TeV$) that the new particles can have an impact on
collider phenomenology.  We find 8 (34) models where this is possible
within class B (C), respectively.  One configuration with
normal hierarchy is model B199\footnote{For the meaning of numerical
  model indices, see Appendix~\ref{sec:fieldsetlist}.}, where we have three
copies of $h$, $F$ and
$\Sigma$ and no $E$. In the inverted case there is a similar model C211, which
has the same spectrum, but three copies of $\Phi$.  The plots are
shown in figure~\ref{fig:plotsB199},~\ref{fig:plotsC208}.

\section{Pati-Salam Models Without Supersymmetry}
\label{sec:non-SUSY}

We now turn to scenarios without supersymmetry.  Obviously, there is
much greater freedom for model building, if we ignore the naturalness
problems that inevitably appears when there are scalar fields in the
spectrum.  To limit this freedom in a meaningful way, we consider the
same classes of models as in the supersymmetric case, but omit the
fermionic superpartners of the additional multiplets. Analogously, we
omit the scalar superpartners of
matter fields and the fermionic superpartners of gauge fields.
This (ad-hoc) restriction allows us to compare supersymmetric and
non-supersymmetric models on the same basis.  The meaning of scales
and symmetry breaking patterns are unchanged.

Since we found that in the non-SUSY case the resulting mass thresholds
tend to be lower, we fixed the GUT scale for the classes B and
C to $M_\text{GUT}\equiv 10^{16}\,\GeV$.  For the value of
$M_\text{GUT}=10^{18.2}\,\GeV$ which we used in the SUSY case, we would have
only one configuration (class C) for both classes that satisfy GCU\@.
Lowering this scale even more, the model space would be less
constrained, but this is disfavored by the proton decay limits.

We considered the same set of 828 models, with the SUSY partners
removed, as in the previous section.  Again, more than half of the
models allow for GCU, given the modification mentioned in the previous
paragraph.  Still it turns out that without SUSY, classes A-D are 
disfavored but not excluded. In contrast to the supersymmetric case, a
considerable set of the successful models fall in class F. Again, the most
belong to class E.  In any case, lower GUT scales tend to be
favored.

As in the SUSY models discussed above, we observe a LR symmetry
breaking scale roughly around $10^{13}\,\GeV$ in a large fraction of
the successful models.  Nevertheless, there are models where this
scale can be much lower, down to below $100\,\TeV$ in some cases.

In class E, there are again many models with the possibility for light
color triplets $F$ in the $\TeV$ regime.  In the non-SUSY setup,
these are scalar particles and obviously cannot mix with
quarks.  We have to assume that there are couplings of either
leptoquark or diquark type that render these particles unstable,
originating from the PS-breaking sector.  Furthermore, in models of
this kind there is a high probability that the MSSM Higgs is not part
of $h$ but of the $h_\Phi$ multiplet (see section~\ref{sec:MSSMHiggs}).
Similar to the SUSY case the number of light bidoublets is
constrained to be one in the classes A to D. For class E we found also a
similar result as in the SUSY case. Again, most configurations prefer
four generations. Six are somehow disfavored, and one to three are equally
likely. In contrast to the previous considerations, there are now plenty of
configurations of class F with three generations of bidoublets, but still one
is in favor here. Again we see, that the multiplicity of these bidoublets
lowers the maximal unification scale.

As in the SUSY case, we find that three generations of light
$SU(2)$ triplets $T$ are excluded.  In particular, in class C the
lower bound for those is $M_{T_L}\gtrsim 10^{8}\,\GeV$.  In class C, the
bound becomes $10^{11}\,\GeV$ and in class B there is no GCU at all.

In the non-supersymetric case we again find in all classes a set of
configurations fixing all mass scales ``exactly''. This usually  corresponds to
degenerate mass scales. A common scale for such classes is 
$M_\text{GUT}\approx10^{14}\,\GeV$.

Again we find that fixing the induced scale $M_\text{IND}$ does not change the
results very much. Thus we see our assumptions justified to include this scale
in our scans. 

A general feature of non-SUSY spectra is the fact that the high-energy
effective values of the gauge couplings are larger than in the SUSY
case.  This is due to the lower number of fields that contribute to
the gauge-coupling running.

\begin{table}
\centering
 \begin{tabular}{|l|c|c|c|c|c|}
 \hline &&&&&\\[-2ex]
  & class B & class C & class E & class F & $\sum$ \\
  \hline &&&&&\\[-2.5ex] \hline &&&&&\\[-1.5ex]
  scanned& 144 & 144 & 324 & 216 & 828 \\[1ex]
  GCU & 10 & 30 & 230 & 201 & 471 \\[1ex]
  $M_\text{GUT}>10^{16}\,\GeV$& 10 & 30 & 111 & 16 & 167 \\[1ex]
  $M_\text{IND}<10\,\TeV \text{ and } M_\text{GUT}>10^{16}\,\GeV$ & 1 & 8 & 110
    & 16 & 135 \\[1ex]
  $M_\text{LR} < 100\,\TeV $ & 0 & 0 & 136 & 0 & 136 \\[1ex]
  $10^{12}\,\GeV < M_{N_R} < 10^{14}\,\GeV $ & 9 & 30 & 211 & 126 & 376 \\
  $ M_\text{IND} \in\,[0.1,10]\;
    \frac{v_{\Phi}^2}{v_{\Sigma}+v_{T}}$& 12 & 36 & 201 & 93 & 342 \\[1ex]
    \hline
 \end{tabular}
 \caption{Number of configurations full filling certain conditions in the
          non-SUSY case}
 \label{tab:NonSUSYOverview}
\end{table}

In the following, we do not repeat the detailed discussion of SUSY
models but pick a few selected models and model types with particular
features.  More generic statistics can be read off
table~\ref{tab:NonSUSYOverview}.

\subsection{Class E}

\begin{figure}[tbp]
\begin{minipage}[b]{.48\textwidth}
 \centering
 \subfigure[Possible scale variation leading to GCU\@. The GUT-scale is shown
in 
  black, the PS-scale in blue and the LR-scale in red. The variation in the
  IND-scale is shown discrete with $M_\text{IND}=10^{4}\,\GeV$ as solid
  lines, $M_\text{IND}=10^{7}\,\GeV$  dashed $M_\text{IND}=10^{10}\,\GeV$
dotted. The
  dots indicate the scales for the exemplary plot shown in (b).]{
 \includegraphics[width=.95\textwidth]{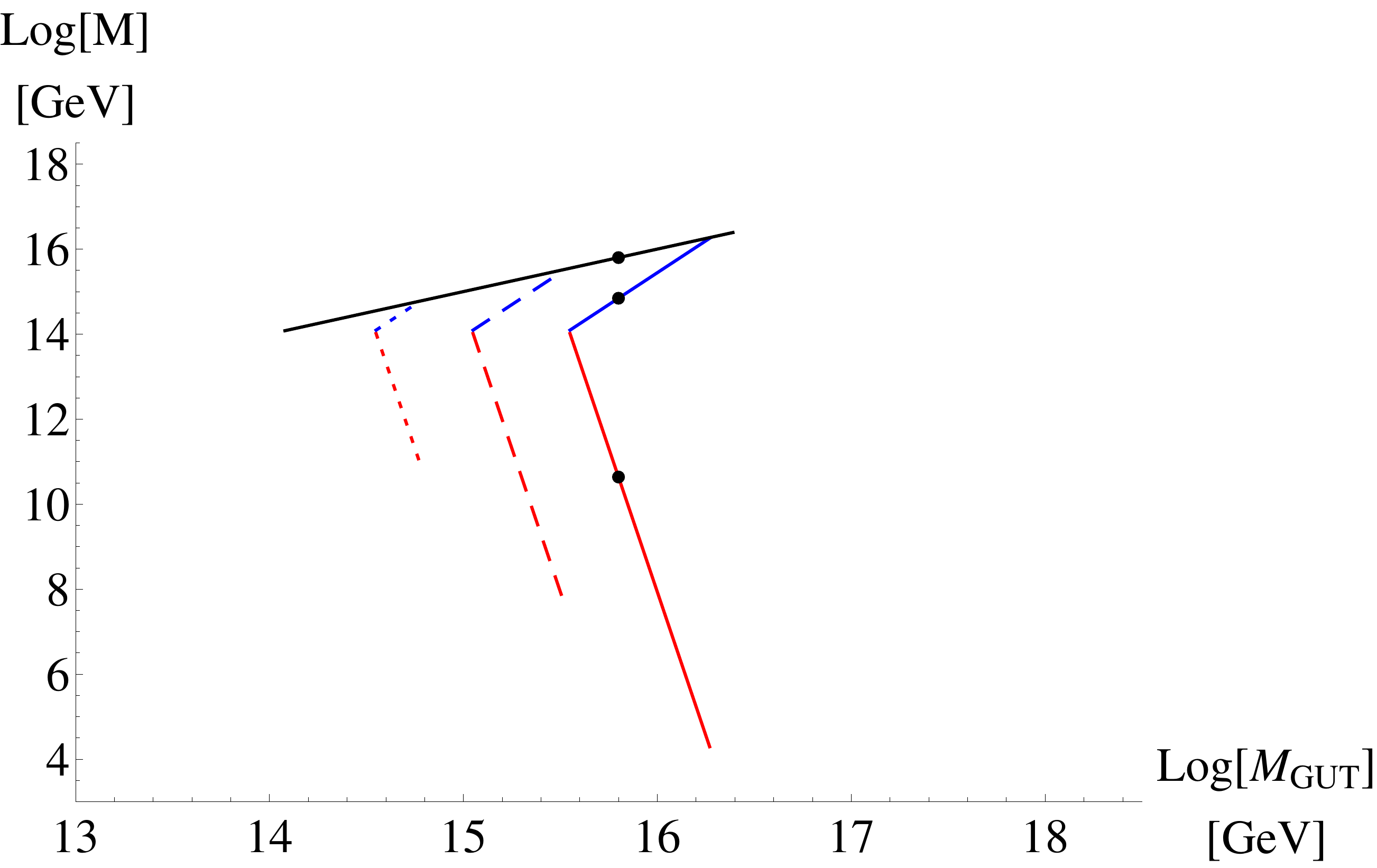}
 }
\end{minipage}\hfill
\begin{minipage}[b]{.48\textwidth}
 \centering
 \subfigure[Exemplary running of the gauge couplings for complete unification
  at $M_\text{GUT}=10^{15.8}\,\GeV$. The hypercharge coupling is shown in red,
  the B-L in green, the weak in blue and the strong coupling in black.]{
 \includegraphics[width=.95\textwidth]{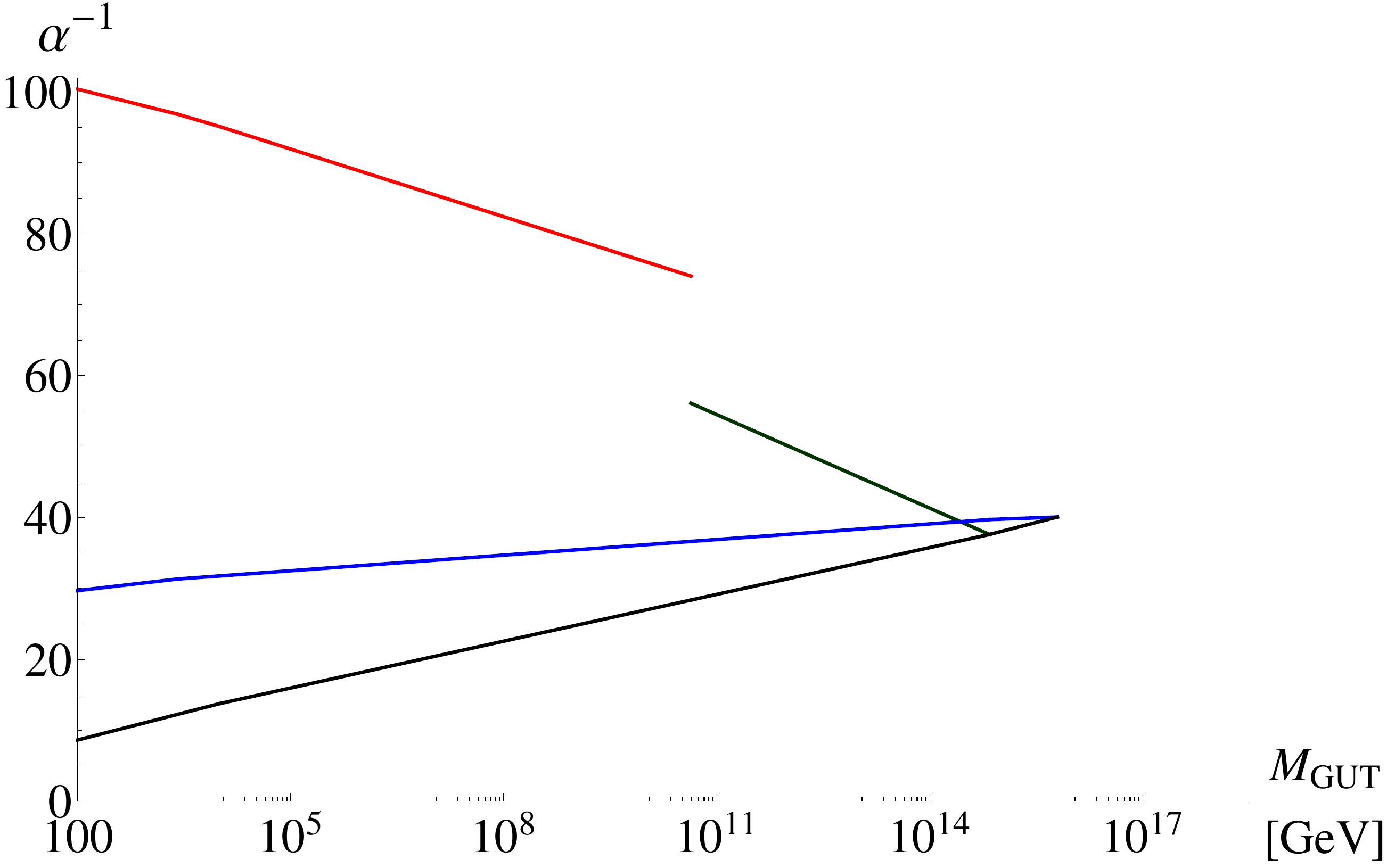}
 }
\end{minipage}
\caption{Variation of the unification scales and exemplary running of the
gauge couplings for the non-SUSY type Ee.}
\label{fig:plotsSMEe}
\end{figure}

For class E it is possible to implement GCU in 230 configurations, of which 23
provide a complete unification near the Planck scale. Similar to the
supersymmetric case, we find 88 configurations where the EWSB Higgs is provided
by $\Phi_L$.
On the other hand, an interesting possibility is the existence of three Higgs
generations (type Ee).  Although, in the non-SUSY case, there is no
direct relation to $E_6$ unification, we may take a look at such
models.  In class E, we find that GCU is possible, and the GUT-scale
can vary between $10^{14}\,\GeV\lesssim M_\text{GUT} \lesssim
10^{17}\,\GeV$.  One possible configuration exhibits three
generations of $F$, one of $\Sigma$ and $\Phi$ each, and no fields $E$
or $T$. For this special configuration the possible variation of the
scales and the unification plot are shown in
figure~\ref{fig:plotsSMEe}. Here we find that the variation of the
LR-scale strongly depends on the GUT-scale. The PS scale varies only
weakly and is always close to the GUT scale.

\subsection{Class F}

In class F, there are plenty of configurations leading to successful GCU\@.
In general, we find that there is not much scope for scale variation. The
GUT scale can be as large as $M_\text{GUT}\approx10^{17}\,\GeV$. 
The scales are
close to each other, since the lightest scale is fixed to be larger than
$M_\text{IND}\gtrsim10^{13}\,\GeV$. In this class there are also models
with GCU where all scales are essentially fixed, and not far from the GUT scale.
Those lead to $M_\text{GUT}\approx 2\times 10^{14}\,\GeV$, which is rather low. 
The LR-scale emerges between $10^{13}\,\GeV \leq M_\text{LR}\leq 2\times
10^{15}\,\GeV$.

\begin{figure}[tbp]
\begin{minipage}[b]{.48\textwidth}
 \centering
 \subfigure[Possible scale variation leading to GCU\@. The GUT-scale is shown
in 
  black, the PS-scale in blue and the QL-scale in red. The variation in the
  IND-scale is shown discrete with $M_\text{IND}=10^{4}\,\GeV$ as solid
  lines, $M_\text{IND}=10^{8.5}\,\GeV$  dashed $M_\text{IND}=10^{13}\,\GeV$
dotted.
  The dots indicate the scales for the exemplary plot shown in (b).]{
 \includegraphics[width=.95\textwidth]{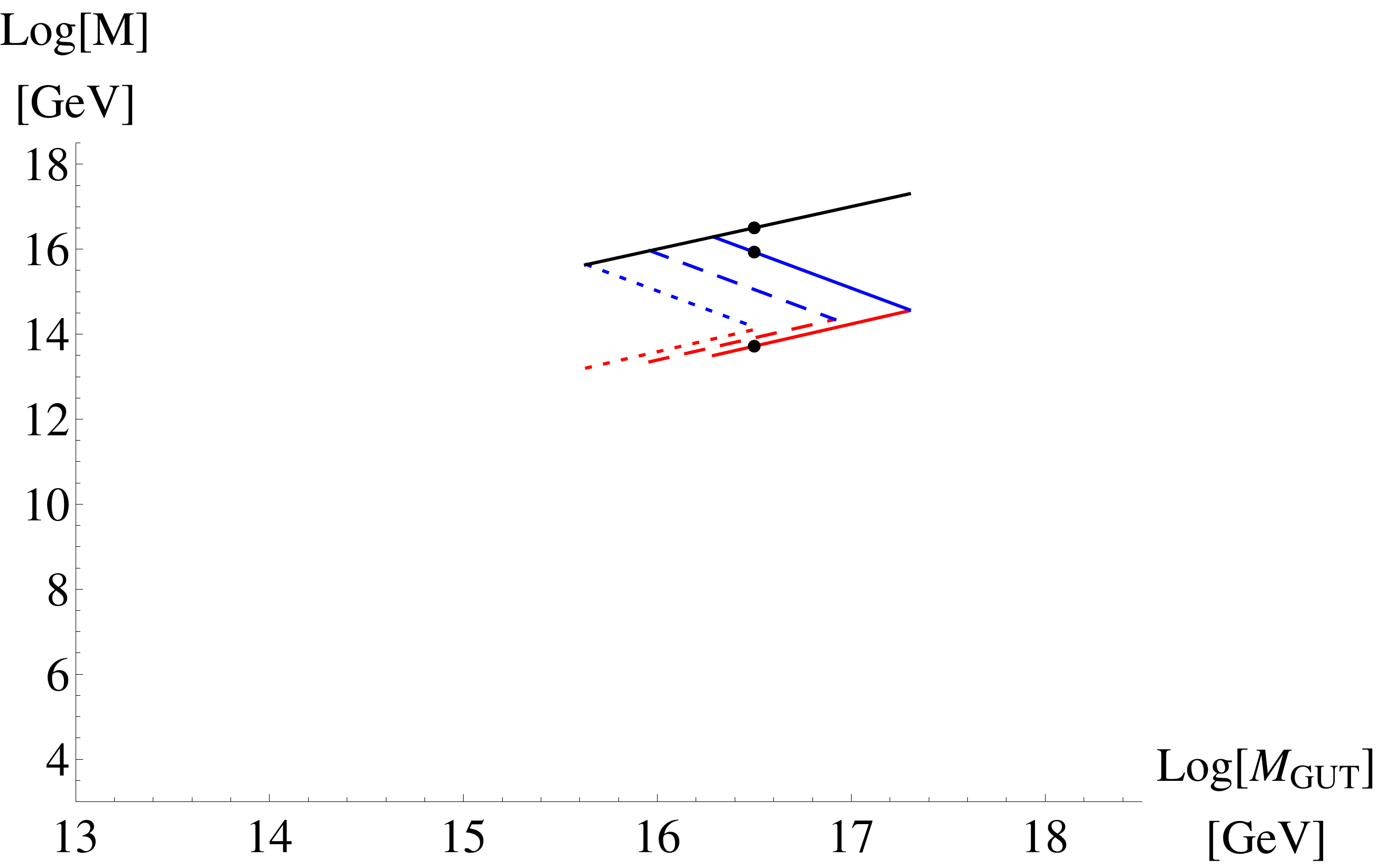}
 }
\end{minipage}\hfill
\begin{minipage}[b]{.48\textwidth}
 \centering
 \subfigure[Exemplary running of the gauge couplings for complete unification
  at $M_\text{GUT}=10^{16.5}\,\GeV$. The hypercharge coupling is shown in red,
  the $U(1)_R$ in brown, the weak in blue and the strong coupling in black.]{
 \includegraphics[width=.95\textwidth]{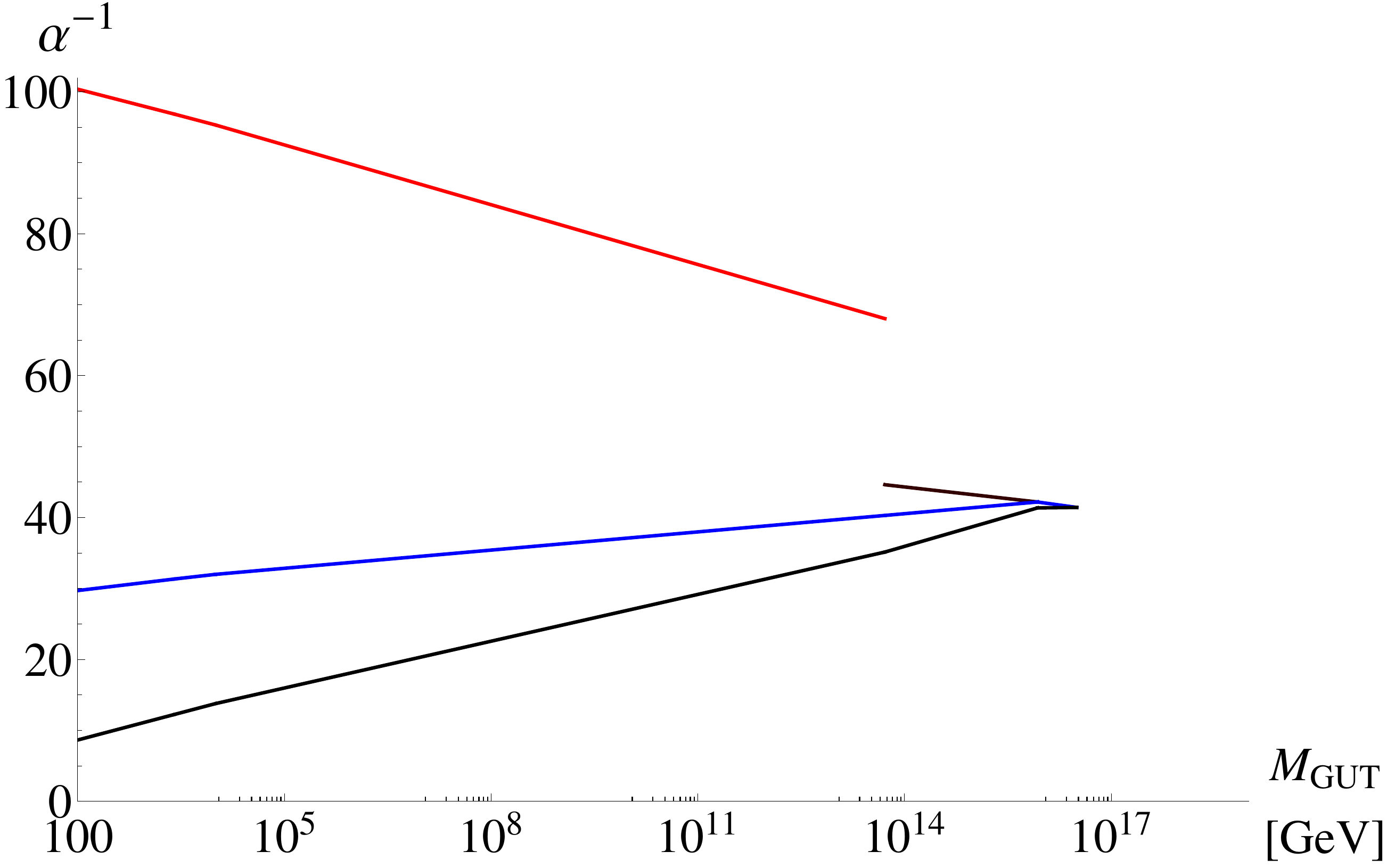}
 }
\end{minipage}
\caption{Variation of the unification scales and exemplary running of the
gauge couplings for the type non-SUSY F213.}
\label{fig:plotsSMF213}
\end{figure}

One exemplary configuration leading to GCU above $10^{16}\,\GeV$ is model
F213. Here, we have three generations of $F$ and one of $T$.  These
scalar particles can be rather light, potentially as low as the SUSY
scale.  In addition, this model contains three generations of the
fields $\Phi$ and $\Sigma$, and one generation of $E$.  We show the
possible scale variation and a sample unification plot for this model
in figure~\ref{fig:plotsSMF213}.

\subsection{Class A to D}

In class C, the QL-scale emerges typically close the PS scale.
Likewise, the PS scale can become as large as the GUT-scale, such that
the energy range with pure PS symmetry may vanish.

Looking at the possibility of three Higgs ($h$) generations, we do not
find any configurations in class B, and a few in class C.  On the
other hand, these model classes favor three as the number of $\Phi$
generations.

\begin{figure}[tbp]
\begin{minipage}[b]{.48\textwidth}
 \centering
 \subfigure[Possible scale variation leading to GCU\@. The PS-scale is shown in 
  black, the LR-scale in blue and the MSSM-scale in red. The variation in the
  IND-scale is shown discrete with $M_\text{IND}=10^{4}\,\GeV$ as solid
  lines, $M_\text{IND}=10^{5}\,\GeV$  dashed $M_\text{IND}=10^{6}\,\GeV$ dotted.
The
  dots indicate the scales for the exemplary plot shown in (b).]{
 \includegraphics[width=.95\textwidth]{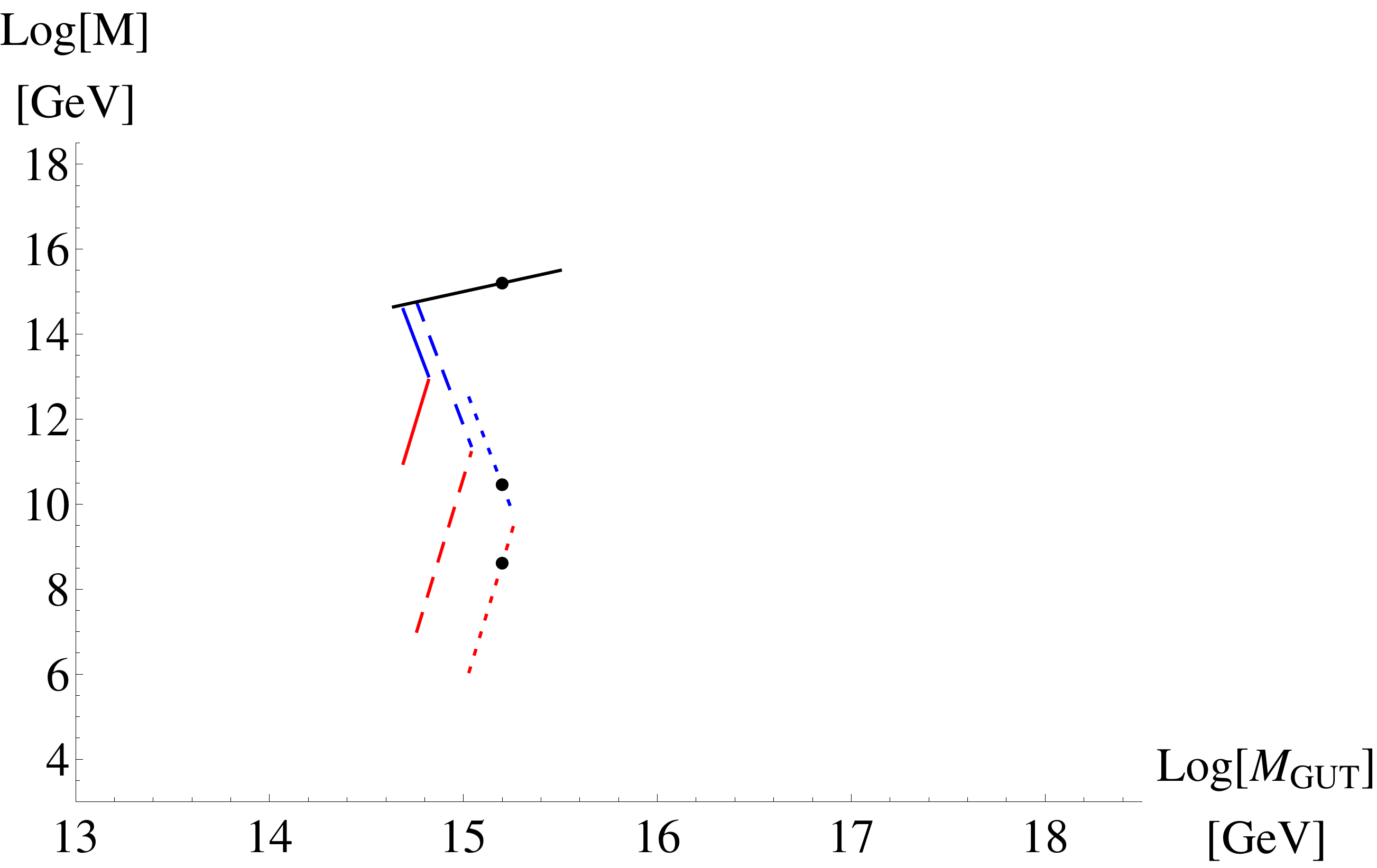}
 }
\end{minipage}\hfill
\begin{minipage}[b]{.48\textwidth}
 \centering
 \subfigure[Exemplary running of the gauge couplings for complete unification
  at $M_\text{GUT}=10^{16}\,\GeV$. The hypercharge coupling is shown in red,
  the $U(1)_R$ in brown, the B-L in green, the weak in blue and the strong
  coupling in black.]{
 \includegraphics[width=.95\textwidth]{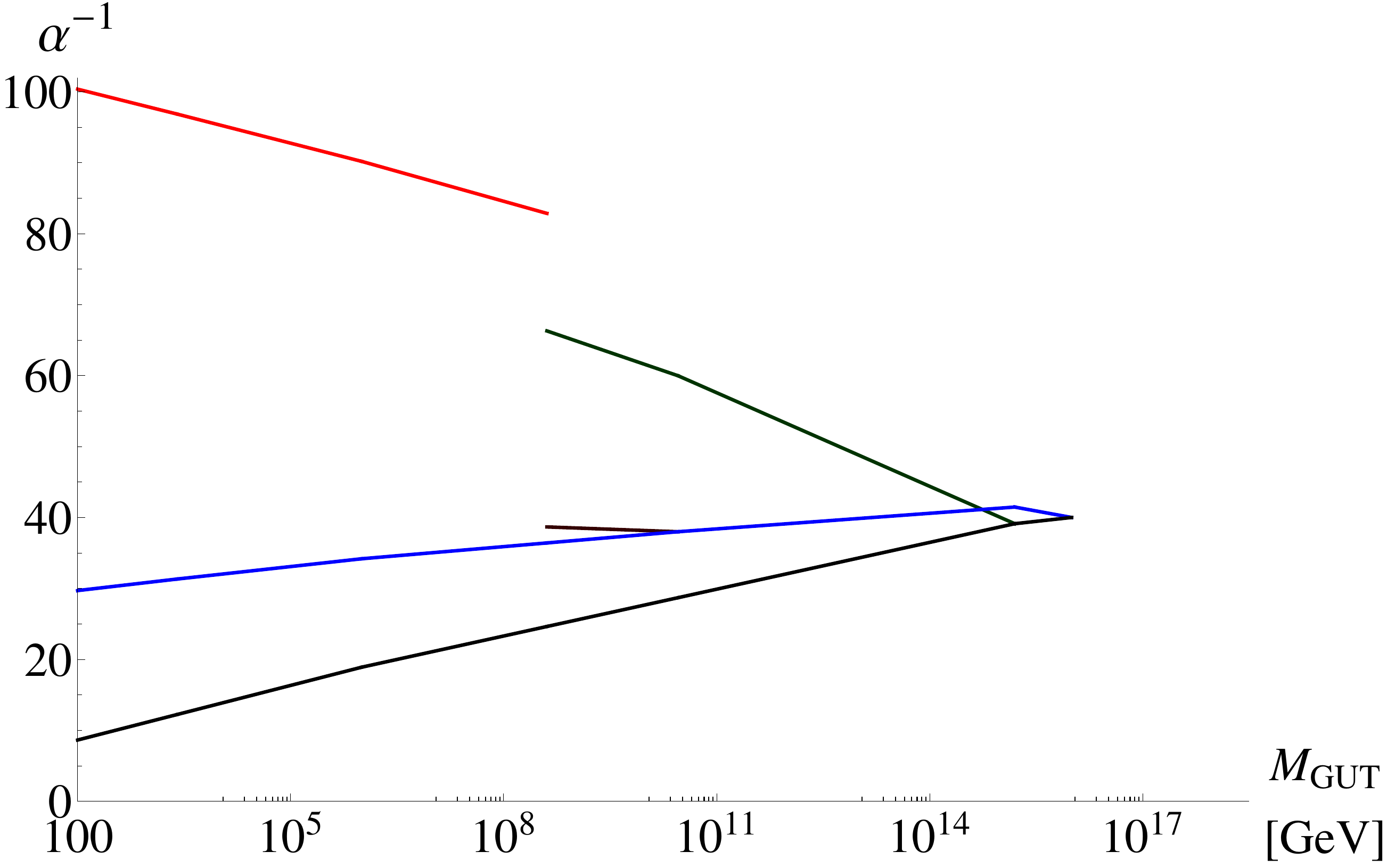}
 }
\end{minipage}
\caption{Variation of the unification scales and exemplary running of the
gauge couplings for the non-SUSY type B25.}
\label{fig:plotsSMB53}
\end{figure}

\begin{figure}[tbp]
\begin{minipage}[b]{.48\textwidth}
 \centering
  \subfigure[Possible scale variation leading to GCU\@. The QL-scale is shown
in 
  black, the PS-scale in blue and the MSSM-scale in red. The variation in the
  IND-scale is shown discrete with $M_\text{IND}=10^{4}\,\GeV$ as solid
  lines, $M_\text{IND}=10^{7}\,\GeV$  dashed $M_\text{IND}=10^{10}\,\GeV$
dotted. The
  dots indicate the scales for the exemplary plot shown in (b).]{
 \includegraphics[width=.95\textwidth]{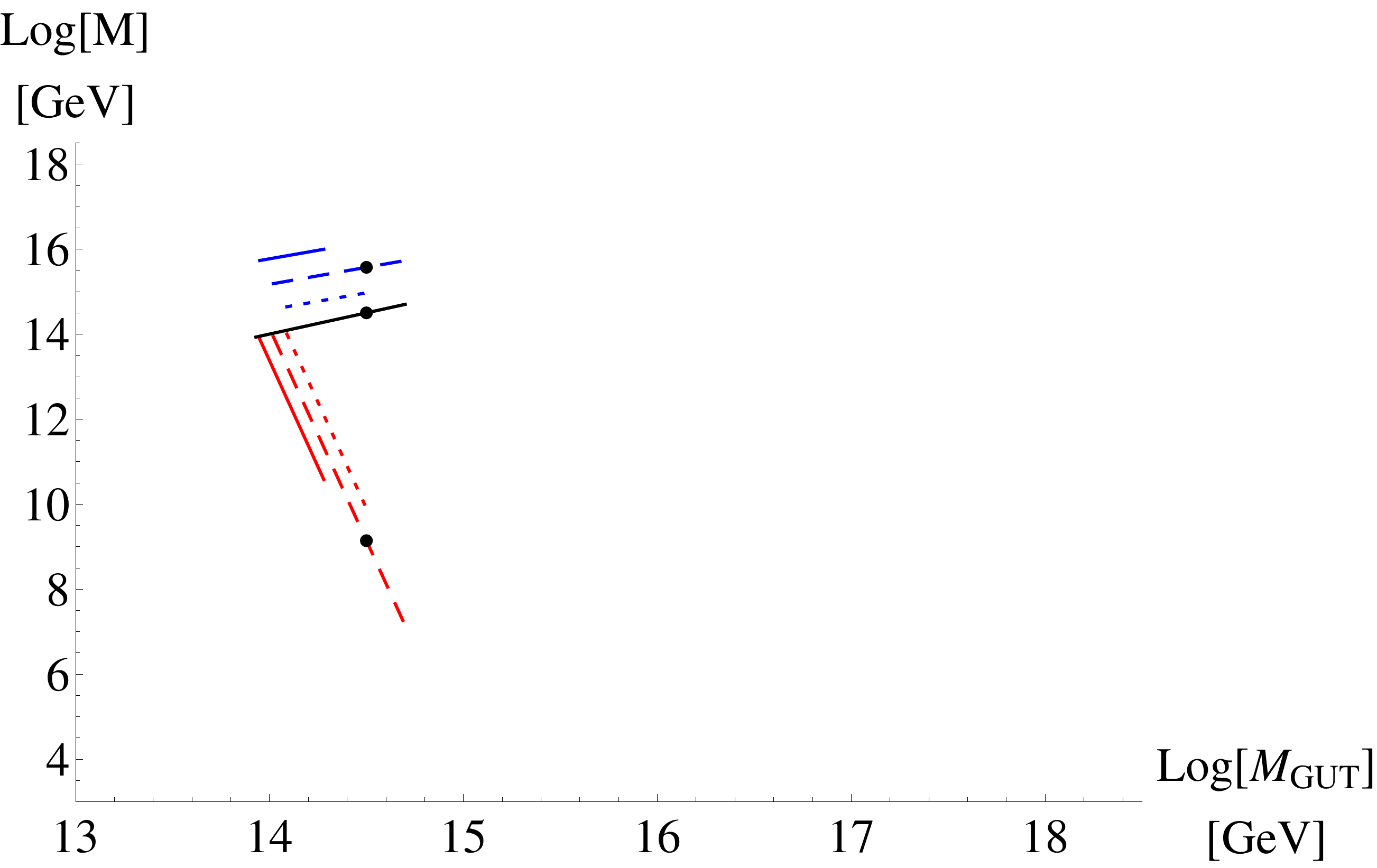}
 }
\end{minipage}\hfill
\begin{minipage}[b]{.48\textwidth}
 \centering
 \subfigure[Exemplary running of the gauge couplings for complete unification
  at $M_\text{GUT}=10^{16}\,\GeV$. The hypercharge coupling is shown in red,
  the $U(1)_R$ in brown, the B-L in green, the weak in blue and the strong
  coupling in black.]{
\includegraphics[width=.95\textwidth]{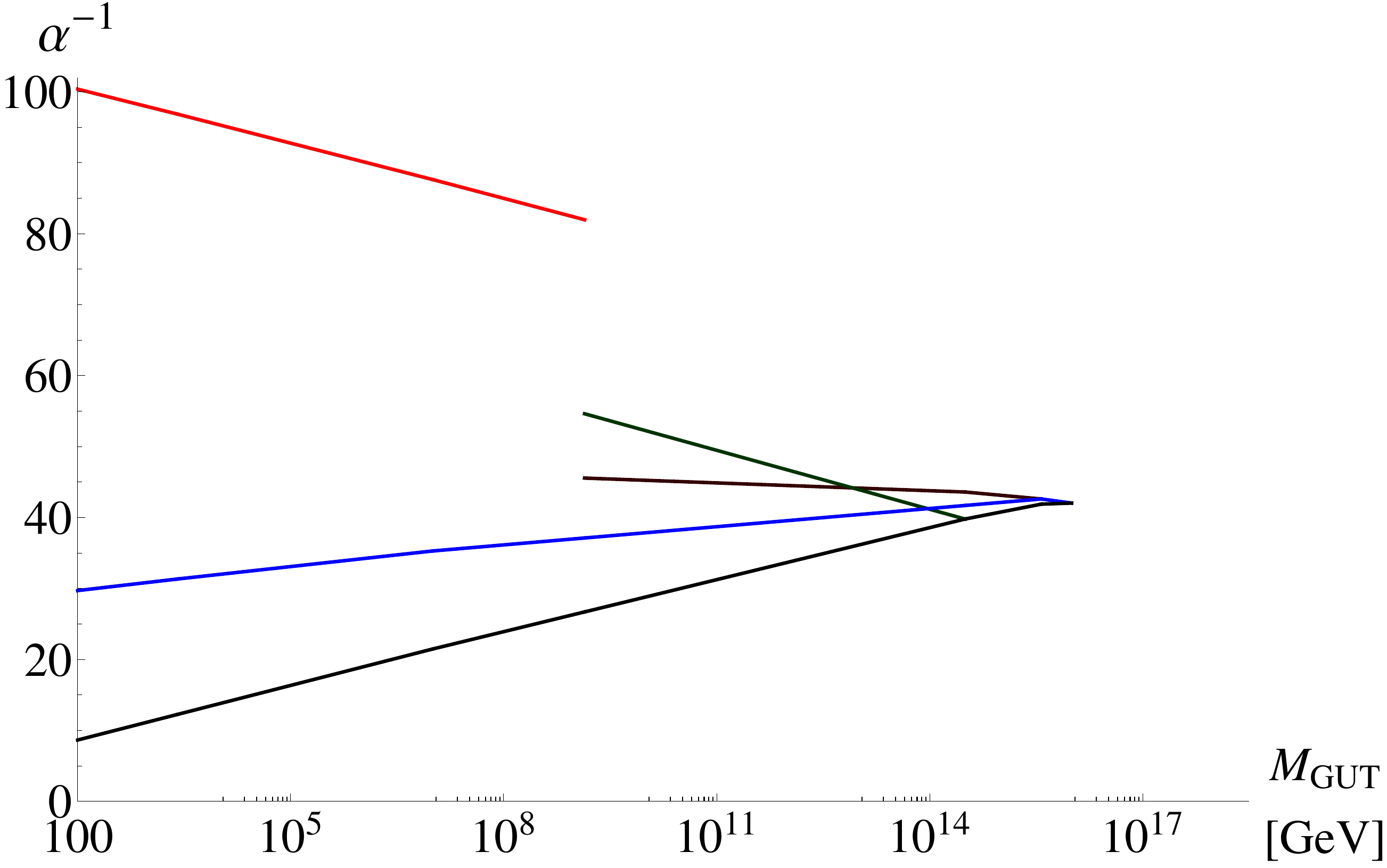}
 }
\end{minipage}
\caption{Variation of the unification scales and exemplary running of the
gauge couplings for the non-SUSY type C27.}
\label{fig:plotsSMC45}
\end{figure}

In figure~\ref{fig:plotsSMB53} and \ref{fig:plotsSMC45}, we display the
scale relations and gauge-coupling unification for two distinct models
B53 and C45\footnote{For the meaning of numerical
  model indices, see Appendix~\ref{sec:fieldsetlist}.}.  The former model
contains three generations of $F$ which can be
light. In addition it contains three generations of $E$ and all other fields
ones. We found quite some range for scale variation. For the largest value of
the PS scale ($M_\text{PS}\approx10^{15}$) we find a degeneracy of
all lower scales. The induced scale can be as light as some $\TeV$.

Model C45 also features new particles at the $\TeV$ scale. Here we include
three generations of $\Phi$ and $\Sigma$ and all other fields ones. We again
see quite some space to vary the scales. 

\section{Summary of Models}
\label{sec:summary}

We have presented a survey of models with gauge-coupling unification
along a path that contains several intermediate scales, corresponding
to left-right symmetry, quark-lepton unification, Pati-Salam symmetry,
and $SO(10)$ or larger GUT symmetry.  We studied both supersymmetric
and non-supersymmetric models, where the latter are derived
from the former by omitting all superpartners.

We have restricted the allowed new chiral superfields (or scalar fields, in the
non-SUSY case) below the GUT scale to a small well-defined set of multiplets,
all of which fit in small representations of $SO(10)$ or
$E_6$.  A large subset of the resulting models is consistent with
gauge-coupling unification, proceeding in several steps.  In supersymmetric
models, there is slightly more freedom in varying the scales than in
non-supersymmetric models.

The assumptions and calculations do not constrain the model space in
such a way that we can get precise numerical predictions, but we can
deduce characteristic patterns in the scale hierarchies that correlate
with specific choices for the spectrum.

In a wide range of models, GUT unification can be pushed up to the
Planck scale.  This fact, together with the properties of Pati-Salam
symmetry, significantly reduces the strain that the non-observation of
proton decay can put on GUT model building.

Additional thresholds in the intermediate range between observable
energies and the GUT scale are likely.  Being associated with
left-right or quark-lepton symmetry breaking, they decouple flavor
physics issues from the requirements of complete unification.  We have
not considered flavor physics in any more detail, but expect that it
can generically be accommodated if non-renormalizable terms are
properly included.  Depending on the particular model and on the
chosen set of flavor-dependent interactions, we expect specific
hierarchies, relations and predictions for the flavor sector.

As an extra feature, the (super)fields $\Sigma$ and $T$ (adjoint of $SU(4)_C$
and $SU(2)_R$) and $\Phi$ (fundamental of $SU(2)_L\times SU(2)_R$) can cooperate
to generate small mass terms for certain particles, including SM-like Higgs
doublets and new exotic (s)quarks, such that they can be accessible at
colliders.  On the other hand, $SU(2)_{L/R}$ triplets as possible
Pati-Salam breaking Higgs fields tend to raise the LR symmetry scale
above the scale of quark-lepton unification and thus may enforce a
direct relation between quark and lepton flavor physics.

Another generic property of the models under consideration is the
scale of left-right symmetry breaking, naturally associated with
neutrino mass generation, in an intermediate mass range.  A neutrino
mass scale significantly below the GUT scale is favored by numerical
estimates of see-saw mechanisms that can explain the small observable
neutrino masses.

We also encounter models where gauge-coupling unification at high
energies implies a multi-Higgs doublet model at low scale, possibly
with flavor quantum numbers.  Furthermore, the observable Higgs
doublet need not be a member of a $(1,2,2)$ representation, as often
assumed, but can also originate from a $(4,2,1) + \text{c.c.}$ representation,
i.e.,
behave as a scalar lepton.  The effective $\mu$ term which sets the
scale for low-energy Higgs physics, can have a see-saw like form and
thus be naturally suppressed with respect to the
higher symmetry-breaking scales.

\section{Conclusions}
\label{sec:conclusions}

In summary, we have studied a range of comparatively simple models
that fit into the framework of GUT theories with intermediate
thresholds.  We have taken a phenomenological viewpoint and specified
the models in form of a chain of effective theories, as far as we
expect that a description in terms of weakly interacting
four-dimensional gauge theory can make sense.

It is remarkable that the most interesting phenomenology, which we may
identify as Planck-scale GUT unification, intermediate PS and LR
scales, and new particles at TeV energies, can be simultaneously
realized in a number of distinct models (cf.\
App.~\ref{sec:spec-models}).  This is not a generic feature.  However,
if not all of these conditions are to be satisfied simultaneously, or
allowing further multiplets or hierarchy patterns in couplings, the
set of interesting models becomes sizable.

One may consider fully specified GUT models that predict the
appearance (or absence) of a PS symmetry and the associated spectrum.
However, the current lack of reliable knowledge about strong and
gravitational interactions which are expected at the highest energies,
denies attempts to ultimately favor or exclude alternative spectra and
symmetry-breaking chains.

In such a situation, it appears worthwhile to rather concentrate on
the implications of a sequence of new thresholds at intermediate
energies, presumably in the context of PS symmetry.  Our survey
demonstrates that intermediate symmetry-breaking scales associated
with flavor mixing and mass generation can emerge in various energy
regions, even if rather specific and simple assumptions about gauge
multiplets are imposed.  The findings suggest that one should study
the hierarchies within flavor observables in relation to hierarchies
in gauge-symmetry breaking, discuss both renormalizable and
non-renormalizable operators, and to combine gauge and flavor
symmetries in a common framework which need not be tied to ultimate
GUT unification.  This program, which has so far been pursued only for
a subset of the possible scenarios, deserves further efforts.

In summary, Pati-Salam models can easily accommodate unification in a
multi-scale framework.  They provide a rich phenomenology and a
promising background for new approaches to the lepton and quark flavor
problem.

\subsection*{Acknowledgments}

This work has been supported by a Coordinated Project Grant of the
Faculty of Science and Technology, University of Siegen.  We thank
T.~Feldmann, C.~Luhn and J.~Reuter for discussions and valuable comments on the
manuscript.

\appendix
\section{Beta-Function Coefficients}
\label{sec:runningcoefficients}

\begin{table}
\hspace{-10ex}
\centering
{\small
\begin{tabular}{|c|l|l|l||c|c|c|c|c|}
\hline &&&&&&&&\\[-2ex]
field	& PS & LR & SM & $\tilde{b}_Y$  &
$\tilde{b}_{B-L}$ &$\tilde{b}_2$ & $\tilde{b}_3$  & $\tilde{b}_4$ \\
\hline  &&&&&&&&\\
$h$ & $(1,2,2)$  & $(1,2,2)_0$ & $(1,2)_{\frac{1}{2}}$ & $\frac{1}{2}$ & 0
    & $\frac{1}{2}$ & 0  & 0 \\[1ex]
    &&& $(1,2)_{-\frac{1}{2}}$ & $\frac{1}{2}$ && $\frac{1}{2}$ & 0&\\[2ex]
$F$ & $(6,1,1)$  & $(3,1,1)_{\frac{2}{3}}$ & $(3,1)_{\frac{1}{3}}$  &
    $\frac{1}{3}$ & $\frac{4}{3}$ & 0 & $\frac{1}{2}$ & 1 \\[1ex]
    && $(\bar{3},1,1)_{-\frac{2}{3}}$ & $(\bar{3},1)_{-\frac{1}{3}}$ &
    $\frac{1}{3}$ & $\frac{4}{3}$ & 0 & $\frac{1}{2}$ & \\[2ex]
$\Phi_R$ & $(\bar{4},1,2)$ & $(\bar{3},1,2)_{\frac{1}{3}}$
    & $(\bar{3},1)_{\frac{2}{3}}$ & $\frac{4}{3}$ & $\frac{2}{3}$ & 0 
    & $\frac{1}{2}$ & 1 \\[1ex]
    &&& $(\bar{3},1)_{-\frac{1}{3}}$ & $\frac{1}{3}$ &   & 0 
    & $\frac{1}{2}$ & \\[1ex]
    && $(1,1,2)_{-1}$ & $(1,1)_0$    & 0  & 2  & 0 & 0  & \\[1ex]
    &&& $(1,1)_{-1}$    & 1  &   & 0 & 0  & \\[2ex]
$\overline{\Phi}_R$ & $(4,1,2)$  & $(3,1,2)_{-\frac{1}{3}}$ &
$(3,1)_{\frac{1}{3}}$ 
    & $\frac{1}{3}$ & $\frac{2}{3}$ & 0 & $\frac{1}{2}$ & 1 \\[1ex]
    &&& $(3,1)_{-\frac{2}{3}}$  & $\frac{4}{3}$ && 0 &$\frac{1}{2}$&\\[1ex]
    && $(1,1,2)_{1}$ & $(1,1)_{1}$ & 1  & 2  & 0 & 0  &  \\[1ex]
    &&& $(1,1)_{0}$ & 0  &   & 0 & 0  & \\[2ex]
$\Phi_L$ & $(4,2,1)$  & $(3,2,1)_{-\frac{1}{3}}$ &
    $(3,2)_{-\frac{1}{6}}$  & $\frac{1}{6}$ & $\frac{2}{3}$ & $\frac{3}{2}$
    & 1 & 1 \\[1ex]
    && $(1,2,1)_{1}$ & $(1,2)_{\frac{1}{2}}$  & $\frac{1}{2}$ & 2 &
    $\frac{1}{2}$ & 0  & \\[2ex]
$\overline{\Phi}_L$ & $(\bar{4},2,1)$ & $(\bar{3},2,1)_{\frac{1}{3}}$ &
    $(\bar{3},2)_{\frac{1}{6}}$ & $\frac{1}{6}$ & $\frac{2}{3}$ &
    $\frac{3}{2}$ & 1  & 1 \\[1ex]
    && $(1,2,1)_{-1}$ & $(1,2)_{-\frac{1}{2}}$  & $\frac{1}{2}$ & 2
    & $\frac{1}{2}$ &  0& \\[2ex]
$\Sigma$& $(15,1,1)$  & $(8,1,1)_0$ & $(8,1)_0$ & 0  & 0  & 0 & 3 & 4 \\[1ex]
    && $(3,1,1)_{-\frac{4}{3}}$ & $(3,1)_{-\frac{2}{3}}$ &
    $\frac{4}{3}$ & $\frac{16}{3}$& 0 & $\frac{1}{2}$ & \\[1ex]
    && $(\bar{3},1,1)_{\frac{4}{3}}$ & $(\bar{3},1)_{\frac{2}{3}}$ &
    $\frac{4}{3}$ & $\frac{16}{3}$& 0 & $\frac{1}{2}$ & \\[1ex]
    && $(1,1,1)_0$   & $(1,1)_0$    & 0  & 0  & 0 & 0 & \\[2ex]
$E$ & $(6,2,2)$  & $(3,2,2)_{\frac{2}{3}}$ & $(3,2)_{\frac{5}{6}}$  &
    $\frac{25}{6}$& $\frac{16}{3}$& $\frac{3}{2}$ & $1$  & $4$ \\[1ex]
    &&& $(3,2)_{\frac{1}{6}}$  & $\frac{1}{6}$&& $\frac{3}{2}$ & $1$ 
&\\[1ex]
    && $(\bar{3},2,2)_{-\frac{2}{3}}$ & $(\bar{3},2)_{-\frac{1}{6}}$ &
    $\frac{1}{6}$& $\frac{16}{3}$ & $\frac{3}{2}$ & $1$  & \\[1ex]
    &&& $(\bar{3},2)_{-\frac{5}{6}}$ & $\frac{25}{6}$&&$\frac{3}{2}$ & $1$ 
    & \\[2ex]
$T^R$ & $(1,1,3)$  & $(1,1,3)_0$ & $(1,1)_1$ & 0 & 0 & 0 & 0  & 0 \\[1ex]
   &&& $(1,1)_0$    & 0  & 0  & 0 & 0  & 0 \\[1ex]
   &&& $(1,1)_{-1}$    & 0  & 0  & 0 & 0  & 0 \\[2ex]
$T^L$ & $(1,3,1)$  & $(1,3,1)_0$ & $(1,3)_0$    & 0  & 0  & $2$ & 0  & 0\\[3ex]
\hline
\end{tabular} }
\caption{Full field content and breaking as well as all contributions to the
beta function}
\label{tab:fullfieldcontribution}
\end{table}

As stated in the paper (cf. section~\ref{sec:unification}) the running of the
gauge couplings can be described by 
\begin{equation}
  \frac{1}{\alpha_i\left(\mu_2\right)} = \frac{1}{\alpha_i\left(\mu_1\right)} -
\frac{b_i}{2\,\pi}\,\ln\left(\frac{\mu_2}{\mu_1} \right) \,.
\end{equation}

The coefficient $b_i$ of the running coupling can be calculated by means of the
representation of the fields contributing at the given mass scale alone 
\cite{Jones:1981we}. For
each set of gauge groups $SU(N)$ with $N\geq2$ the contribution of an field with
representation $(\re{I}_1,\ldots,\re{I}_n)$ to the running coefficient
$\tilde{b}_i$ is given as
\begin{equation}
 \tilde{b}_i^\mathfrak{R} = T(\re{I}_i)\,\prod_{k\neq i} d(\re{I}_k) \,.
\end{equation}
were $d(\re{I}_i)$ is the dimension and the normalization
of the representation $T(\re{I}_i)$. These can be calculated using the
representing matrices $R^a$
\begin{equation}
\text{tr}\,R^a R^b \,=\, T(\mathfrak{R})\, \delta^{ab} \,,
\end{equation}
with $T(\re{N})=\frac{1}{2}$.

For a $U(1)$ the contribution is up to an consistent rescaling:
\begin{equation}
 \tilde{b}_{U(1)}^\mathfrak{R} = Y^2 \,\prod_{k} d(\re{I}_k) \,.
\end{equation}

The complete running coefficient depends now on whether we work in a
supersymmetric theory or not. For the non-supersymetric case one has to divide
the fields in scalar and fermionic contributions:
\begin{equation}
  b_i^{\text{SM}}= 
  \frac23 \sum_{\mathfrak{R}_\text{ferm.}} \tilde{b}_i^\mathfrak{R}  
  + \frac13 \sum_{\mathfrak{R}_\text{scalar}}
  \tilde{b}_i^\mathfrak{R}
  - \frac{11}{3}\,C_2(G_i) \,,
 \label{eq:runninggcoefficientsSM}
\end{equation}
where $C_2(\re{G}_i) = \text{dim}(G_i)$ is the quadratic Casimir operator.

Since in the supersymmetric case there is a superpartner for each
scalar/fermionic field, there is no need to divide the fields in such a way.
Thus, the relation simplifies to
\begin{equation}
  b_i^{\text{SUSY}}= 
  \sum_{\mathfrak{R}} \tilde{b}_i^\mathfrak{R}  
  - 3\,C_2(G_i) \,.
 \label{eq:runninggcoefficientsSUSY}
\end{equation}

Table~\ref{tab:fullfieldcontribution} displays the contribution of each
field and its complete decomposition w.r.t. the subgroups  of PS symmetry.

\section{Vacuum Expectation Values and Mass Scales}
\label{sec:notationswitch}

In the first part of this paper we calculate the
superpotential and the masses of all superfields. Therefore we use the vevs as
natural scales. Since in this part the breaking associated to the vev is not
the most important thing but the fields they are related to, we name them after
those. 

In the second part, we are primarily interested in the scales present in the
running of the gauge couplings. Hence, we switch our
notation to the mass scales. These are labeled by a subscript that indicates
the symmetry which is broken at this stage. Nevertheless, these are of course
related to the vevs discussed before, but this relation depends on the class one
is discussing. We show this relation explicitly in
table~\ref{tab:notationswitch}.

\begin{table}[tbp]
 \centering
 \begin{tabular}{|ccccc|}
  \hline
  vev & class B & class C & class E & class F \\
  \hline &&&& \\[-2ex]
  $v_\Sigma$ & $M_\text{PS}$ & $M_\text{QL}$ & $M_\text{PS}$ & --- \\
  $v_T$ & $M_\text{LR}$ & $M_\text{PS}$ & --- & $M_\text{PS}$ \\
  $v_\Phi$ & $M_\text{MSSM}$ & $M_\text{MSSM}$ & $M_\text{LR}$ & $M_\text{LR}$
     \\
  $\frac{v_\Phi^2}{v_\Sigma+v_T}$ & $M_\text{IND}$ & $M_\text{IND}$ &
     $M_\text{IND}$ & $M_\text{IND}$ \\[1ex]
  \hline
 \end{tabular}
 \caption{Relation between the vevs and the mass scales for the different
          classes. In class A there is no hierarchies in the vevs and thus all
          scales are equal to $M_\text{PS}$. In class D there are only the
	  scales $M_\text{PS}$ and $M_\text{LR}$}
\label{tab:notationswitch}
\end{table}

\section{Model Naming Scheme}
\label{sec:fieldsetlist}

The global naming convention is laid out at the end of section~\ref{sec:model}.
For
all configurations of type g we use a numerical naming scheme. The numbers
follow an internal numbering given by the structure of our Mathematica file.
This file  is available by the authors upon request.

Table~\ref{tab:configurationlist} displays the connection between the
multiplicities of the different fields present below the Planck scale and the
model names used in this paper.

\begin{table}[tbp]
\centering
\begin{tabular}{|ll|cccccc|}
\hline
&name & $\#h$ & $\#F$ & $\#\Phi$ & $\#\Sigma$ & $\#E$ & $\#T_{L/R}$ \\
\hline&&&&&&& \\[-1ex]
SUSY models: 
&Em & 0 & 0 & 1 & 1 & 0 & 0 \\
&Fm & 0 & 0 & 1 & 0 & 0 & 1 \\
&Es/Fs & 1 & 1 & 1 & 1 & 1 & 1 \\
&Ee/Fe & 3 & 3 & 3 & 3 & 1 & 1 \\
&Ff & 1 & 3 & 1 & 1 & 1 & 1 \\
&Bm/Cm & 0 & 0 & 1 & 1 & 0 & 1 \\
&Bs/Cs & 1 & 1 & 1 & 1 & 1 & 1 \\
&Be/Ce & 3 & 3 & 1 & 1 & 1 & 1 \\
&B199 & 3 & 3 & 1 & 3 & 0 & 1 \\
&C211 & 3 & 3 & 3 & 3 & 0 & 1 \\[1ex]
\hline &&&&&&& \\[-1ex]
Non-SUSY models:
&E289 & 3 & 3 & 1 & 1 & 0 & 0 \\
&F213 & 1 & 3 & 3 & 3 & 1 & 1 \\
&B53 & 1 & 3 & 1 & 1 & 3 & 1 \\
&C45 & 1 & 1 & 3 & 3 & 1 & 1 \\[1ex]
\hline
\end{tabular}
\caption{Field content and multiplicities for the discussed models in this
paper.}
\label{tab:configurationlist}
\end{table}

\clearpage
\section{Specific Models}
\label{sec:spec-models}

Among the models that are consistent with unification, there is a
subset where further interesting conditions are met simultaneously.
In this Appendix, we list all models within our framework that (i)
show complete GCU at the Planck scale, taken as $M_\text{Planck} =
10^{18.2}\,\GeV$; (ii) predict new ``exotic'' particles at accessible
energies $M_\text{low}\sim 10\,\TeV$; and (iii) exhibit a
right-handed neutrino scale in the range $10^{12}\,\GeV<
M_{N_R}<10^{14}\,\GeV$.

\subsubsection*{Class: F (supersymmetric)}
\begin{center}\footnotesize
 \begin{tabular}{|c|cccccc|cc|}
  \hline &&&&&&&&\\[-2ex]
  Model & \#h & \#F & \#$\Phi$ & \#$\Sigma$ & \#E & \#T & $M_\text{LR}$ [GeV] &
$M_\text{PS}$ [GeV] \\[1.5ex]
   F41  & 0 & 1 & 1 & 1 & 1 & 1 & $\sim9\times10^{11}$ & $\sim6\times10^{14}$ \\
   F186 & 1 & 3 & 1 & 1 & 1 & 3 & $\sim7\times10^{12}$ & $\sim8\times10^{14}$ \\
   F262 & 3 & 1 & 1 & 3 & 0 & 0 & $\sim1\times10^{11}$ & $\sim3\times10^{14}$ \\
   \hline
  \end{tabular}
\end{center}
\vspace{\baselineskip}

\subsubsection*{Class: B (supersymmetric)}
%

\begin{center}\footnotesize
  \begin{tabular}{|c|cccccc|ccc|}
  \hline &&&&&&&&&\\[-2ex]
  Model & \#h & \#F & \#$\Phi$ & \#$\Sigma$ & \#E & \#T & $M_\text{MSSM}$ [GeV]
& $M_\text{LR}$ [GeV] &  $M_\text{PS}$ [GeV] \\[1.5ex]
   B19  & 0 & 0 & 3 & 3 & 0 & 1 & $3\times10^{10}\,-\,3\times10^{13}$ &
$31\times10^{13}\,-\,3\times10^{15}$ & $2\times10^{15}\,-\,8\times10^{15}$\\
   B115  & 1 & 1 & 3 & 3 & 0 & 1 & $1\times10^{11}\,-\,3\times10^{13}$ &
$3\times10^{13}\,-\,3\times10^{15}$ & $2\times10^{15}\,-\,6\times10^{15}$\\
   B199 & 3 & 3 & 1 & 3 & 0 & 1 & $7\times10^{11}\,-\,1\times10^{12}$ &
$2\times10^{13}\,-\,2\times10^{14}$ & $3\times10^{15}\,-\,4\times10^{15}$\\
\hline
 \end{tabular}
\end{center}
\vspace{\baselineskip}

\subsubsection*{Class: C (supersymmetric)}

\begin{center}\footnotesize
  \begin{tabular}{|c|cccccc|ccc|}
  \hline &&&&&&&&&\\[-2ex]
  Model & \#h & \#F & \#$\Phi$ & \#$\Sigma$ & \#E & \#T & $M_\text{MSSM}$ [GeV]
&$M_\text{PS}$ [GeV] &  $M_\text{QL}$ [GeV] \\[1.5ex]
   C43  & 1 & 1 & 3 & 3 & 0 & 1 & $4\times10^{10}\,-\,3\times10^{12}$ &
$7\times10^{15}\,-\,3\times10^{16}$ & $7\times10^{15}\,-\,7\times10^{17}$ \\ 
   C53 & 1 & 3 & 1 & 1 & 3 & 1& $2\times10^{12}\,-\,1\times10^{14}$ &
$3\times10^{16}\,-\,6\times10^{16}$ & $1\times10^{17}\,-\,2\times10^{18}$\\
   C127 & 3 &3 & 1 & 3 & 0 & 1 & $4\times10^{11}\,-\,4\times10^{12}$ &
$5\times10^{15}\,-\,8\times10^{15}$ & $5\times10^{15}\,-\,5\times10^{17}$\\
\hline
 \end{tabular}
\end{center}
\vspace{\baselineskip}

\subsubsection*{Non-supersymmetric}

\begin{center}\footnotesize
  \begin{tabular}{|c|cccccc|ccc|}
  \hline &&&&&&&&&\\[-2ex]
  Model & \#h & \#F & \#$\Phi$ & \#$\Sigma$ & \#E & \#T & $M_\text{SM}$ [GeV]
&$M_\text{PS}$ [GeV] &  $M_\text{QL}$ [GeV] \\[1.5ex]
  E46 & 0 & 1 & 1 & 3 & 0 & 0 & $4\times10^{12}$ & $3\times10^{14}$ & ---\\
  E73 & 0 & 3 & 1 & 1 & 0 & 0 & $9\times10^{10}$ & $6\times10^{14}$ & ---\\
  E87 & 0 & 3 & 1 & 3 & 1 & 3 & $9\times10^{10}$ & $6\times10^{14}$ & ---\\
  E89 & 0 & 3 & 1 & 3 & 3 & 1 & $9\times10^{10}$ & $6\times10^{14}$ & ---\\
  C193 & 1 & 3 & 1 & 3 & 0 & 1 & $6\times10^{10}\,-\,3\times10^{13}$ &
$2\times10^{13}\,-\,1\times10^{14}$ & $3\times10^{16}\,-\,4\times10^{17}$\\
\hline
 \end{tabular}
\end{center}
\vspace{\baselineskip}

\end{document}